\documentclass[useAMS,usenatbib]{mn2e}

\usepackage{psfig, epsf, epsfig}
\usepackage{graphicx}
\usepackage{amssymb}
\usepackage{pifont} 

\title[Globular cluster formation]{
Globular cluster formation with multiple stellar populations:
A comprehensive overview of a star-cloud interaction scenario}
\author[K. Bekki,  M. McKenzie, ...]
{Kenji Bekki${}^1$\thanks{E-mail:
kenji.bekki@uwa.edu.au} and
Madeleine McKenzie${}^{2}$\thanks{NASA Hubble Fellow} \\
${}^1$ICRAR M468
The University of Western Australia
35 Stirling Hwy, Crawley
Western Australia 6009, Australia \\
${}^2$
Carnegie Science institution,
813 Santa Barbara Street, Pasadena, CA 91101, USA
}

\begin{document}

\date{Accepted, Received 2005 February 20; in original form }

\pagerange{\pageref{firstpage}--\pageref{lastpage}} \pubyear{2005}

\maketitle

\label{firstpage}

\begin{abstract}

We present a new scenario of globular cluster (GC) formation with multiple
stellar populations (MPs) in which  both the first and second populations
(1P and 2P, respectively) of stars form from giant molecular clouds
(GMCs) polluted by asymptotic giant branch (AGB)  stars within and around the GMCs.
Unlike previous GC formation scenarios with AGB stars being the primary polluters,
the new scenario alleviates tensions with the mass-budget and dilution-timing problems
The principal results based on idealized analytic models of the formation scenario
are as follows.
The observed fraction of 1P stars and the helium abundance
spreads between the 1P and 2P as a function of GC masses can be well reproduced.
The modelled GCs show O$-$Na, C$-$N, and Mg$-$Al anticorrelations and
Si$-$Al, $^{25}$Mg/Mg, $^{26}$Mg/Mg correlations, which are qualitatively
similar to observations.
The observed Mg-K anticorrelation can be reproduced,
only if super-AGB stars make a significant contribution
to chemical enrichment within GMCs.
The lack of correlations of Li abundances with [Na/Fe] and [Al/Fe] can be reproduced,
only if about 20\% of the polluting AGB stars produce Li-rich ejecta, 
which disfavours scenarios with polluters incapable of Li production.
Iron-complex, Type-II GCs with two distinct populations having different chemical abundances
of $s$-process elements
can be formed through merging of two GCs formed from two GMCs within a host 
dwarf galaxy  at different epochs.
The new scenario predicts
young massive clusters formed in galaxy environments with surface star formation
rate densities well below $1 {\rm M}_{\odot}$ yr$^{-1}$ kpc$^{-2}$ are unlikely to evolve
into GCs with MPs. 
It also predicts  low $^{12}$C/$^{13}$C ratios of 2P ($\approx 5$),
a [Na/Fe]$-$[F/Fe] anticorrelation, and 
P-rich star formation with [P/Fe]$>0.5$ and [N/Fe]$>0.5$.
These predictions are tested against more than 30 observed properties of GCs with MPs,
representing one of the most comprehensive observational benchmarks against a specific GC formation scenario to date.

\end{abstract}


\section{Introduction}
Globular cluster (GC) formation represents a crucial 
crossroad between stellar nucleosynthesis, star formation feedback, and galaxy 
evolution. For instance, mass-loss processes within star-forming gas clouds dictate 
whether newly born clusters remain tightly bound as GCs or ultimately disintegrate 
(e.g., Hills 1980; Geyer \& Burkert 2001; Baumgardt \& Kroupa 2007).
Furthermore, a remarkably high pressure of interstellar medium in galaxies  ($P \ge 10^5 {\rm k}_{\rm B}$)
is a necessary condition for GC formation
within isolated and merging  galaxies 
(e.g., Elmegreen \& Efremov 1997; Bekki et al. 2002;  Pfeffer et al. 2024).
The observed correlations and anticorrelations between
chemical abundances of various elements in GCs,
such as C-N anticorrelations (e.g., Norris \& Cottrell 1979; Suntzeff 1981; Lardo et al. 2014),
O-Na anticorrelations (e.g., Kraft et al. 1993; Gratton et al. 2001; Carretta et al. 2009a,b, C09a, C09b),
and Mg-K anticorrelations
(e.g., Mucciarelli et al. 2012; Alvarez Garay et al. 2022, 2025)
can be used to provide theoretical constraints  on the possible nucleosynthetic sites
producing these elements 
(e.g., Ventura et al. 2012, V12;  Iliadis et al. 2016, I16).

One of the most remarkable breakthroughs
in  GC research  over the last 25 years is that
almost all  of the Galactic GCs harbor more than one stellar population
 (i.e., multiple stellar populations,
``MPs'';  Bedin et al. 2004; Piotto et al. 2005; see Gratton, Sneden, \& Carretta 2004 and
Gratton et al. 2019, G19 for reviews).
Spectroscopic studies of GC stars have also revealed  anticorrelations between
light elements, such as O-Na, C-N, and  Mg-Al anticorrelations
(e.g., Ramirez \& Cohen 2002;  C09a; Yong et al. 2009; Nataf et al. 2019; M\'esz\'aros et al. 2020; ME20).
It is commonly assumed that GC stars with chemical abundances similar to those of  the 
Galactic halo field stars with similar metallicities are ``first generation'' of ``first population'' stars
(referred to as ``1P'' from now on) formed earlier whereas
Na-rich (O-depleted etc) stars are ``second generation'' or ``second population''
stars (2P) formed later.
Consequently, numerous observational and theoretical studies
have been conducted to reveal the physical properties of
1P and 2P stars of the Galactic GCs
(see G19;  Bastian \& Lardo 2019, BL19 for recent reviews
on this issue).

The physical  
properties of GCs hosting MPs, as revealed by these observations, 
have provided valuable constraints on the GC formation scenarios.
For example, the fraction of 1P stars is observed to be anticorrelated with the present-day
and initial GC masses (e.g., C09a; Milone \& Marino 2022,  MM22), which implies 
that 2P formation is more efficient in more massive GCs.
Clusters exhibiting MPs are broadly divided
into two categories
(e.g., Marino et al. 2015; Milone et al. 2017), Type I and II, depending on
whether they have large [Fe/H] spreads of more than 0.05 dex
(Type II) or not (Type I),
and Fe-rich  stars  in Type II GCs
also have  significantly enhanced abundances of $s$-process elements 
(``$s$-rich'' population).
The observed abundance patterns of these $s$-rich populations are thought to be consistent
with  pollution by low-mass AGB stars (e.g., Shingles et al. 2014; McKenzie et al. 2022, 2024).
Because existing theories of cluster formation have yet to 
fully explain this diverse array of empirical properties, this paper addresses these phenomena in a comprehensive manner.

\begin{table}
\centering
\begin{minipage}{90mm}
\caption{ Description of physical meanings for symbols
often used in the present study. A symbol with ``0'' in its subscript means
the initial value of a physical property. For example, $M_{\rm gc, 0}$ is
the initial stellar mass of a GC  at its birth ($M_{\rm gc}$ is the present-day one).
Likewise, $F_{\rm 1P,0}$, which is not shown in this table, is the initial
1P fraction.
1P and 2P instead of FG (1G) and SG (2G)  are used  to represent first and second populations
of stars formed within GCs just for convenience  in the present study.}
\begin{tabular}{ll}
Symbol &  Physical meaning  \\
$M_{\rm gc}$  &  GC mass  \\
$M_{\rm gc,0}$  &  Initial GC mass  \\
$M_{\rm 1P}$  &  Total mass of 1P stars  \\
$M_{\rm 2P}$  &  Total mass of 2P stars  \\
$M_{\rm gmc}$  &  Initial GMC mass  \\
$M_{\rm g}$  &  Total mass of pristine gas (within a GMC)  \\
$M_{\rm agb}$  &  Total mass of AGB stars (around a GMC)  \\
$M_{\rm ej}$  &  Total mass of AGB ejecta  \\
$M_{\rm o}$  &  Original  mass required to form a GC with MPs  \\
$M_{\rm ns}$  &  Total mass of new stars around/within a GMC  \\
$t_{\rm life}$  &  GMC lifetime \\
$F_{\rm 1P}$  &  1P fraction  ($=M_{\rm 1P}/M_{\rm gc}$) \\
$F_{\rm 2P}$  &  2P fraction ($=1-F_{\rm 1P}$)   \\
$F_{\rm dil}$  & Dilution factor ($=M_{\rm g}/M_{\rm ej}$   \\
$F_{\rm b}$  & Mass budget factor ($=M_{\rm o}/M_{\rm gc}$)  \\
$R_{\rm s}$  &  Mass ratio of new stars to GMC mass   \\
$\rho_{\rm agb}$  & Mass density of AGB stars    \\
$\rho_{\rm agb,th}$  & Threshold $\rho_{\rm agb}$ for GCs with MPs   \\
$\Sigma_{\rm agb}$  & Surface mass density of AGB stars    \\
$\Sigma_{\rm SFR}$  & Star formation rate (SFR) surface density   \\
$\Sigma_{\rm SFR,th}$  & Threshold $\Sigma_{\rm SFR}$ for GCs with MPs  \\
$\delta Y$  &  Helium abundance difference between 1P and 2P  \\
$\alpha$  & IMF slope  \\
$\epsilon_{\rm sf}$  & Star formation efficiency  \\
$m_{\rm agb}$  & Total mass of an AGB star   \\
$m_{\rm agb,l}$  & Lower mass cut-off of polluting AGB stars   \\
$m_{\rm agb,u}$  & Higher mass cut-off of polluting AGB stars   \\
\end{tabular}
\end{minipage}
\end{table}

\begin{figure*}
\psfig{file=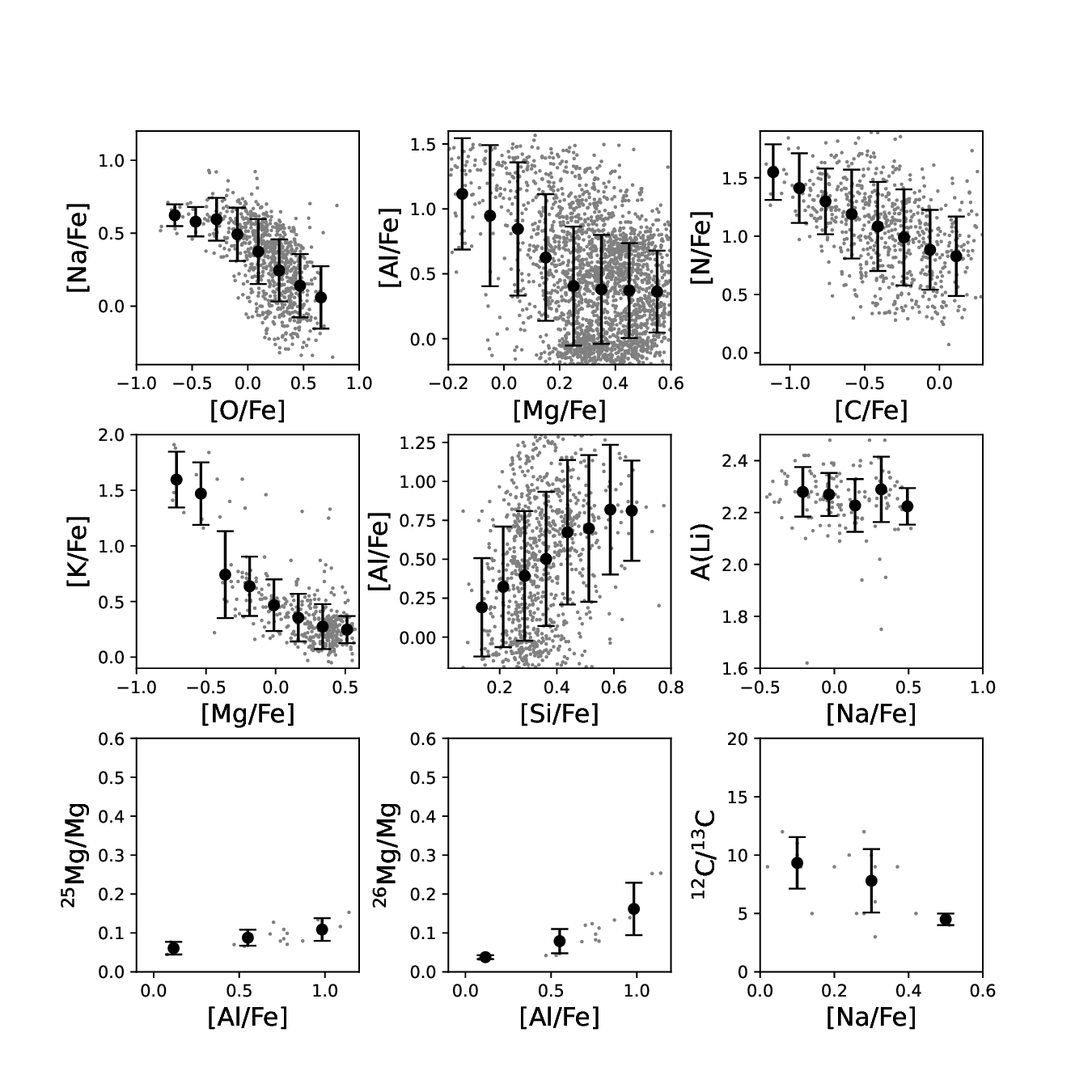,width=18cm} 
\caption{
Observed correlations and anticorrelations between various chemical abundances of GC stars
that this paper discusses in the context of the SCI scenario.
The mean and 1$\sigma$ dispersion of chemical abundances at different abundance
bins are shown by
filled circles and error bars, respectively.
The details of these observational data sets are given in the main text.
{\it These are suggested to be the nine benchmark tests for any theory of GC formation in this paper.}
}
\label{Figure. 1}
\end{figure*}

\begin{figure*}
\includegraphics[width=18.0cm]{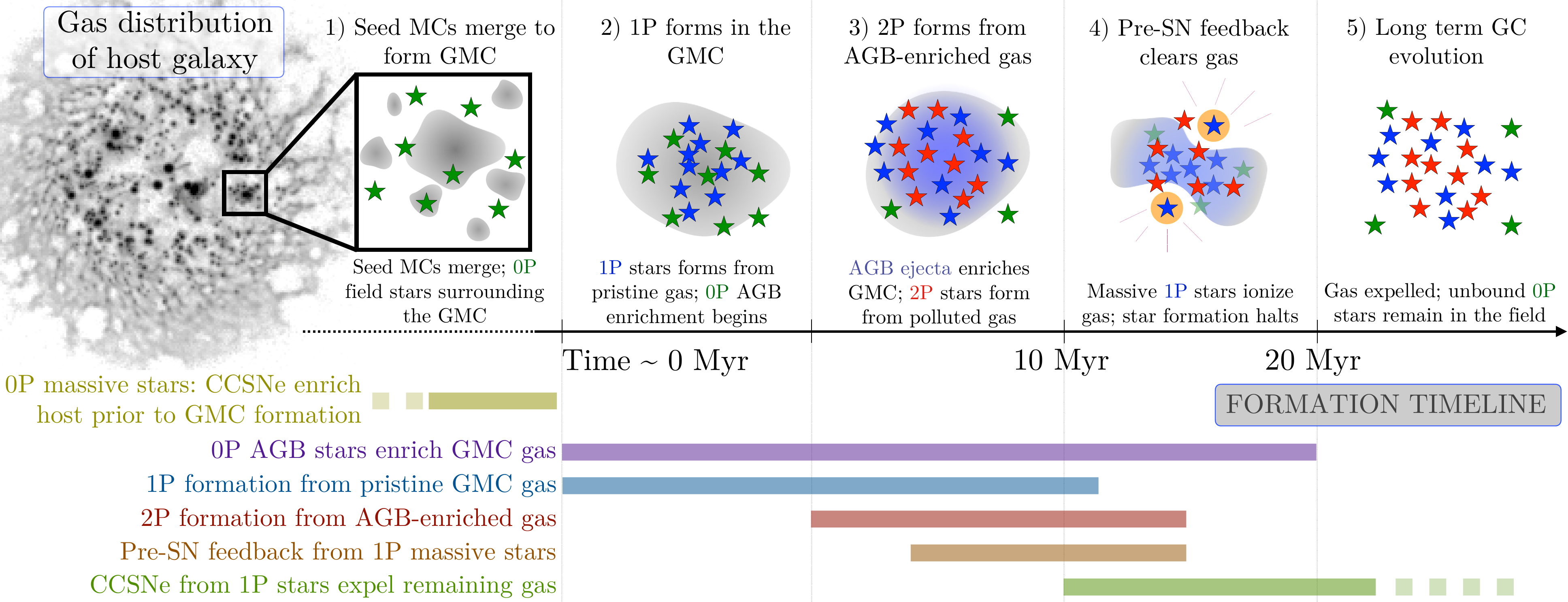}
\caption{
An illustration of the SCI scenario.
In this scenario,  seed small molecular clouds (MCs) can grow through merging and accretion
of other small MCs to become a GMC (Step 1). New stars (``1P'') can form
during this growth, and the growing GMC can be also polluted  
by pre-existing field stars (``0P'')  within and around GMCs (2).
As star formation accelerates within the GMC,  
new stars  (``2P'') start to form from the polluted (thus
enriched) gas  (3).
Soon after the formation of a cluster of massive stars ($m > 10 {\rm M}_{\odot}$),
their strong stellar feedback ionizes the gas
and subsequently truncates further star formation and expel all remaining gas
from the GMC (4). Finally, a massive star cluster (SC) with MPs forms, which can be later identified as
as a GC with MPs. Thus, MPs can be formed before stellar feedback effects (winds and core collapse supernovae, 
CCSNe) of clusters of massive stars completely
destroy SC-hosting GMCs: the chemical enrichment timescale of a GMC corresponds to the lifetime ($t_{\rm life}$).
Although this paper 
focuses exclusively on the chemical enrichment by pre-existing intermediate-mass
stars, other stars can possibly enrich GMCs.
For example, if ONe novae pollute forming GMCs, then
stars formed from pristine gas mixed with nova ejecta can have rather high [P/Fe] ($>1$),
which can be observed as P-rich stars.
Mergers between neutron stars (NSs) around GMCs can introduce large scatters in the chemical abundances
of $r$-process elements within GMCs.
The number densities of pre-existing
intermediate-mass stars within and around GMCs can determine
whether or not SCs can finally have MPs in the scenario.
GMCs in the present-day galaxies are unlikely to host SCs with MPs due to 
very low densities of pre-existing stars around and within them.
}
\label{Figure. 2}
\end{figure*}

The most popular scenario for globular cluster (GC) formation with MPs posits that 2P stars form from pristine gas 
mixed with (diluted by) gaseous ejecta from ``polluters'' originating from the 1P. Several 
candidate polluters have been proposed, such as Asymptotic Giant Branch (AGB) stars 
(e.g., Cottrell \& Da Costa 1981; D'Antona \& Caloi 2004), massive stars (e.g., Prantzos \& Charbonnel 2006), fast-rotating massive stars (e.g., Decressin et al. 2007), supermassive stars (e.g., Denissenkov \& Hartwick 2014), very massive stars (e.g., Gieles et al. 2025), and massive interacting binaries (e.g., de Mink et al. 2009; Bastian et al. 2013; Nguyen \& Sills 2024).

Among the GC formation frameworks involving these polluters, the ``AGB scenario''—where massive AGB stars ($m \ge 4~{\rm M}_{\odot}$) act as the primary polluters—has been the most extensively investigated using 
one-zone chemical evolution models or idealized analytical ones 
(e.g., Fenner et al. 2004; Conroy \& Spergel 2010; D'Ercole et al. 2010, D10) and numerical simulations of cluster formation (e.g., D'Ercole et al. 2008, D08; Bekki 2010, 2011, 2019a; 
Calura et al. 2019; McKenzie \& Bekki 2021a,b; Lacchin et al. 2022, 2026; Yaghoobi et al. 2022).

The AGB framework, referred to here as the ``classic AGB scenario'' for convenience, assumes that: (i) original GCs 
initially consist entirely of 1P stars, 
and (ii) pristine gas later mixes with the gaseous ejecta of these 1P AGB stars to ultimately form the 2P population (e.g., D08). These previous investigations have provided specific, verifiable predictions regarding differences in binary star fractions (e.g., Vesperini et al. 2011), spatial distributions (e.g., Bekki 2011; Vesperini et al. 2013, 2021; Mastrobuono-Battisti et al. 2021), and global rotation or anisotropy in stellar velocity dispersions (e.g., Bekki 2010; Mastrobuono-Battisti et al. 2016) between the 1P and 2P stars of GCs.

Bekki et al. (2007, B07) demonstrated that if GCs form from the interstellar gas of host dwarf galaxies that has been enriched by the field AGB stars of those hosts, they can naturally display internal star-to-star abundance variations. In their models, both 1P and 2P stars can form during a continuous star-formation event, which bypasses the requirement for the exceptionally massive initial stellar systems invoked by the classic AGB scenario.

Using high-resolution simulations of dwarf galaxies containing GCs, Maxwell et al. (2014) first demonstrated that if GCs are located within the central regions of dwarf galaxies, they can accrete both pristine interstellar gas and gaseous ejecta from field AGB stars. This processed material subsequently triggers a secondary episode of star formation from the accreted gas, yielding multiple populations. These two pioneering studies clearly demonstrated that field AGB stars can play a vital role in the formation of GCs with MPs.

Nonetheless, the classic AGB scenario faces two potentially severe challenges. The first is that the primordial stellar systems, consisting initially of only 1P stars, must be at least $\approx 10$ times more massive than the present-day GC masses (widely known as the ``mass-budget problem''). Using $N$-body simulations tracking the long-term dynamical evolution of Galactic GCs, Webb \& Leigh (2015) demonstrated that: (i) proto-GCs can typically be a factor of 4.5 more massive than present-day GCs, and (ii) the initial masses of at least three specific GCs were up to 10 times larger than their current masses.

Recent studies exploring the potential initial masses of GCs have also shown that the mass fractions of 1P and 2P stars lost from clusters can be highly diverse (e.g., Leitinger et al. 2023). Therefore, the mass-budget problem might not be an insurmountable issue for the classic AGB framework.

The second challenge is the dilution timing problem: the dilution of AGB ejecta by pristine gas must occur at the exact epoch when massive AGB stars are actively shedding their stellar winds into the intracluster environments. This problem has yet to be fully resolved, although recent hydrodynamical simulations of GC formation and evolution demonstrate that GCs can efficiently accrete interstellar gas (McKenzie \& Bekki 2021a), and that secondary star formation from AGB ejecta mixed with accreting gas can successfully take place within pre-existing GCs (e.g., Calura et al. 2019).

If star-forming giant molecular clouds (GMCs) encounter nearby field AGB stars or
intermediate-age SCs, 
the dilution of AGB ejecta with pristine gas immediately following wind ejection becomes inevitable. 
Furthermore, if GMCs assemble through the accretion and merger of small, low-mass molecular clouds (e.g., Kobayashi et al. 2017), and if these small precursor clouds already harbor low- and intermediate-mass stars—akin to populations observed in environments like the Taurus Molecular Cloud (e.g., Cohen \& Kuhi 1979)—the resulting GMC will naturally contain AGB stars within and around its volume.

Consequently, new star clusters (SCs) forming inside GMCs polluted by AGB stars will display distinct chemical abundances dictated by the local degree of enrichment. This AGB–GMC interaction naturally resolves the dilution timing problem inherent to the classic AGB scenario, although the detailed mechanism of star formation from pristine GMC gas mixed with AGB winds warrants further investigation.

Because the AGB stars polluting a GMC originate from older stellar generations formed earlier within the host galaxy, their collective mass can be substantially larger than that of the newborn GC inside the cloud. Since most of these polluting stars are not gravitationally bound to the GMC (or the newborn cluster), they disperse into the field after the GMC is disrupted by stellar feedback. As a result, 
this framework avoids the mass-budget problem entirely. This GC formation model, based on the physical interaction between GMCs and stellar populations (specifically intermediate-mass 
AGB stars here), is termed the ``star-cloud interaction (SCI) scenario'' to clearly distinguish it 
from the classic AGB framework.
Within the framework presented here, we adopt the definition that stellar systems only qualify as GCs if 
they host MPs arising from the chemical enrichment processes described above. 
Massive stellar systems that form through similar pathways, but in environments 
where the number density of AGB stars  falls below a threshold value for MP formation,
and therefore lack MPs, are designated as star clusters (SCs) rather than GCs.

The primary objective of this study is to present a comprehensive overview of the SCI scenario by exploring the results of idealized analytical models. We focus specifically on: (i) the correlations of present-day GC masses ($M_{\rm gc}$) with the number fractions of 1P stars ($F_{\rm 1P}$) and internal helium abundance spreads ($\delta Y$); (ii) the correlations and anticorrelations among various light-element abundances (e.g., the O–Na and Mg–K anticorrelations); (iii) the varying degrees of lithium depletion across different clusters; (iv) the physical origin of the Type~I and Type~II GC dichotomy; and (v) the potential minimum metallicity threshold ($[{\rm Fe}/{\rm H}]_{\rm min}$) required for the formation of GCs with MPs.

Fig. 1 summarizes the nine observed chemical abundance relations whose physical origins are investigated in detail using this new scenario. Because this work relies on analytical modeling rather than full numerical hydrodynamics, we do not address the observed structural and kinematic differences between 1P and 2P stars, even though such properties are vital for testing cluster formation theories. These crucial structural and kinematic aspects will be addressed in future papers.

The plan of this paper is organized as follows. In \S 2, we present the overview of the SCI scenario and explain its advantages over the classic model in accounting for various empirical properties of GCs. We describe the analytical framework capable of predicting the properties of GCs across different mass regimes ($M_{\rm gc}$) in \S 3. The primary results and their underlying physical mechanisms are presented in \S 4. The broader physical implications of these findings for observational studies of GCs with MPs in both the local and high-redshift universe are discussed in \S 5. Finally, we summarize our key conclusions and model predictions in \S 6.

\begin{figure*}
\psfig{file=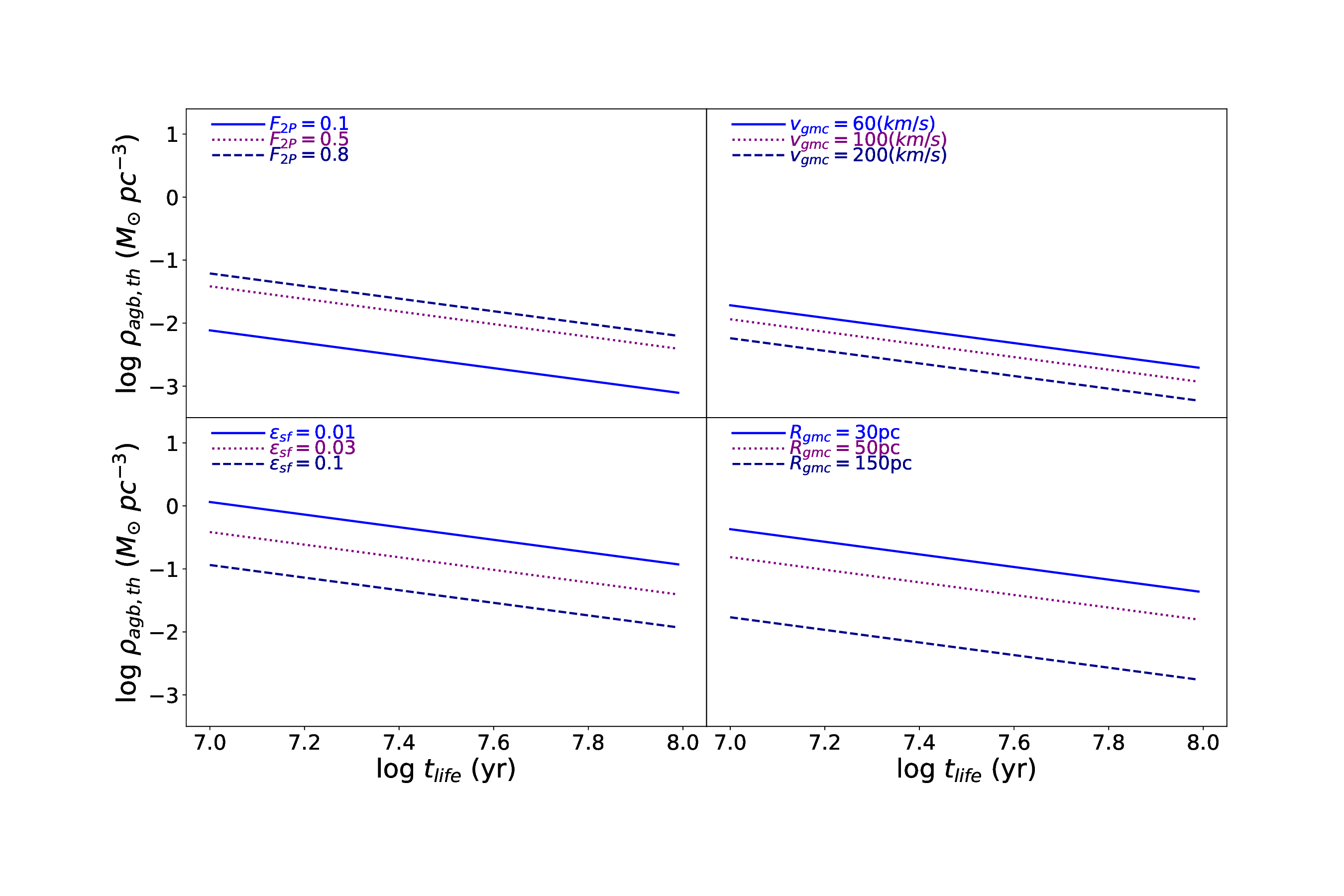,width=18cm} 
\caption{
Threshold mass densities of AGB stars ($\rho_{\rm agb, th}$) above which
GCs with MPs can  finally form from GMCs in gas-rich galaxies as a function of GMC lifetimes ($t_{\rm life}$)
for different model parameters: dependence on $F{\rm 2P}$ in the upper left panel,
on $v_{\rm gmc}$ in the upper right,  on $\epsilon_{\rm sf}$ in the lower left,
and on $R_{\rm gmc}$ in the lower right).
}
\label{Figure. 3}
\end{figure*}

\begin{figure}
\psfig{file=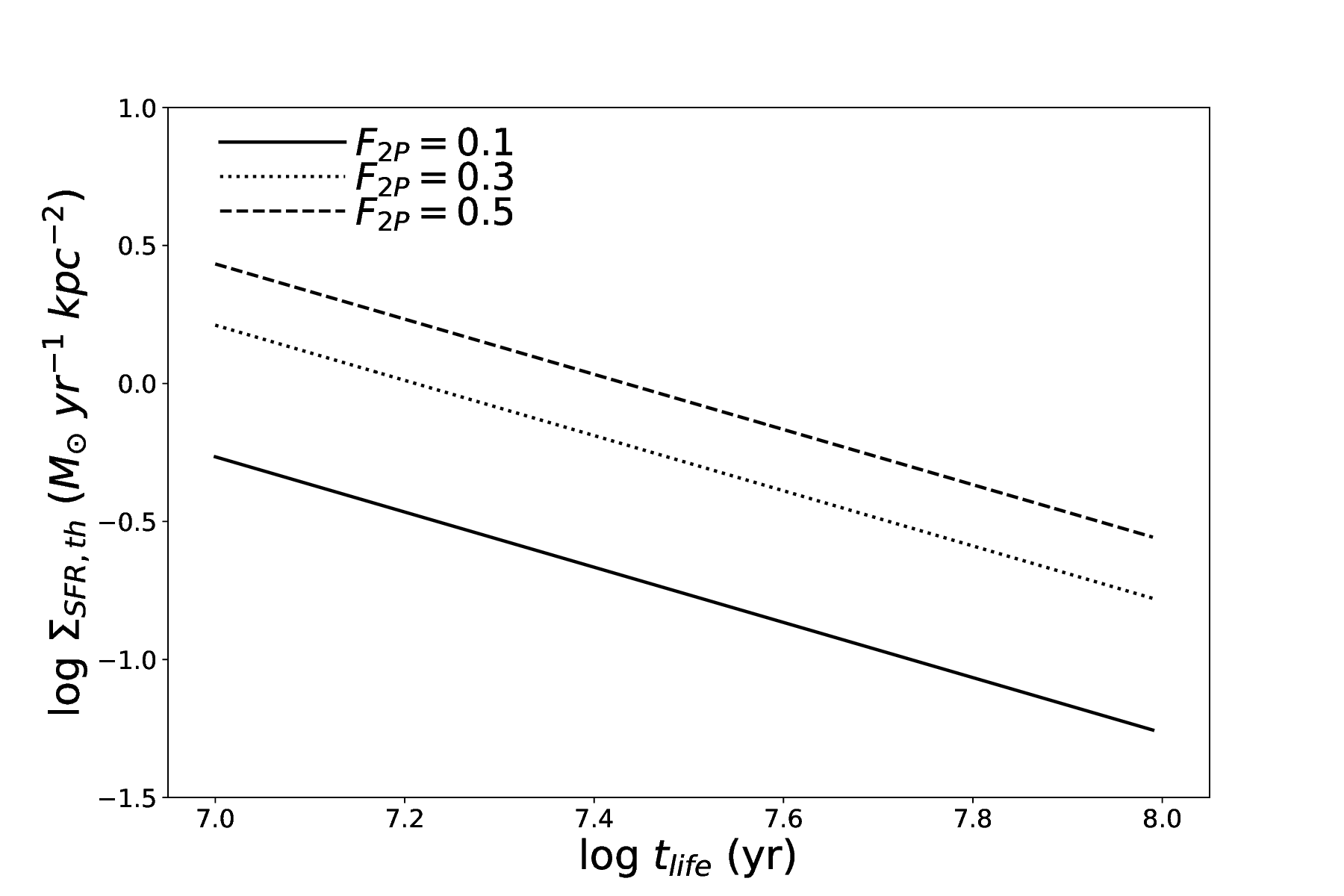,width=8.5cm} 
\caption{
Threshold SFR  densities ($\Sigma_{\rm SFR, th}$) above which
GCs with MPs can finally form from GMCs in gas-rich galaxies 
as a function of $t_{\rm life}$ for $F_{\rm 2P}$=0.1 (solid), 0.3 (dotted),
and 0.5 (dashed).  Clearly, 
$\Sigma_{\rm SFR, th}$ should be higher for higher $F_{\rm 2}$ for a given $t_{\rm life}$.
}
\label{Figure. xi43}
\end{figure}

\section{Overview of the new scenario}

\subsection{New scenario}

\subsubsection{AGB stars within and around GMCs as polluters}

Harris \& Pudritz (1994) first discussed how GCs are formed from high-density cores of supergiant molecular clouds (`SGMCs'') to explain the observed properties of GCs, such as GC numbers as a function of GC masses. We also consider that GCs can be formed from GMCs in gas-rich dwarf galaxies in the SCI scenario. Massive gas clouds that can form GCs with typically [Fe/H]$\approx -1.6$ could not be rich in molecular hydrogen due to the low metallicity and dust abundances of the clouds. Nevertheless, 
we use the term ``GMCs'' for GC-forming gas clouds just for convenience in this paper.
In the new scenario, star formation is assumed to continue within GMCs during their lifetimes ($t_{\rm life}$), which can typically be [1-3]$\times 10^7$ yr (e.g., Kawamura et al. 2009; Kruijssen et al. 2019; Chevance et al. 2020).
Clusters of massive stars with masses ($m$) of more than $10 {\rm M}_{\odot}$ are assumed to quench star formation through their pre-supernova (pre-SN) feedback effect soon after their formation (e.g., Kim et al. 2021).
Thus, it is assumed in the scenario that gaseous ejecta from very massive and supermassive stars, fast-rotating massive interacting binaries, and colliding stars cannot be converted into new stars after the pre-SN feedback effect starts to clear GMC gas.

Various stellar sources surrounding GMCs, such as AGB and super AGB stars (sAGB), 
CO and ONe novae, CCSNe, 
Type Ia supernovae (SNIa), etc., can possibly interact with GMCs to eventually influence the chemical and dynamical evolution of GMCs through their stellar winds and feedback effects.
We here focus exclusively on interactions between AGB stars and GMCs among various star-cloud interactions.
AGB and sAGB stars with masses ($m_{\rm agb}$) ranging from $\approx 4 {\rm M}_{\odot}$ to $\approx 10 {\rm M}_{\odot}$ are assumed to be the major polluters that can enrich the pristine gas of growing GMCs in the new scenario.
The lower and upper mass cutoffs of $m_{\rm agb}$ ($m_{\rm agb, l}$ and $m_{\rm agb, u}$, respectively) could be different in GMCs with different physical properties.
Hydrodynamical interactions of the much less energetic winds from these AGB stars with GMCs 
can cause efficient mixing with (and dilution of) the pristine gas of the GMCs to form new (2P) stars from the mixed gas.

In present-day luminous star-forming galaxies, $\rho_{\rm agb}$
is too low for AGB-GMC interactions to significantly alter GMC chemical abundances.
However, the new scenario assumes that $\rho_{\rm agb}$ can be high enough to alter the original chemical abundances of GMCs in forming compact dwarf galaxies.
An outcome of this scenario is  that massive star clusters formed in galaxies with lower $\rho_{\rm agb}$ do not form MPs.
Fig. 2 illustrates the key ingredients of the new scenario, and Table 1 summarizes the physical meanings of the symbols often used in discussing the scenario.
We can make a rough estimation of the $\rho_{\rm agb}$ required for GC formation with MPs ($\rho_{\rm agb, th}$) as follows.
We first assume that (i) 2P stars are formed from AGB ejecta mixed with GMC gas, (ii) the total mass of 2P stars ($M_{\rm 2P}$) is a significant fraction of the total mass ($M_{\rm gc}$), and (iii) all young massive AGB stars approaching within GMC sizes ($R_{\rm gmc}$) can pollute GMCs within GMC lifetimes ($t_{\rm life}$).
The present-day fraction of 2P stars ($F_{\rm 2P}$) in a GC is a key parameter in this $\rho_{\rm agb, th}$:
\begin{equation}
F_{\rm 2P} = \frac{ M_{\rm 2P} }{ M_{\rm gc} }.
\end{equation}

We assume that a GMC can capture stellar winds from AGB stars within a cylindrical volume of $\pi R_{\rm gmc}^2v_{\rm gmc}t_{\rm life}$, where $v_{\rm gmc}$ is the 3D velocity of the GMC.
The total mass of AGB ejecta ($M_{\rm ej}$) mixed with the pristine gas of a GMC is therefore estimated as follows:
\begin{equation}
M_{\rm ej} = \pi f_{\rm ej} R_{\rm gmc}^2v_{\rm gmc}t_{\rm form}  \rho_{\rm agb},
\end{equation}
where $f_{\rm ej}$ is the mass fraction of AGB ejecta in AGB stars.
We here consider that a fraction ($f_{\rm lost}$) of the initial GC mass ($M_{\rm gc, i}$) can be lost due to internal two-body relaxation effects and external tidal stripping.
If AGB ejecta mixed with pristine gas with a total mass of $M_{\rm g}$ is converted into new stars with a star formation efficiency ($\epsilon_{\rm sf}$), then $M_{\rm ej}$ can be related to $M_{\rm gc}$ as follows:
\begin{equation}
F_{\rm 2P}M_{\rm gc} = \epsilon_{\rm sf} (1+F_{\rm dil})(1-f_{\rm lost}) M_{\rm ej},
\end{equation}
where $F_{\rm dil}$ is the ratio of pristine gas mass to AGB ejecta mass (the ``dilution factor'') and is defined as:
\begin{equation}
F_{\rm dil} =\frac{ M_{\rm g} }{ M_{\rm ej} } .
\end{equation}
Therefore, $F_{\rm 2P}$ depends on multiple parameters including $\rho_{\rm agb}$:
\begin{equation}
F_{\rm 2P} = \frac{ \pi \epsilon_{\rm sf} (1+F_{\rm dil})(1-f_{\rm lost})
f_{\rm ej} R_{\rm gmc}^2v_{\rm gmc}t_{\rm life}  \rho_{\rm agb} }
{ M_{\rm gc} }.
\end{equation}
Thus, it is possible that $F_{\rm 2P} \approx 0$ in GMCs formed within galaxies with very low $\rho_{\rm agb}$.
Using the above equations, the threshold $\rho_{\rm agb, th}$ can be derived for a given set of these variables (e.g., $M_{\rm gc}$):
\begin{equation}
\rho_{\rm agb,th}=\frac{ F_{\rm 2P}M_{\rm gc} }{ \pi \epsilon_{\rm sf}
f_{\rm ej} R_{\rm gmc}^2v_{\rm gmc} t_{\rm life} (1+F_{\rm dil})(1-f_{\rm lost}) }.
\end{equation}
Thus, $\rho_{\rm agb,th}$ depends both on GMC properties (e.g., $R_{\rm gmc}$) and galactic dynamical parameters (e.g., $v_{\rm gmc}$).
For example,
\begin{equation}
\rho_{\rm agb,th}= 2.0 \times 10^{-2}  {\rm M}_{\odot} {\rm pc}^{-3},
\end{equation}
for a fiducial set of variables: $ M_{\rm gc}=2\times 10^5 {\rm M}_{\odot} $, $ F_{\rm 2P}=0.5$, $\epsilon_{\rm sf}=0.3$, $F_{\rm dil}=0.5$, $f_{\rm lost}=0.5$, $f_{\rm ej}=0.9$, $R_{\rm gmc}=100$ pc, $v_{\rm gmc}=30$ km s$^{-1}$, and $t_{\rm life}=2 \times 10^7$ yr.
As shown in Fig. 3, $\rho_{\rm agb, th}$ depends on the parameters of GCs, GMCs, and their host galaxies.
It is clear that $\rho_{\rm agb, th}$ is larger for shorter $t_{\rm life}$, larger $F_{\rm 2P}$, higher $v_{\rm gmc}$, lower $\epsilon_{\rm sf}$, and smaller $R_{\rm gmc}$.
Fig. 3 clearly demonstrates that galaxies need to have more than $10^{-3}$ ${\rm M}_{\odot}$ ${\rm pc}^{-3}$ for their SCs to have MPs.
If a luminous disk galaxy like our own Milky Way (MW)
has a disk radius ($R_{\rm d}$) of 15 kpc, a disk thickness ($z_{\rm d}$) of 0.2 kpc, a SFR of $1 {\rm M}_{\odot} {\rm yr}^{-1}$, and an AGB mass fraction ($f_{\rm agb}$) of 0.085 among all stars, then the mean $\rho_{\rm agb}$ of the galaxy is only $6 \times 10^{-5}$ ${\rm M}_{\odot}$ ${\rm pc}^{-3}$, which is too low for the galaxy's SCs to have MPs.
A LMC-type dwarf galaxy with $R_{\rm d}=5$ kpc, $z_{\rm d}=0.2$ kpc, and SFR$=0.1$ ${\rm M}_{\odot} {\rm yr}^{-1}$ can have only $8 \times 10^{-5}$ ${\rm M}_{\odot}$ ${\rm pc}^{-3}$.
Even if the SFRs of these galaxies are enhanced by a factor of 10, $\rho_{\rm agb}$ would still be less than $10^{-3}$ ${\rm M}_{\odot}$ ${\rm pc}^{-3}$.
These low $\rho_{\rm agb}$ values imply that the present-day MW and LMC are highly unlikely to have SCs with MPs.
This $\rho_{\rm agb,th}$ is currently unable to be derived directly from the observed surface brightness and colors of a star-forming galaxy.
We accordingly consider that the surface density of a star formation rate ($\Sigma_{\rm SFR}$) in a star-forming galaxy can be used to discuss whether its massive young star clusters (SCs) can possess MPs.
If all young stars are formed in the thin disk of a galaxy, then the surface mass density of young AGB stars with masses ranging from $4 {\rm M}_{\odot}$ to $10 {\rm M}_{\odot}$ can be approximated using $\Sigma_{\rm SFR}$:
\begin{equation}
\Sigma_{\rm agb}= f_{\rm agb}t_{\rm agb}\Sigma_{\rm SFR},
\end{equation}
where $f_{\rm agb}$ is the mass fraction of the AGB stars and $t_{\rm agb}$ is a timescale within which all stars with $4 \le m/{\rm M}_{\odot} \le 10$ can evolve into AGB phases.
Although the SFR of a galaxy could rapidly change 
on a timescale of an order of $10^8$ yr, we here ignore such time variations.
Using $\Sigma_{\rm agb}=z_{\rm d}\rho_{\rm agb}$ (where $z_{\rm d}$ is the disk thickness of a galaxy) and the above two equations, the threshold SFR density ($\Sigma_{\rm SFR, th}$) over which GCs with MPs can be formed in a galaxy can be estimated:
\begin{equation}
\Sigma_{\rm SFR, th}= \frac{ F_{\rm 2P} M_{\rm gc} z_{\rm d} }{ \pi \epsilon_{\rm sf}
f_{\rm ej} f_{\rm agb} R_{\rm gmc}^2v_{\rm gmc} t_{\rm life} t_{\rm agb}
(1+F_{\rm dil})(1-f_{\rm lost}) }.
\end{equation}

This $\Sigma_{\rm SFR, th}$ in units of ${\rm M}_{\odot}$ ${\rm kpc}^{-2}$ ${\rm yr}^{-1}$ can be expressed as a function of the physical parameters of GCs as follows:
\begin{equation}
\Sigma_{\rm SFR, th}= 1.4
( \frac{ M_{\rm gc} }{ 2 \times 10^5 {\rm M}_{\odot}  })
( \frac{ \epsilon_{\rm sf} }{0.3})^{-1}
( \frac{ 1 + F_{\rm dil} }{ 1.5 } )^{-1}
( \frac{ 1 - f_{\rm lost} }{ 0.5 }
)^{-1}
\end{equation}
for $F_{\rm 2P}=0.5$, $f_{\rm agb}=0.1$, $R_{\rm gmc}=100$ pc, $v_{\rm gmc}=30$ km s$^{-1}$, $t_{\rm life}=2 \times 10^7$ yr, $t_{\rm agb}=10^8$ yr, and $z_{\rm d}=300$ pc.
Like $\rho_{\rm agb, th}$, this $\Sigma_{\rm SFR,th}$ depends on multiple model parameters in a complicated way, the details of which we do not discuss in this paper.
One example of these parameter dependencies is shown in Fig. 4, describing a higher $\Sigma_{\rm SFR,th}$ for a larger $F_{\rm 2P}$.
Since $f_{\rm agb}$ depends strongly on the power-law slope and the upper and lower mass cutoffs of the stellar
initial mass function (IMF)
 in dwarf galaxies, $\Sigma_{\rm SFR,th}$ also depends on the IMF (e.g., lower $\Sigma_{\rm SFR,th}$ for a more top-heavy IMF).
It would be safe to claim that $\Sigma_{\rm SFR, th}$ should be as high as $\approx 1$ ${\rm M}_{\odot}$ ${\rm kpc}^{-2}$ ${\rm yr}^{-1}$ in GC formation with $F_{\rm 2P} \approx 0.5$ from GMCs with typical $t_{\rm life}$.
$\Sigma_{\rm SFR, th}$ can be used to assess whether young massive SCs in galaxies can have MPs or not, as discussed later in \S 5.

\subsubsection{Physical properties of GC-hosting GMCs}

Recent theoretical studies of GMC formation and evolution have demonstrated that (i) massive GMCs ($\ge 10^6 {\rm M}_{\odot}$) can form through continuous gas accretion and the merging of other GMCs, (ii) the formation timescales can be longer than $10^7$ yr, and (iii) collisions or merging of massive GMCs can trigger massive star formation and enhance star formation efficiencies (e.g., Kobayashi et al. 2017).
The longer $t_{\rm life}$ ($>10^7$ yr) 
 is a crucial condition required for the chemical enrichment of GMCs by AGB stars in the SCI scenario.
The bursty star formation during GMC formation triggered by GMC collisions can 
correspond to 2P formation in the scenario and also explain the high SFEs required 
for bound cluster formation (e.g., Baumgardt \& Kroupa 2007).

Low-mass GMCs in the Galaxy, such as the Taurus Molecular Cloud, are observed to form preferentially low- and intermediate-mass stars (e.g., Cohen \& Kuhi 1979).
Accordingly, the accretion and merging of such low-mass GMCs dominated by low- and intermediate-mass stars can be a key physical process for massive GMCs hosting GCs.
If a large number of pre-existing intermediate-mass stars from low-mass GMCs can eventually become centrally concentrated within a GC-forming GMC, then they can pollute the central region of the GMC to a much larger extent after they enter into AGB phases.
The observed extreme 2P population with a rather large [Na/Fe] in GCs could be due to chemical enrichment by clusters of pre-existing AGB stars in the central regions of GC-forming GMCs.

Grasha et al. (2019) investigated the distances ($R$) of SCs of various ages ($T_{\rm age}$) from  GMCs in M51. 
 From this, they derived the $T_{\rm age}$ distributions of SCs within three spatial 
zones: $R\le R_{\rm gmc}$, $R_{\rm gmc} < R \le 2R_{\rm gmc}$, and $2R_{\rm gmc} < R \le 3R_{\rm gmc}$ 
(see their Fig. 8). Although they found a high fraction ($\approx 0.4$) of 
young SCs ($T_{\rm age} \approx 10^6$ yr) within $R \le R_{\rm gmc}$, there is a noticeable trend of 
increasing SC numbers with age for $T_{\rm age} > 10^7$ yr across all three investigated areas around 
the GMCs. The peak around $T_{\rm age} = 2 \times 10^8$ yr is particularly intriguing, suggesting that GMCs 
could potentially be influenced by stellar winds originating from these intermediate-age SCs.
However, the seemingly low spatial densities of SCs around GMCs in 
M51 (e.g., 129 SCs around 112 GMCs for $R\le R_{\rm gmc}$) indicate that these GMCs 
cannot be significantly enriched chemically by nearby intermediate-age SCs; 
consequently, new SCs MPs cannot form from them. Conversely, these observations also 
imply that intermediate-age stars within SCs could enrich nearby GMCs to a greater extent in environments where the densities of such stars are substantially higher, providing supportive observational evidence for the SCI scenario.

\subsubsection{Type II GC formation via GC merging}

Carretta et al. (2010) proposed that NGC 1851 was formed from the merging of two SCs to have two different populations with 
a [Fe/H] spread of 0.08 dex.
Lee (2015) also proposed that the observed chemical abundance patterns of M22 can be nicely explained by the merging of two GCs within their host dwarf galaxy.
Using one-zone chemical evolution models of dwarf galaxies, 
Bekki \& Tsujimoto (2016, BT16) demonstrated that two GCs formed from two distinct GMCs at different epochs within their host dwarf galaxy can have different [Fe/H] and $s$-process element abundances.
These two merge together later within their host dwarf galaxy to form a new GC with two distinct [Fe/H] and $s$-process element abundances (e.g., [Ba/Fe]) due to the lower velocity dispersion of stars in the host dwarf; the new GC can be classified as a Type II GC (BT16).
This merger scenario can nicely explain not only the two distinct $s$-poor and $s$-rich populations observed in some Type II GCs (e.g., M22; Marino et al. 2015) but also the mass-budget problem in $s$-rich populations, as discussed later in \S 5.

The mass ratio ($m_2$) of the two merging GCs can possibly determine the mass fractions of Fe-rich and $s$-rich populations in these GCs.
It would be possible that the two merging GCs can merge with other very small SCs ($m_2 \ll 0.1$) before or after their merging; however, the chemical fingerprint of such minor merger events might be hard to find observationally.
Our recent numerical simulations of GC formation in gas-rich galaxy mergers have shown that massive nuclear SCs can be formed through multiple merging between a number of SCs with different ages and [Fe/H] (Matsui et al. 2025).
If these nuclear SCs are stripped from their host galaxies to become 
complex GC-like objects like  $\omega$ Centauri, then they should show wide [Fe/H] spreads.
Thus, in this GC merger scenario, pair merging can create Type II GCs with two distinct $s$-poor and $s$-rich populations, whereas multiple merging can lead to the formation of nuclear SCs that finally become massive GCs with wide age and [Fe/H] spreads like M54 and $\omega$ Centauri.
Lower mass AGBs (1-3 ${\rm M}_{\odot}$),
which are not viewed as the dominant driver of MP formation, 
can be responsible for establishing the s-process differences in Type II GCs, 
thus it is expected that there is some [Fe/H] enhancement due to 
galactic chemical evolution between the formation of s-poor and s-rich populations.

\subsection{Solutions for the potential problems of the classic AGB scenarios}

\subsubsection{The mass budget problem}

If 2P stars in a GC are formed from gas ejected from polluters that evolved from 1P stars of the 
original GC, then the total mass ($M_{\rm o}$) of the original GC can be inferred from the present-day GC mass ($M_{\rm gc}$), the chemical abundances of 2P stars, and the chemical yields of the polluters 
(e.g., Bekki \& Norris 2006). We here define the mass budget factor as follows:
\begin{equation}
F_{\rm b} = \frac{ M_{\rm o} }{ M_{\rm gc} }.
\end{equation}
This $F_{\rm b}$ needs to be rather large ($>10$) to explain the observed fractions of
2P stars in GCs for a canonical IMF (the so-called mass budget problem). 
In the classic AGB scenario, the progenitor systems are assumed to be GC-like objects that are considerably more massive than present day GCs, which exacerbates rather than alleviates the problem. In the SCI scenario, massive progenitors are not required as any AGB stars passing through or near a growing GMC can enrich the pristine gas. The total mass of polluting AGB stars, which depends on the broader field star population of the host galaxy, can naturally exceed Mgc without invoking an unrealistically large stellar mass. The mass-budget problem is therefore not a fundamental concern within our new framework.
\subsubsection{Timing of dilution by pristine gas}

The classic AGB scenario needs to assume that the right amount of pristine gas is accreted onto existing GCs through some physical mechanisms (e.g., Bondi-type gas accretion) just when intermediate-mass stars enter into AGB phases (e.g., D10).
It also needs to fine-tune the time evolution of gas accretion rates in order to dilute AGB ejecta to the right degree and thereby reproduce well the observed anticorrelations between light elements (e.g., D10).
While the ejection of pristine gas from intermediate mass close binaries and subsequent mixing with AGB ejecta has been proposed as a workaround within the classic AGB framework (e.g., Vanbeveren 2012; Bekki 2023), a comprehensive solution is yet to emerge. In the SCI scenario, however, this issue does not pose any challenges as stellar winds from AGB stars directly mix with the pristine gas of the surrounding GMC immediately following ejection, making the precise timing of gas accretion onto an existing cluster less crucial.

\subsubsection{Feedback effects from delayed CCSNe and type Ia SNe}

The classic AGB scenario does not have the ``supernova avoidance problem'', which would be one of the serious problems in other scenarios (e.g., Renzini et al. 2015), because chemical pollution of pristine gas by AGB stars can proceed after CCSNe have exploded.
However, delayed CCSNe (e.g., Zapartas et al. 2017) and prompt Type Ia SNe could possibly expel pristine gas and AGB ejecta within GC-forming gas clouds due to their explosive energy  in the scenario.
Although the dynamical influences of Type Ia SNe on the intra-cluster gas of GCs have been investigated using hydrodynamical simulations of forming GCs (e.g., Lacchin et al. 2026), previous theoretical studies have not investigated whether intra-cluster gas can be completely expelled by the feedback effects of delayed CCSNe.
Since delayed CCSNe explosions can start to occur $\sim 40$ Myr after the initial 
bursty formation of stars in GC formation (e.g., Zapartas et al. 2007), they can synchronize the gas ejection from massive AGB stars.
Therefore, it would be likely that delayed CCSNe expel all AGB ejecta and intracluster gas from GCs to completely truncate secondary star formation,
representing a serious challenge for the classic AGB scenario that has yet to be fully addressed.

If AGB ejecta can be retained even after mixing with ejecta from delayed CCSNe, new 2P stars formed from AGB ejecta might have significantly larger [Fe/H] than 1P stars.
No such [Fe/H] enhancement  is  observed in Type I GCs.
Therefore, it is a pressing issue for the classic AGB scenario to investigate how delayed CCSNe influence the formation processes and the chemical abundances of 2P stars.
This possibly serious problem in the classic AGB scenario can be avoided in the new scenario, because the chemical enrichment of GMCs by AGB stars is completed well before the onset of CCSN and delayed CCSN explosions.
Gaseous ejecta from these SNe can enrich the ISM of GC-hosting dwarf galaxies, if the gas can be trapped by the gravitational potentials of the galaxies.
Gaseous ejecta from low-mass AGB stars ($m \le 3 {\rm M}_{\odot}$) formed within GCs would be immune to these feedback effects of delayed CCSNe; however, they are highly likely to be influenced by the feedback effects of numerous SNIa.
It is thus possible that AGB ejecta after GC formation can all be expelled from GCs in the SCI scenario.
\subsubsection{Star formation suppression by existing stars}

Many previous simulations based on the classic AGB scenario assumed that secondary star formation from AGB ejecta retained in young SCs is possible from high-density gas (e.g., Bekki 2010).
However, these simulations did not investigate how existing stars in young massive SCs (i.e., GC progenitors) influence the collapsing protostellar cores due to limitations in the spatial resolution of the simulations.
Bobrick et al. (2025) have recently shown that secondary star formation directly from retained AGB ejecta can be severely suppressed (or even not possible) due to multiple disruptive encounters of protostellar cores with existing stars in young massive SCs.
Bekki (2019b) also demonstrated that the interaction of small clouds with existing stars and the deep gravitational potential of SCs combine to severely suppress the formation of more massive stars ($m > 10 {\rm M}_{\odot}$).
So it is currently not so clear whether secondary star formation from retained AGB ejecta is really possible in dense stellar systems.
The SCI scenario does not have such a potential problem in secondary star formation, 
because all stellar populations form during the assembly and evolution of the parent GMC, 
prior to its dispersal by stellar feedback, rendering secondary star formation within an already-formed cluster unnecessary.

\section{Analytic models}

\subsection{A model for mass-dependent GC properties}

Fig. 2 briefly summarizes several key points of the analytic model for the new scenario.
In the new GC formation scenario, a present-day GC consists of (i) stellar populations formed from pristine gas (`1P''), (ii) those formed from the gas of a GC-forming GMC chemically polluted by surrounding AGB stars (`2P''), and 
(iii)  stars that pre-dated  before 1P and 2P formation.
Our recent simulations of GC formation from GMCs have already shown that GMCs can have a ``precursor population'' (or ``0P'') of stars that were gravitationally trapped by GMCs (McKenzie \& Bekki 2021b).
These 0P stars of a GC have chemical abundance patterns of the field stars of their host galaxy, so their abundances can be quite distinct from those of the 1P and 2P stars of the GC.
For example, their metallicities can be slightly lower than those of the 1P and 2P stars of the GC due to their earlier formation epochs.

If these stars can continue to be gravitationally trapped by GCs until the present, they are highly likely to be observed in the outer parts of GCs (i.e., stellar halos).
It is thus an important observational question whether such 0P stars with a peculiar chemical abundance pattern can be found in the outer parts of GCs.
Although pre-existing AGB stars chemically pollute a GC-forming GMC, the present-day GC might have lost most of these stars due to long-term stripping processes.
Our previous numerical simulations of GC formation showed that only a small fraction of field stars in dwarf galaxies can be gravitationally trapped by GCs (Bekki \& Yong 2012; BY12).
Accordingly, the total GC mass is as follows:
\begin{equation}
M_{\rm gc} = M_{\rm 1P} +M_{\rm 2P} + M_{\rm s},
\end{equation}
where $M_{\rm s}$ is the total mass of the pre-existing stars gravitationally trapped by
GMCs  and could be much smaller than $M_{\rm 1P}$ and $M_{\rm 2P}$.
The present-day 1P fraction is therefore defined as follows:
\begin{equation}
F_{\rm 1P} = \frac{ M_{\rm 1P} }{ M_{\rm gc} } .
\end{equation}
The pre-existing stars could have chemical abundances similar to those of GMCs because they are formed only slightly before GMC formation.
If we distinguish between these stars based on their chemical abundances, and not on their formation epochs, 1P and the pre-existing stars have similar chemical abundances so that they can be observationally identified as `1P''. 
The total mass of observationally identified 1P stars is therefore the sum of $M_{\rm 1P}$ and $M_{\rm s}$. 
We suggest that pre-existing stars could be observed as loosely bound stellar halos around GCs. The initial GC masses ($M_{\rm gc,i}$) should be significantly larger than $M_{\rm gc}$, firstly because $M_{\rm gc}$ is the total mass of low-mass 
stars ($m \le 0.8 {\rm M}_{\odot}$), and secondly because GCs can lose their stars through internal dynamical 
evolution and tidal stripping. 
\begin{equation} M_{\rm gc} = (1-f_{\rm lost})  M_{\rm gc,0},
\end{equation} 
where $f_{\rm lost}$ is the fraction of initial stars that are lost from a GC.
This $f_{\rm lost}$ depends on $f_{\rm low}$ and $f_{\rm strip}$, where $f_{\rm low}$ is the fraction of low-mass stars with $0.1 \le m/{\rm M}_{\odot} \le 0.8 {\rm M}$ and $f_{\rm strip}$ is the fraction of stripped stars. Although these $f_{\rm low}$ and $f_{\rm strip}$ could be different in different GCs, we use a fixed $f_{\rm lost}=0.5$ for all stellar populations. The original gas mass ($M_{\rm o}$) required for the formation of a GC is calculated for a given star formation efficiency ($\epsilon_{\rm sf}$): \begin{equation} M_{\rm o} =\frac{ M_{\rm gc,0} }{ \epsilon_{\rm sf} } \end{equation} 
If all stars in GCs are formed from the gas of GMCs, 
then $M_{\rm o}$ can be the total mass of the GMC ($M_{\rm gmc}$). 
If $M_{\rm gc}=2 \times 10^5 {\rm M}_{\odot}$, $f_{\rm lost}=0.75$, and $\epsilon_{\rm sf}=0.1$ are adopted, then $M_{\rm gmc}=8 \times 10^6 {\rm M}_{\odot}$ is expected; this would be a conservative value due to the adopted high $\epsilon_{\rm sf}$.

Since 2P stars are formed from AGB ejecta ``diluted'' by (i.e., mixed with) pristine GMC gas, the fraction of 2P stars ($F_{\rm 2P}$) is determined by the dilution factor ($F_{\rm dil} =\frac{ M_{\rm g} }{ M_{\rm ej} }$).
We here adopt the following linear relation between $F_{\rm dil}$ and the initial fraction of 2P stars ($F_{\rm 2P, i}$):
\begin{equation}
F_{\rm 2p,0} = a_{\rm dil} F_{\rm dil} + b_{\rm dil},
\end{equation}
where $a_{\rm dil}=-0.1$ and $b_{\rm dil}=1.0$.
Appendix B describes how the above relation is derived for $0 \le F_{\rm dil} \le 10$.
The initial 1P fraction ($F_{\rm 1P,i}$) can be simply derived from $F_{\rm 2P}$ ($F_{\rm 1P,0}=1-F_{\rm 2P,0}$).
It should be noted here that this $F_{\rm 1P, 0}$ can be different from $F_{\rm 1P}$ due to the long-term dynamical evolution of GCs.
Since it is assumed that pre-existing stars can chemically pollute GMCs, $M_{\rm ej}$ is estimated as follows:
\begin{equation}
M_{\rm ej} = f_{\rm agb} f_{\rm ej} M_{\rm s,0},
\end{equation}
where $f_{\rm agb}$ is the mass fraction of intermediate-mass stars among all new stars, $f_{\rm ej}$ is the mass fraction of AGB ejecta in AGB stars, and $M_{\rm s,0}$ is the total mass of new stars that can interact with a GMC during the GMC's lifetime.
In order to estimate $f_{\rm agb}$, we adopt the following power-law function for the IMF of stars:
\begin{equation}
\Psi (m) = C_{\rm imf}m^{-\alpha},
\end{equation}
where $m$ is the initial mass of a star and $\alpha$ is the power-law slope.
The IMF with $\alpha =2.35$ corresponds to the canonical Salpeter IMF (Salpeter 1955).
The normalization factor $C_{\rm imf}$ is determined by $\alpha$, $m_{\rm l}$ (lower mass cutoff), and $m_{\rm u}$ (upper mass cutoff).
We assume that $m_{\rm l}$ and $m_{\rm u}$ are fixed at $0.1 {\rm M}_{\odot}$ and $50 {\rm M}_{\odot}$, respectively, for all models.
The lower and upper mass cutoffs for AGB (sAGB) stars that can chemically enrich GMCs are referred to as $m_{\rm agb, l}$ and $m_{\rm agb, u}$, respectively.
We adopt $m_{\rm agb, l}=4 {\rm M}_{\odot}$ and $m_{\rm agb, u}=10 {\rm M}_{\odot}$, unless specified.
We adopt $f_{\rm ej}=0.9$ as a fiducial value, though these depend on stellar masses and the IMF slope (e.g., Weidemann 2000).
We also adopt $f_{\rm agb}=0.12$, which is reasonable for the standard Salpeter IMF with $\alpha=2.35$.
Only a very small fraction of existing stars polluting a GMC can be trapped by the GMC and subsequently by its new SC (BY12):
\begin{equation}
M_{\rm s} = f_{\rm trap} M_{\rm s,0},
\end{equation}
where we adopt $f_{\rm trap}=0.01$ as a fiducial value.
The ratio of $M_{\rm s,0}$ to $M_{\rm gmc}$ is a crucial parameter that determines the chemical enrichment processes within GMCs in the new scenario:
\begin{equation}
M_{\rm s,0} = R_{\rm s} M_{\rm gmc}.
\end{equation}
Physically, $R_{\rm s}$  represents the mass fraction of new stars (including intermediate-mass stars)
residing within and around a GMC relative to the total GMC mass, integrated over the GMC lifetime. 
A larger $R_{\rm s}$ therefore implies a 
higher local density of AGB stars available to pollute the GMC gas, 
and consequently a more chemically enriched 2P population.
In order to understand how $R_{\rm s}$ depends on $M_{\rm gmc}$, we have performed new hydrodynamical simulations of disk galaxies with a revised model of molecular gas formation on dust grains; the details of the simulation and the model for GMC formation are briefly described in Appendix A.

In particular, we have investigated the total mass of newly formed stars ($M_{\rm ns}$) with ages ranging from 0.02 Gyr to 0.1 Gyr around each GMC in simulated galaxies at different time steps, and found that $M_{\rm ns}$ depends strongly on $M_{\rm gmc}$ in a simulated compact dwarf disk galaxy with many star-forming GMCs (see Appendix A for the results). 
Based on these results, we adopt the following relation between $R_{\rm s}$ and $M_{\rm gmc}$:
\begin{equation}
R_{\rm s} = a_{\rm s} \log (\frac{ M_{\rm gmc} }{ 10^7 {\rm M}_{\odot} }) + b_{\rm s},
\end{equation}
where $a_{\rm s}$ and $b_{\rm s}$ are set to be 0.15 and 0.01, respectively.
This $R_{\rm s}-M_{\rm gmc}$ relation can depend on galactic properties such as their sizes and masses, as demonstrated by our galaxy-scale simulations.
We adopt a reasonable and realistic assumption that only a fraction of GMC gas is mixed with AGB ejecta to form new stars:
\begin{equation}
M_{\rm g} = f_{\rm g} M_{\rm gmc},
\end{equation}
where $f_{\rm g}$ is a parameter that controls $M_{\rm g}$ and thus $F_{\rm dil}$ for a given $M_{\rm ej}$.
Since pristine gas polluted by AGB stars is converted into new 2P stars, the initial total mass of 2P stars ($M_{\rm 2P,0}$) is estimated as follows:
\begin{equation}
M_{\rm 2P, 0} = F_{\rm 2P,0}M_{\rm gmc}(R_{\rm s}f_{\rm agb}f_{\rm ej}
+f_{\rm g}).
\end{equation}
Significant fractions of these initial 2P stars can be lost due to long-term dynamical processes and stellar mass loss; we estimate the present-day total mass of 2P stars as follows:
\begin{equation}
M_{\rm 2P} = (1-f_{\rm lost}) M_{\rm 2P,0},
\end{equation}
where $f_{\rm lost}=0.75$ is adopted, though $f_{\rm lost}$ could be different between initial 1P and 2P stars.
In the present scenario, 1P stars can be formed before and after chemical enrichment by AGB stars. Therefore, the initial total mass of 1P stars ($M_{\rm 1P,0}$) is a function of multiple model parameters:
\begin{equation}
M_{\rm 1P, 0} = f_{\rm sf}M_{\rm gmc}
+M_{\rm gmc}(1-F_{\rm 2P,0})(R_{\rm s}f_{\rm agb}f_{\rm ej}
+f_{\rm g}),
\end{equation}
where $f_{\rm sf}$ is the fraction of new stars steadily formed before chemical pollution by AGB stars.
It should be stressed here that a fraction of stars formed from pristine gas polluted by AGB stars can be classified as 1P due to their chemical abundances (e.g., lower [Na/Fe]).
We show the results of the model with $f_{\rm sf}=0.02$ only, because the present results do not depend on $f_{\rm sf}$.
We can estimate $M_{\rm 1P}$ from $M_{\rm 1P,0}$ as done for $M_{\rm 2P}$:
\begin{equation}
M_{\rm 1P} = (1-f_{\rm lost})M_{\rm 1P,0},
\end{equation}
where $f_{\rm lost}$ is fixed at 0.75.

Our future simulations need to be run to find a reasonable range of $f_{\rm lost}$ for 1P and 2P stars, because $f_{\rm lost}$ could be significantly different due to possibly different initial distributions of these stars, as shown in previous and recent numerical simulations of GCs with MPs based on the classic AGB scenario (e.g., Vesperini et al. 2011).
Having established the relative mass fractions between 1P and 2P stars, 
we now estimate the helium abundance of the 2P, which is a key observational diagnostic of the chemical 
enrichment by AGB stars.
We estimate the helium abundances of 2P stars ($Y_{\rm 2P}$) as follows:
\begin{equation}
Y_{\rm 2P} = \frac{ Y_{\rm gmc} M_{\rm p} + Y_{\rm agb} M_{\rm ej} }{
M_{\rm p}+M_{\rm ej} } ,
\end{equation}
where $Y_{\rm gmc}$ and $Y_{\rm agb}$ are the helium abundances of GMC gas and AGB stars, respectively.
We adopt the following definition of the helium abundance spread between 1P and 2P stars, where $Y_{\rm 1P}$ is assumed to be equal to $Y_{\rm gmc}$:
\begin{equation}
\delta Y =  Y_{\rm 2P}- Y_{\rm 1P},
\end{equation}
We adopt $Y_{\rm 1P}=Y_{\rm gmc}=0.25$ and $Y_{\rm agb}=0.34$ (e.g., Ventura et al. 2013, V13)
as fiducial values in order to estimate $\Delta Y$ for GCs with different $M_{\rm gc}$.
We investigate 300 GCs for $5 \times 10^5 {\rm M}_{\odot} \le  M_{\rm gmc} \le 5 \times 10^7 {\rm M}_{\odot}$ 
(initial GMC mass)
and $0.02 \le f_{\rm g} \le 0.05$.
Also, we consider that the $1-\sigma$ dispersion of $R_{\rm s}$ for a given $M_{\rm gmc}$ is 0.2 to simulate the possible diverse 1P fractions among GCs with different $M_{\rm gmc}$.
We use a random number generator to determine $M_{\rm gmc}$, $f_{\rm g}$, and $R_{\rm s}$ for each GC.

\begin{figure}
\psfig{file=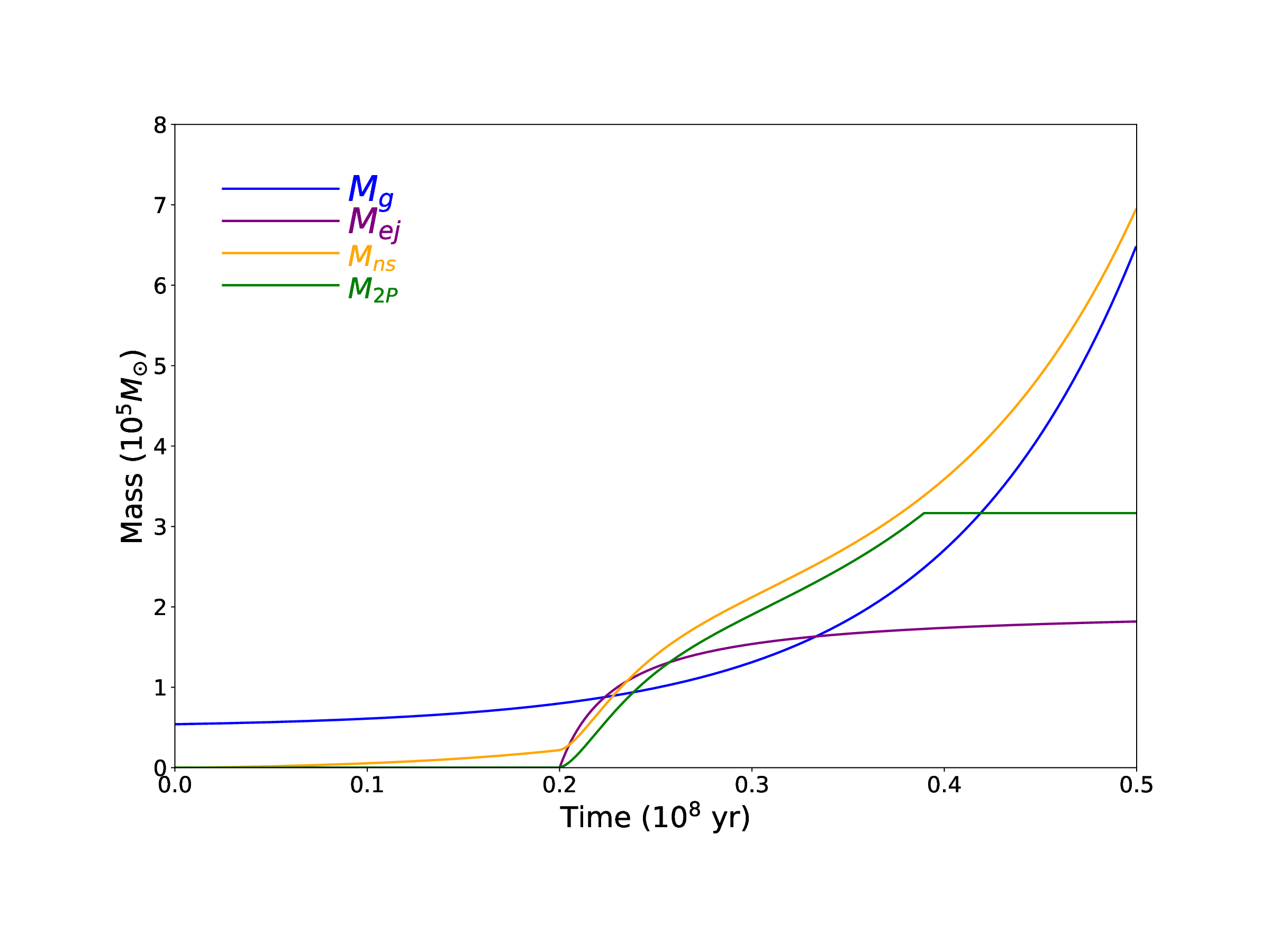,width=8.5cm} 
\caption{
Time evolution of the total masses of
pristine GMC gas ($M_{\rm g}$, blue),
AGB ejecta ($M_{\rm ej}$, purple),
all new stars ($M_{\rm ns}$, orange),
and 2P stars ($M_{\rm 2P}$, green) in the fiducial model.
}
\label{Figure. 5}
\end{figure}

\begin{figure}
\psfig{file=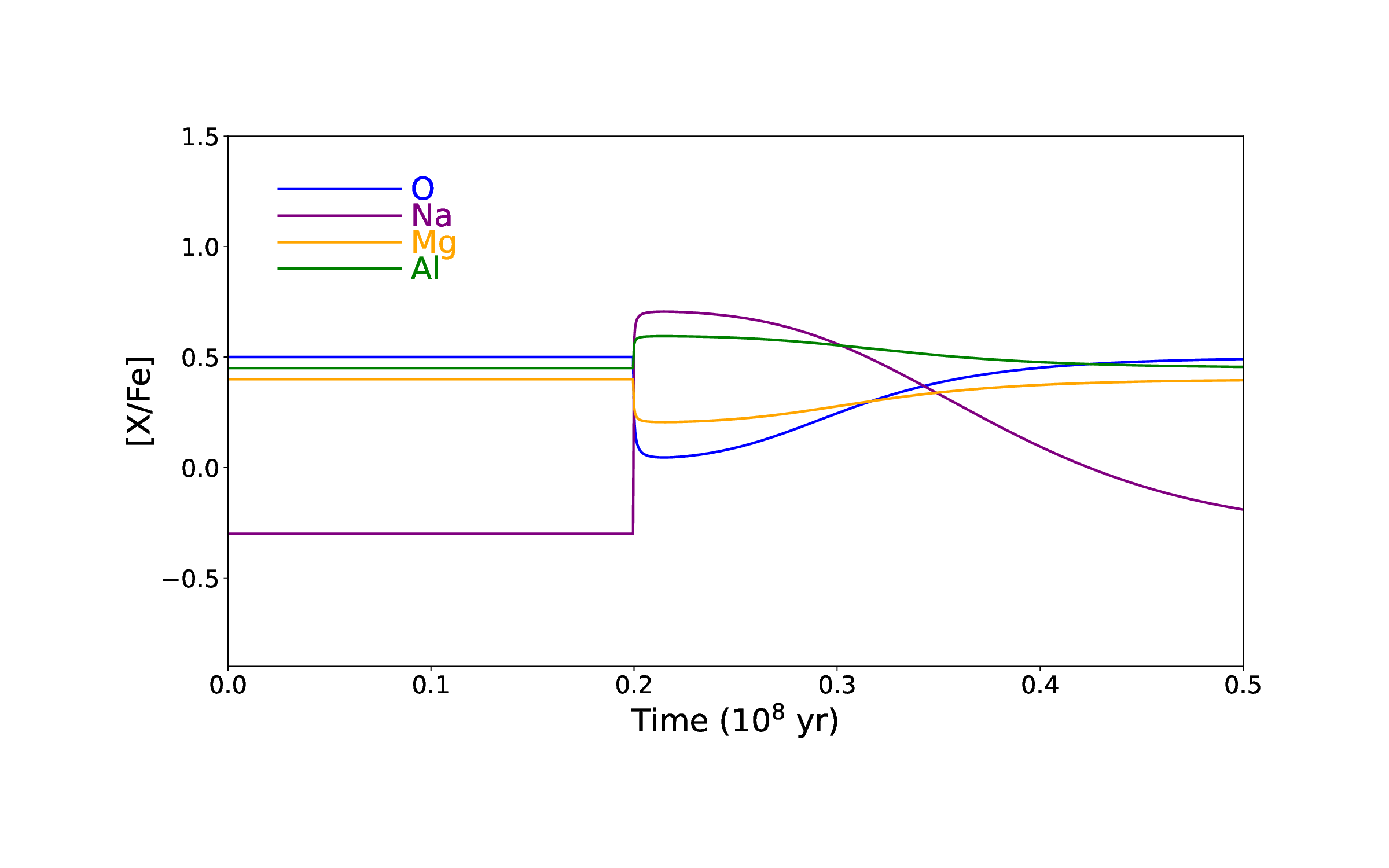,width=8.5cm} 
\caption{
Time evolution of 
[O/Fe] (blue),
[Na/Fe] (purple),
[Mg/Fe] (orange),
[Al/Fe] (green) for gas  in the fiducial model.
}
\label{Figure.6}
\end{figure} 

\subsection{One-zone models  for chemical enrichment in growing GMCs}

We investigate the time evolution of chemical abundances of new stars and gas
within GMCs
in order to discuss the observed abundance spread in various elements.
We use the ``one-zone'' model in which gas and metals from AGB stars
can be instantaneously mixed and converted into new stars.
The mass and time units in the model is $10^6 {\rm M}_{\odot}$
and $10^8$ yr, respectively, and all model parameters are given
in these units.
The total masses of pristine GMC gas ($M_{\rm g}$) and AGB ejecta
($M_{\rm ej}$) are assumed to be time-dependent, and
a GMC is assumed to grow exponentially within its lifetime ($t_{\rm life}$):
\begin{equation}
M_{\rm g}(t) = M_{\rm g,0}+ c_{\rm g,1} \exp (c_{\rm g,2} (t-t_{\rm grow})/t_{\rm grow}) ),
\end{equation}
where 
$M_{\rm g, 0}$ is the initial mass for the GMC (``seed mass''),
$t_{\rm grow}$ corresponds to the growth timescale,
and  $c_{\rm g, 1}$ and $c_{\rm g, 2}$ are parameters that control
the details of the mass evolution.
The ratio of the final mass of pristine gas ($M_{\rm g}$) at $t=t_{\rm life}$
to the total mass of AGB ejecta ($M_{\rm ej}$)
corresponds to $F_{\rm dil}$, and 
is a key parameter  that determines the abundance patterns of 2P stars.
The value of $c_{\rm g,1}$ is chosen
such that $M_{\rm g}(t)$ at $t=t_{\rm life}$ 
can be $M_{\rm g}$.
We mainly show the results of the fiducial model
with $M_{\rm g}=6 \times 10^5 {\rm M}_{\odot}$ 
($M_{\rm gmc}=6 \times 10^6 {\rm M}_{\odot}$ for $f_{\rm g}=0.1$),
$M_{\rm ej}=2 \times 10^5 {\rm M}_{\odot}$,
$t_{\rm grow}=10^7$ yr,
$t_{\rm life}= 5 \times 10^7$ yr,
and $c_{\rm g, 2}=1.0$ (in model unit).

The total mass of accumulated AGB ejecta can be time-dependent too,
given that AGB stars with different initial masses eject gas at different times.
We accordingly adopt the following simple analytic model to describe the time evolution:
\begin{equation}
M_{\rm ej}(t) = c_{\rm ej,1} \frac{ (1 - c_{\rm ej,2}t_{\rm agb, min}) }{
( t-t_{\rm agb,min}+c_{\rm ej,2}t_{\rm agb, max} ) },
\end{equation}
where $t_{\rm agb, min}$ and $t_{\rm agb,max}$ are the times at which
AGB stars with $m=10 {\rm M}_{\odot}$ and $4 {\rm M}_{\odot}$ 
start to eject gas, respectively,
and $c_{\rm ej, 1}$ and $c_{\rm ej,2}$ control the details of the mass evolution. 
The value of $c_{\rm ej,1}$ is chosen
such that $M_{\rm ej}(t)$ at $t=t_{\rm life}$ 
can be $M_{\rm ej}$.
We consider $t_{\rm agb, min}=2\times 10^7$yr and $t_{\rm agb, max}=3\times 10^8$
yr in the present study.
We adopt the above formula in order
to mimic the time evolution of $M_{\rm ej}$ estimated in our
previous calculations for the standard IMF (Bekki 2011).
$c_{\rm ej,2}$ is fixed at 0.01  (in model unit) in the fiducial model.

The star formation rate (SFR, $\phi(t)$) is assumed to be proportional to
the total gas mass within a GMC:
\begin{equation}
\phi(t) = c_{\rm sf} (M_{\rm g}(t)+M_{\rm ej}(t) ),
\end{equation}
where $c_{\rm sf}$ is the coefficient that controls the rapidity of star formation.
We mainly show the results of the models with $F_{\rm dil}=3$ in the present study,
though the results depend on this parameter.
Fig. 5 describes the time evolution of $M_{\rm g}$, $M_{\rm ej}$,
$M_{\rm 1P}$, and $M_{\rm 2P}$ in the fiducial model. The details of
these mass evolution depend on the adopted model parameters in
the present one-zone models.

The mass evolution of $i$-th element ($i$=H, He, C, etc) in a GMC
is  described by the following equation:
\begin{equation}
\frac{d M_{i} }{dt}= \frac{ (d A_{i, \rm g} M_{\rm g})  }{ dt }
+\frac{ d (A_{i, \rm  ej} M_{\rm ej})  }{dt},
\end{equation}
where $A_{i, \rm g}$ and $A_{i, \rm ej}$ are the mass fractions of
$i$-th element in pristine gas and AGB ejecta, respectively.
Since we use an idealized model for the mass evolution of AGB ejecta,
we adopt an assumption that
$A_{i \rm g}$ and $A_{i, \rm ej}$ are fixed (i.e., time-independent).
This assumption would be over-simplified,
because AGB stars with different stellar masses 
have different chemical yields and pollute GC-forming GMCs at different times.
Nevertheless, we adopt the above assumption, which enables us to 
more clearly elucidate the roles of AGB stars in the SCI scenario.
We will discuss the relative contributions of AGB stars with different masses 
using a more fully self-consistent chemical evolution models in our future papers.
Fig. 6 describes
the time evolution of O, Na, Mg, and Al abundances of stars in the fiducial model
over $ 5 \times 10^7$ yr, i.e., the lifetime of the GMC.

We adopt the chemical yields used by D10, V13, and Dell’Agli et al. (2018, D18) for AGB stars and by
I16 and V12
for sAGB stars. 
It should be stressed here that AGB yields for elements investigated in this paper
are different from those predicted from other groups  including
Karakas (2010, K10), Cristallo et at (2011), and Straniero et al. (2014).
Therefore, if we adopt AGB yields from other groups, then
the results can be significantly changed.
The yields from D18 are adopted in discussing how the observed anticorrelations between
light elements depend on metallicities. 
The nucleosynthesis model in I16 is 
fine-tuned to reproduce the observed chemical abundance
patterns of the low-metallicity GC NGC 2419 with [Fe/H]$\approx -2.1$.
Understanding the metallicity dependence of sAGB yields, particularly for Mg and K, 
represents an important avenue for future stellar evolution calculations.
Since K yields of sAGB stars with different masses and metallicities
are not given in Doherty et al. (2015),
we use only sAGB yields by I16 and V12.
D'Antona et al. (2012, D12) proposed
that Li production in massive AGB stars due to Cameron-Fowler effects is crucial
in reproducing the Li abundances of GC stars. We here introduce a parameter
($P_{\rm Li}$) for the fraction of massive 
AGB stars producing Li among all polluting AGB stars
in order to discuss the correlations between Li abundances
(A(Li)) and light element  abundances (e.g., [Na/Fe]).
We adopt $^{12}$C/ $^{13}$C ratios  and F yields  from K10 in order to briefly discuss
the observed C-isotope ratios and [Na/Fe]-[F/Fe] anticorrelations 
in GCs.

We adopt the following initial elemental abundances for pristine gas:
[Fe/H]=$-1.6$, $Y=0.24$,  [O/Fe]=0.5,  [Mg/Fe]=0.4, [Na/Fe]=-0.3,
[Al/Fe]=0, [K/Fe]=0, ${}^{25}$Mg/Mg=0.05, ${}^{26}$Mg/Mg=0.02,
and A(Li)=2.3.
Using the standard Salpeter IMF with the slope of $\alpha=-2.35$,
we estimate the IMF-averaged yields for O, Mg, Na, and Al
abundances of AGB winds. 
For comparison, we also investigate the models with 
different $\alpha$ to understand how the IMF influences the chemical abundances
of GCs.
We consider that the Mg isotope ratios of AGB ejecta, which are represented by
$R({}^{25}$Mg/Mg) and R(${}^{26}$Mg/Mg),
are free parameters,
because the models  with
$R({}^{25}$Mg/Mg)$\approx 0.2$ and $R({}^{26}$Mg/Mg)$\approx 0.6$ predicted
from Ventura et al. (2018) for $m=5 {\rm M}_{\odot}$ AGB stars
fail to reproduce the observed correlations between heavy Mg isotopes and [Al/Fe].
We thus  consider that the observed relations between Mg isotope abundances
can provide direct astrophysical constraints on the nuclear reaction rates governing the Mg-Al cycle in 
AGB and sAGB stars.

\subsection{Observations to be compared with models}

For this work, we adopt a set of elemental and isotopic abundance constraints to test the predictions of our 
GC formation model. The primary spectroscopic constraints are drawn from large, homogeneous abundance studies, 
including the optical GIRAFFE and UVES analysis of C09a and C09b, 
and the APOGEE-based analysis of ME20. 
Together, these studies provide broad coverage of the light-element abundance patterns that define 
MPs in GCs. C09a and C09b provide a large and homogeneous reference sample using optical wavelengths, 
with strong constraints on Na and O abundances. Additionally, ME20 derive Mg, Al, and Si abundances using 
infrared APOGEE spectra and a specialised abundance pipeline suitable for RGB GC stars. 

We also include selected 
follow-up studies to constrain additional abundance patterns, including Li and K, 
which can be derived using optical wavelengths, and P and F abundances, 
which are primarily measured in the infrared. 
Isotopic abundances are more observationally challenging, often requiring very 
high signal-to-noise and high-resolution spectra. In this work, we compare to both carbon isotopic ratios 
from stars on the main sequence and Mg isotopic ratios from cool RGB stars. Although N 
enhancements are one of the most notable features of GC 2P populations, C and N 
abundances are often derived from molecular bands, making them more susceptible to 
degeneracies in the abundance determination. These elements and isotopes provide 
critical diagnostic constraints on the temperatures and nucleosynthetic pathways used by the cluster 
polluter. In addition to spectroscopy, we use photometric constraints on multiple stellar populations, 
primarily from Hubble Space Telescope-based studies. These include photometric estimates of the first-population fraction and He abundance spreads from Milone et al. (2017, 2018, M18).

We emphasise that the adopted literature constraints are not expected to lie on a single absolute abundance scale. Differences in wavelength coverage, line selection, spectral resolution, signal-to-noise, stellar parameter determination approaches, abundance pipelines, and the treatment of 3D and non-LTE effects can all introduce systematic offsets between studies. Additionally, estimates of spectroscopic errors and uncertainties are not uniformly defined across the literature. For this reason, 
we place an emphasis on the relative trends between abundances, rather than the absolute abundances themselves. 
Overall, this compilation represents one of the most extensive observational benchmarks assembled for 
testing a specific GC formation scenario, placing strong constraints on whether the model can simultaneously 
reproduce the diverse chemical, isotopic, and photometric signatures of MPs.
Full observational references for each diagnostic are provided in the corresponding subsections of Section 4 and 5 where the model predictions are compared directly with the data.

\begin{figure}
\psfig{file=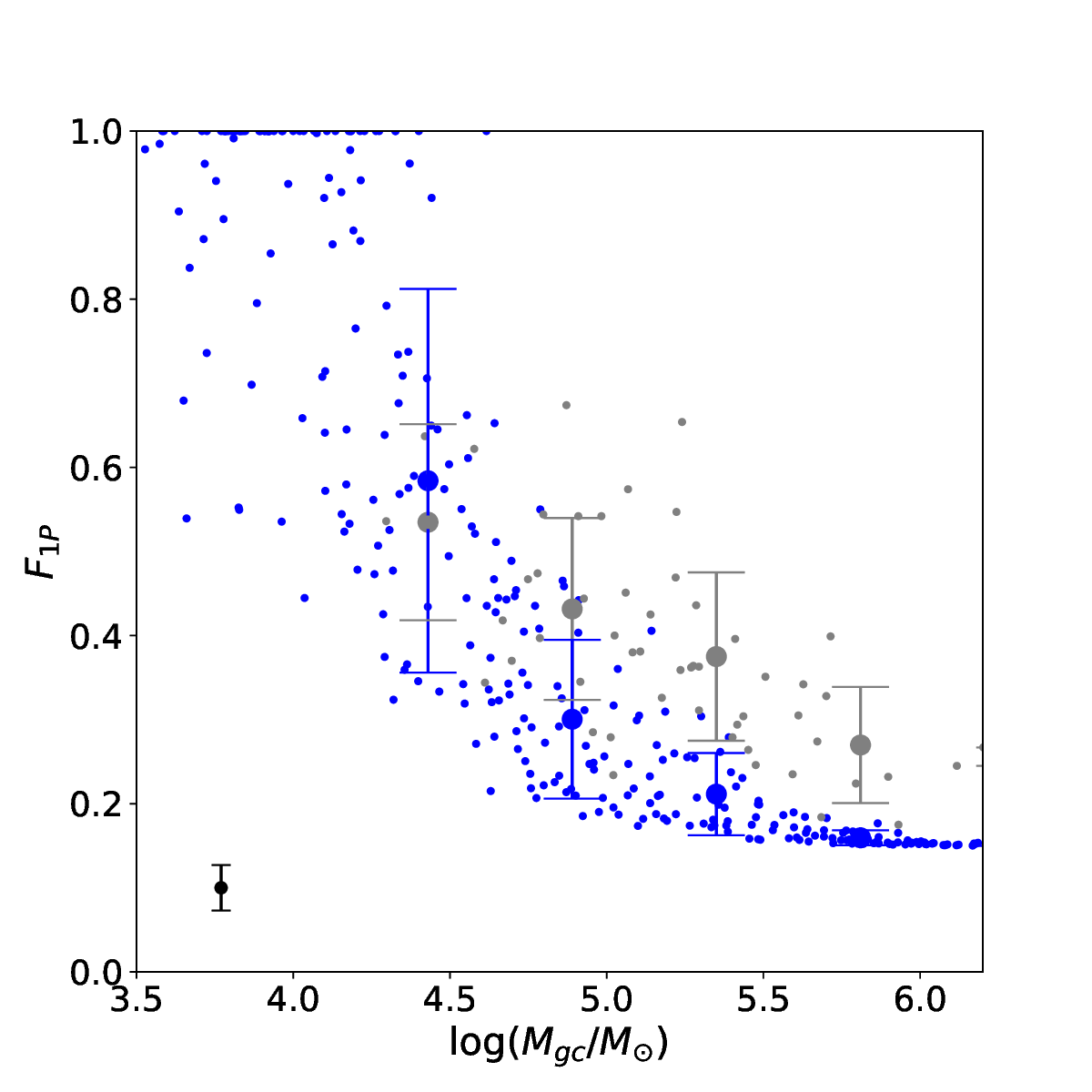,width=8.5cm} 
\caption{
Mass fractions of 1P stars ($F_{\rm 1P}$) as a function
of the present-day GC masses ($M_{\rm gc}$) for simulated 300 GCs (blue)
and observations (gray).
Observational data from M18 is used here.
}
\label{Figure. 7}
\end{figure}

\begin{figure}
\psfig{file=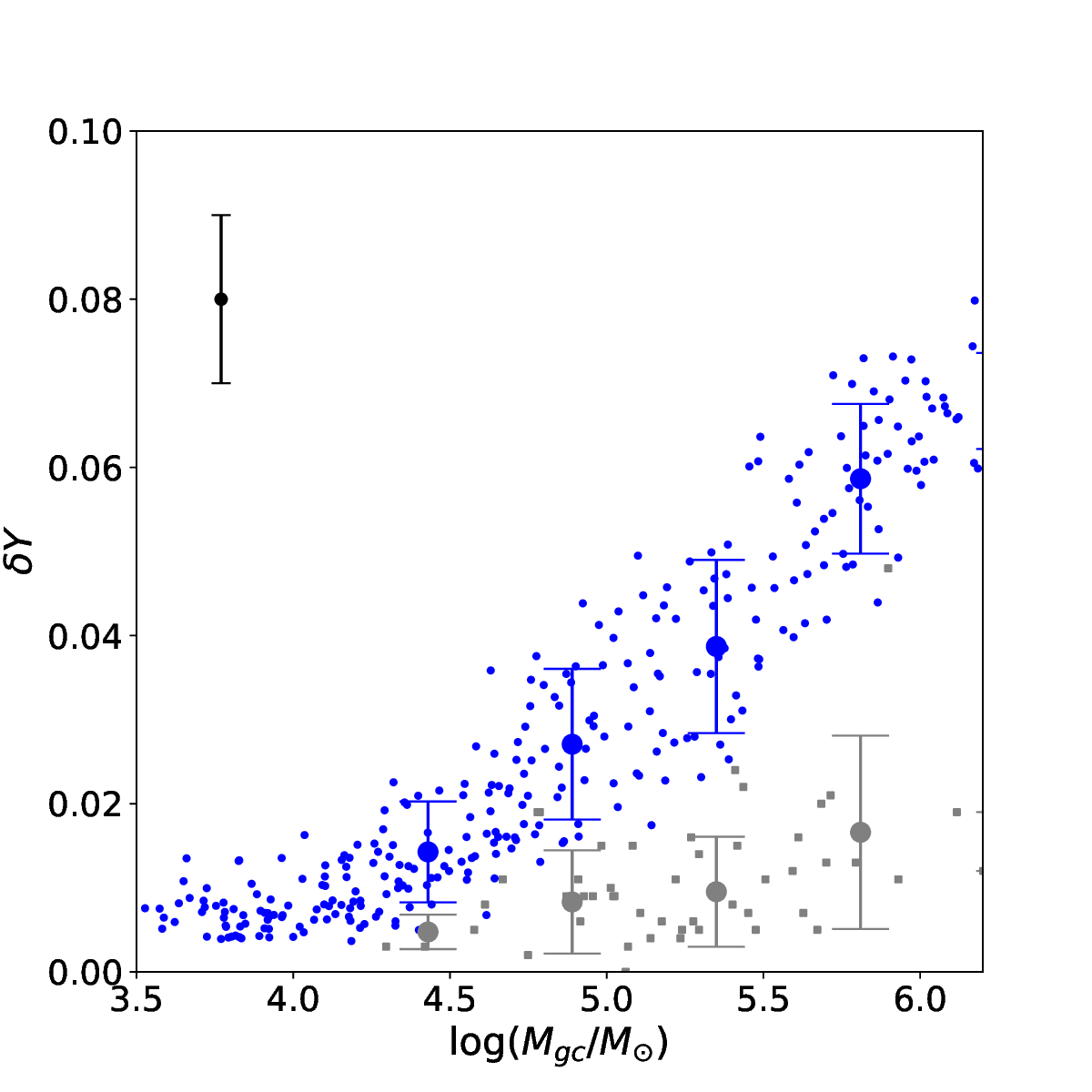,width=8.5cm} 
\caption{
Differences in helium abundances ($\delta Y$) between 1P and 2P stars 
as a function of $M_{\rm gc}$.
for simulated GCs (blue)
and observations (gray). Observational data from M18 is used here.
}
\label{Figure. 8}
\end{figure} 

\section{Results}
\subsection{$F_{\rm 1P}$ and $\delta Y$ dependent on $M_{\rm gc}$}

Fig. 7 shows the distribution of the simulated 300 GCs on the $M_{\rm gc}$-$F_{\rm 1P}$ plane for the fiducial set of model parameters.
Clearly, more massive GCs can have lower $F_{\rm 1P}$ in this model, which is qualitatively consistent with the corresponding observations by MM22.
The main reason for this $M_{\rm gc}$-dependent $F_{\rm 1P}$ is that more massive GMCs can have larger fractions of AGB stars within and around them to be chemically enriched to a greater extent.
More massive GMCs with larger sizes can interact with larger numbers of AGB stars (per unit mass) and gravitationally trap stellar winds from these stars more efficiently. 
We note that unlike the classic AGB scenario, where the gravitational potential of the 1P plays a critical role in retaining AGB ejecta for 2P star formation, this is not required for the SCI scenario.
The dispersions in $F_{\rm 1P}$ for a given $M_{\rm gc}$ can be significant due to the adopted initial dispersion in $R_{\rm s}$ and $F_{\rm dil}$.
Since these results are for a model with a fixed standard IMF, the dispersions in $F_{\rm 1P}$ could be even more pronounced if IMF variations across GMC masses are introduced.
The present model does not include the effects of the dynamical evolution of GCs on $F_{\rm 1P}$, which is suggested to be important (e.g., Parmentier 2025).
Therefore, if such effects are properly incorporated into the present model, scatters in the $M_{\rm gc}$-$F_{\rm 1P}$ relation could be even more pronounced.

The flat distribution of GCs for $M_{\rm gc} > 5 \times 10^5 {\rm M}_{\odot}$ is largely due to the adopted assumption that pre-existing stars are identified as 1P.
Although the fractions of stars with low [Na/Fe] are small, $F_{\rm 1P}$ cannot become lower than a certain value due to the presence of pre-existing stars ($M_{\rm s}=0.01 M_{\rm gmc}$ in this model).
If these pre-existing stars are not counted as 1P, then a steeper anticorrelation between $M_{\rm gc}$ and $F_{\rm 1P}$ can be obtained for $M_{\rm gc} > 5 \times 10^5 {\rm M}_{\odot}$.
It is unclear how $M_{\rm s}$ depends on $M_{\rm gmc}$ without performing numerical simulations on the dynamical evolution of pre-existing stars within and around GC-forming GMCs.
We also note that the observed F1P values themselves carry a systematic uncertainty 
arising from the operational definition of 1P and 2P membership, 
which differs between photometric and spectroscopic studies. 
Chromosome maps, pseudo two colour diagrams constructed using 
Hubble space telescope photometry, tend to produce apparently discrete population grouping 
whereas spectroscopic surveys typically reveal a continuous distribution along the 
Na-O anticorrelation with no unambiguous gap between populations (e.g., C09a). 
The placement of the 1P/2P boundary therefore introduces a study dependent offset in the f1P that likely contributes to some observed scatter.

It should be stressed here, however, that the GC distribution on the $M_{\rm gc}$-$F_{\rm 1P}$ plane depends strongly on the adopted $M_{\rm gmc}$-$R_{\rm s}$ relation.
If a flatter $M_{\rm gmc}$-$R_{\rm s}$ relation is adopted, the simulated $M_{\rm gc}$-$F_{\rm 1P}$ relation can also become rather flat, which is inconsistent with the observed one (MM22).
It would also be possible that low-mass GCs formed in galaxies with rather high $R_{\rm s}$ ($\approx 1$) can have low $F_{\rm 1P}$ ($<0.4$).
Since global galactic properties such as sizes and gas mass fractions can determine the $M_{\rm gmc}$-$R_{\rm s}$ relation, GC systems in galaxies with different galactic properties might have different $M_{\rm gc}$-$F_{\rm 1P}$ relations.
While high resolution spectroscopy of individual stars sufficient to directly measure 
light element abundance variations remains limited to GCs within and immediately surrounding the 
Local Group, integrated light spectroscopic techniques have the potential to detect abundance variations characteristic of MPs in GCs at greater distances (e.g., Sakari et al. 2016; Larsen et al. 2018). 
Coupled with the advent of 30-meter class telescopes, there is a possibility of extending the $M_{\rm gc}-F_{\rm 1P}$ 
 relation to 
more diverse extragalactic GC systems.

Fig. 8 describes how helium abundance differences between 1P and 2P stars ($\delta Y$) depend on $M_{\rm gc}$ for the fiducial set of model parameters.
More massive GCs can have larger $\delta Y$ in this model, which is at least qualitatively consistent with the corresponding 
observations (M18).
The derived $M_{\rm gc}$-$\delta Y$ relation results from larger $R_{\rm s}$ in higher $M_{\rm gmc}$, as seen in Fig. 7. Helium-enriched stellar winds from massive AGB stars can mix well with the pristine gas of GC-forming GMCs so that 2P stars formed from the mixed gas can have higher $Y$.
As a result of this, $\delta Y$ can be larger for GC-forming GMCs with larger $R_{\rm s}$ (i.e., larger numbers of more massive AGB stars interacting with GMCs).

The dispersion in $\delta Y$ is significant ($\approx 0.01$) for a given $M_{\rm gc}$, which is due largely to the initial dispersion in $R_{\rm s}$ and $F_{\rm dil}$.
The simulated $M_{\rm gc}$-$\delta Y$ correlation is steeper than the observed one, which could be a potentially serious problem in the new scenario.
This inconsistency is due to the adopted large (0.34) $Y$ yields of AGB ejecta in the present models.

Previous works have already claimed that the predicted large $Y$ in 2P stars (thus large $\delta Y$) in GC formation scenarios with polluters being AGB stars is a long standing tension  (e.g., Bastian et al. 2015).
Doherty et al. (2017) suggested that the primary processing sites of helium and Na in 
AGBs and sAGBs are physically distinct, with helium enrichment predominantly occurring through second 
dredge up during the early AGB, while Na and O are modified through hot bottom burning at the base of the 
convective envelope during the thermally pulsing AGB phase. Thus, helium and light element yields could 
in principle be decoupled.
Clearly, if a smaller $Y$ ($\approx 0.28$) can be adopted, the simulated $M_{\rm gc}$-$\delta Y$ correlation can better match the observed one.
GC distributions on the $M_{\rm gc}-\delta Y$ plane also depend strongly on the adopted $M_{\rm gmc}$-$R_{\rm s}$ relation, as seen in the $M_{\rm gc}$-$F_{\rm 1P}$ plane; these $M_{\rm gc}$-$F_{\rm 1P}$ and $M_{\rm gc}$-$\delta Y$ relations have a common physical origin.
Since it is beyond the scope of this paper to discuss why $R_{\rm s}$ and $F_{\rm dil}$ can be diverse in GC-forming GMCs with different $M_{\rm gmc}$, we will investigate this point in our forthcoming study based on hydrodynamical simulations of GMC environments in galaxies.
We note that helium abundances in RGB stars cannot be measured directly through 
optical spectroscopy as cool giants lack photospheric He lines in the optical wavelength range. The $\delta Y$
 values must be derived indirectly from multiband photometric comparisons of population sequences using synthetic spectra and isochrone fitting (see M18).

\begin{figure}
\psfig{file=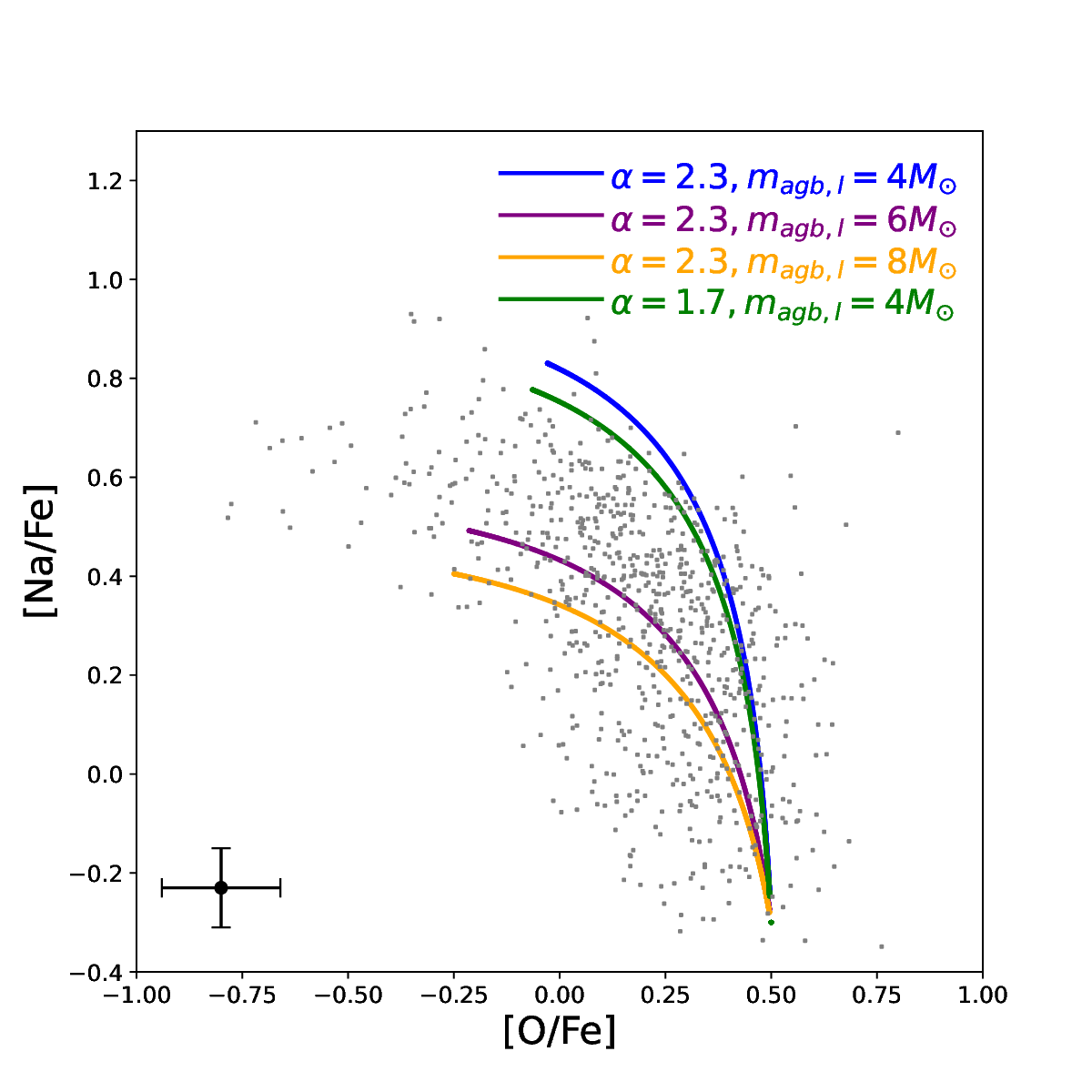,width=8.5cm} 
\caption{
Anticorrelations between [O/Fe] and [Na/Fe] in three models
with $\alpha=2.3$ and $m_{\rm agb, l}=4 {\rm M}_{\odot}$ (blue)
$\alpha=2.3$ and $m_{\rm agb, l}=6 {\rm M}_{\odot}$ (purple)
and $\alpha=1.7$ and $m_{\rm agb, l}=4 {\rm M}_{\odot}$ (orange)
and observations from C09a and C09b (gray).
[O/Fe] and [Na/Fe] at 1000 time steps are shown for each model in this figure.
The AGB models  from D10 are used to calculate the IMF-averaged AGB yields in these models.
}
\label{Figure. 9}
\end{figure}

\begin{figure}
\psfig{file=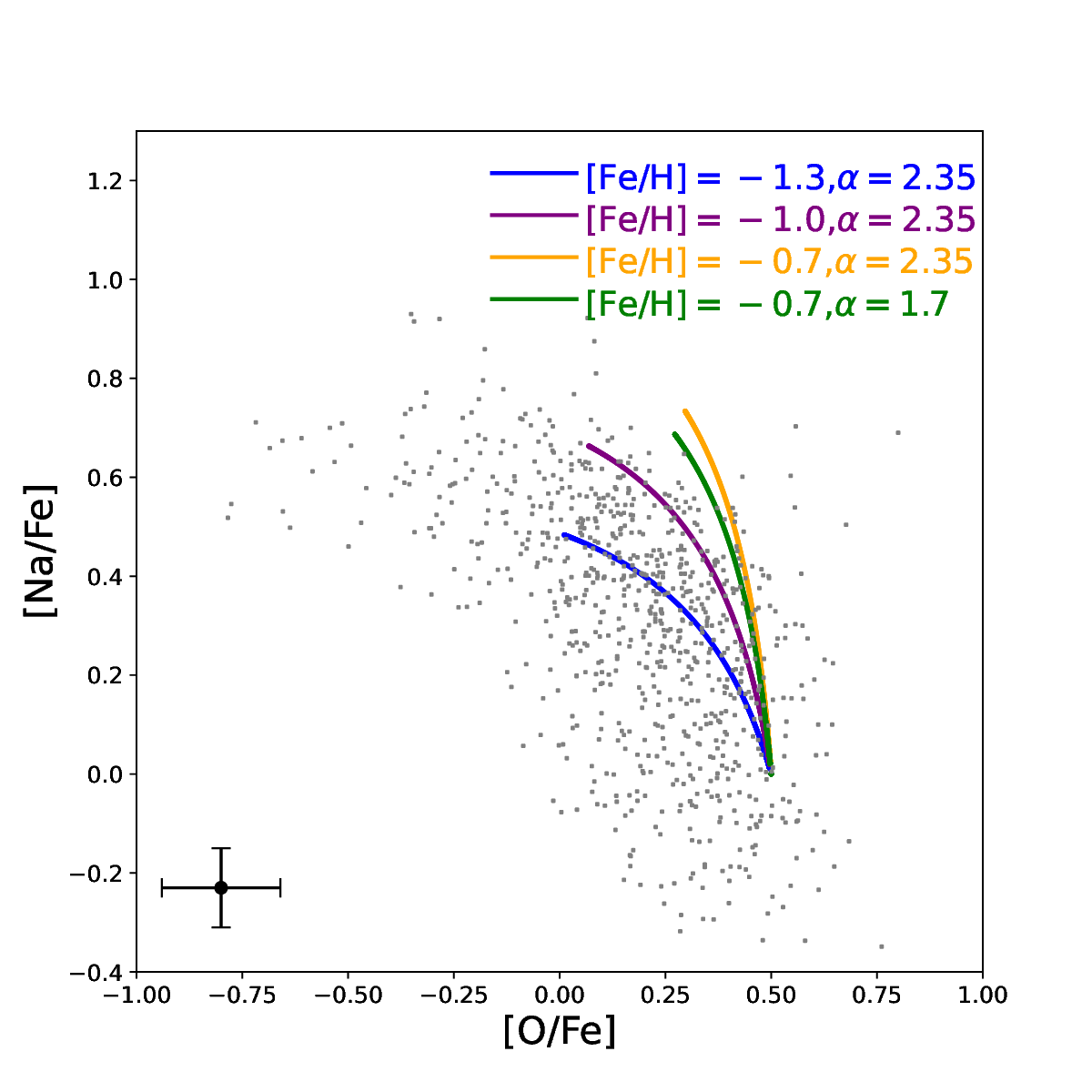,width=8.5cm} 
\caption{
Same as Fig. 9 but for different [Fe/H] and $\alpha$.
}
\label{Figure. 10}
\end{figure}

\subsection{O-Na anticorrelation}

The O-Na anticorrelation is a direct signature of hydrogen burning at high temperatures. 
Oxygen depletion occurs through the ON cycle, while Na is simultaneously enhanced through proton capture on 
$^{22}$Ne via the Ne-Na cycle (e.g., Langer et al. 1993). 
Both cycles are activated during hot bottom burning (HBB), 
the process by which the base of the convective envelope in massive AGB stars reaches temperatures 
sufficient to activate proton-capture nucleosynthesis.
Fig. 9 shows the time evolution of the present chemical
evolution models on the [O/Fe]-[Na/Fe] plane
for three different IMF-averaged AGB yields.
The IMF slope ($\alpha$) and the lower mass cutoff of AGB stars ($m_{\rm agb, l}$)
are different between the three models so that the chemical abundances of star-forming
GMCs can be quite different after mixing with AGB ejecta.
Clearly, O-Na anticorrelations can be seen in the three models;
however, the evolutionary locus and the final O and Na abundances are
significantly different among the three.
Although the IMF slope $\alpha$ does not significantly influence the time evolution
of O and Na abundances,
the evolutionary locus depends strongly on $m_{\rm agb, l}$,
reflecting that AGB yields are different between different $m_{\rm agb}$.
The present models cannot simply explain the observed extreme (``E'') populations
with rather low [O/Na] $<-0.9$ due to the adopted IMF-averaged AGB yields.
The formation of E populations might be a result of star formation directly from
the gaseous ejecta of very massive AGB stars.

C09a pointed out that a simple dilution model with the same 
AGB yields
cannot reproduce the observed distributions of stars
on the [O/Fe]-[Na/Fe] plane for NGC 2808 and M4 simultaneously.
We suggest that the IMF-averaged yields can be quite different between
NGC 2808 and M4, and this yield difference can account for the 
different  evolutionary loci observed
for these two GCs.
Possibly, only high-mass AGB stars ($m_{\rm agb} \ge 7 {\rm M}_{\odot}$) 
might have polluted
a GMC forming NGC 2808 so that the 2P stars can have very low [O/Fe] and [Na/Fe].
NGC2808 represents an extreme case in this context, hosting some of the most chemically 
anomalous populations of any Galactic GC, including a Mg-K anticorrelation (see Section 4.1.1), 
suggesting that it's parent GMC was polluted by unusually massive sAGB stars operating at 
exceptionally high HBB temperatures, beyond what IMF-averaged yields for a standard mass range can reproduce.
Although it is beyond the scope of this paper to discuss what determines
the mass ranges of AGB stars that can pollute GMCs,
$t_{\rm life}$ (GMC lifetime) could be one of the key factors for this.

Fig. 10 describes how the shapes of the simulated O-Na anticorrelations depend on [Fe/H]
and $\alpha$ in the models based on AGB yields from D18.
Clearly, the slopes are steeper for higher [Fe/H] in the three models
with [Fe/H]=$-1.3$, $-1.0$, and $-0.7$
for a given $\alpha$ (=2.35).
The 1$\sigma$ dispersions in [Na/Fe] ($\sigma$([Na/Fe])) 
can therefore be smaller for lower [Fe/H]: 0.22, 0.19, and 0.14 for [Fe/H]$=-1.3$,
$-1.0$, and $-0.7$, respectively. Therefore,
\begin{equation}
\Delta \sigma( {\rm [Na/Fe]} )/\Delta {\rm [Fe/H]} \approx -0.13
\end{equation}
in these models. 
Since the efficiency of the Ne-Na cycle during HBB is sensitive to the metallicity through its 
effects on the temperatures reached at the base of the convective envelope (e.g., V13), 
the present models predict a negative correlation between $\sigma$([Na/Fe]) and 
[Fe/H] that represents a direct observational test of the temperature-dependent nucleosynthesis described above.
The IMF slope $\alpha$ does not have a major effect on the shape of the
O-Na anticorrelations in the high-metallicity models with [Fe/H]=$-0.7$.

\begin{figure}
\psfig{file=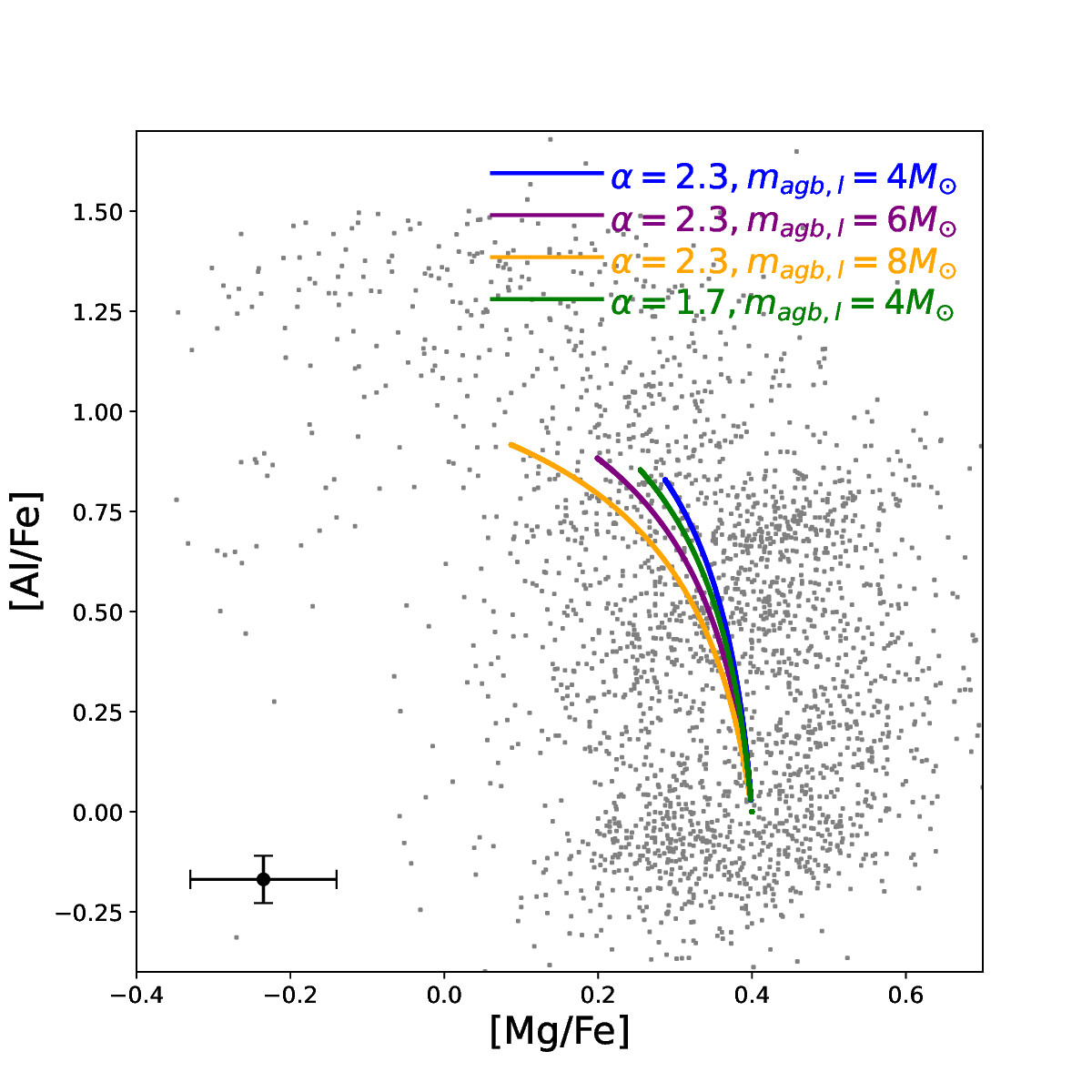,width=8.5cm} 
\caption{
Same as Fig. 9 but for Mg-Al anticorrelations. Observational data from ME20
 is used here.
}
\label{Figure. 11}
\end{figure} 

\begin{figure}
\psfig{file=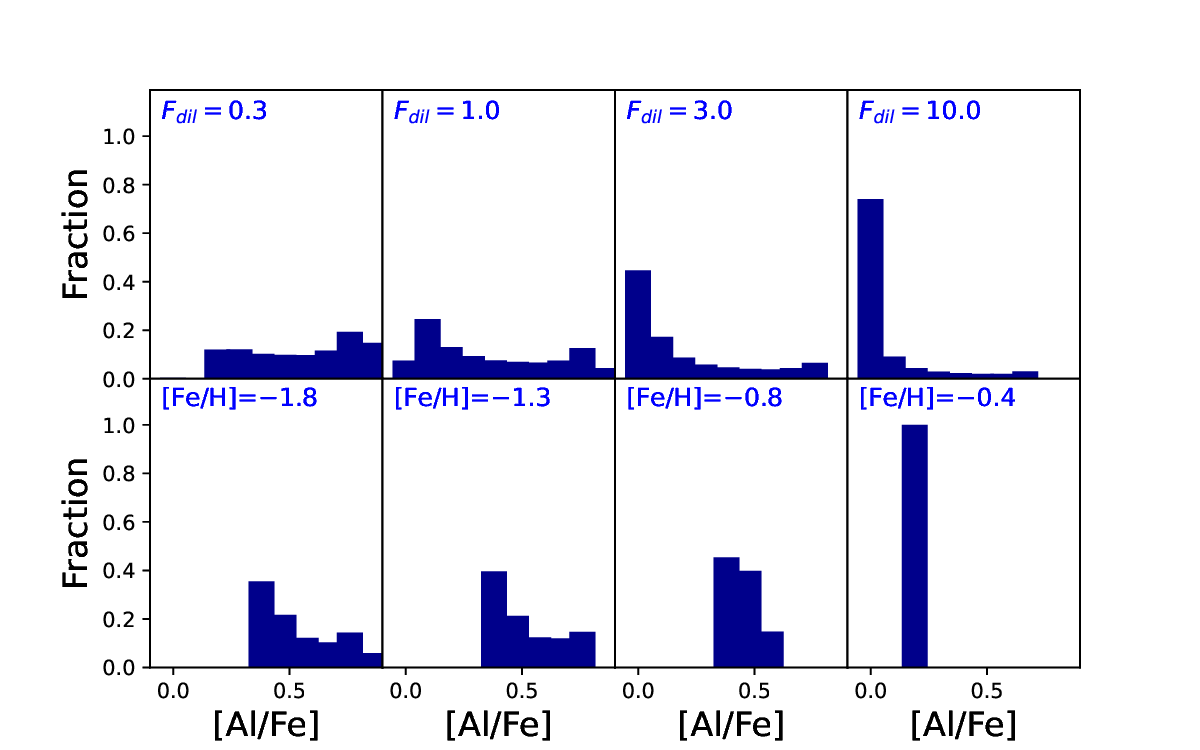,width=8.5cm} 
\caption{
Upper  and lower four panels shows how [Al/Fe] distributions of stars
depend on $F_{\rm dil}$ and [Fe/H] in GCs, respectively.
}
\label{Figure. 12}
\end{figure}

\subsection{Mg-Al anticorrelation}

The Mg-Al anticorrelation requires more extreme HBB conditions than the 0-Na anticorrelation, 
becoming active only when temperatures at the base of the convective envelope are sufficient to drive proton 
capture on $^{24}$Mg through the Mg-Al chain to produce $^{27}$Al (V13). 
The extent and variations in the strength of this anticorrelation relative to O-Na reflects this higher temperature threshold.
Fig. 11 describes how the evolutionary loci of the three models
on the [Mg/Fe]-[Al/Fe] plane
are determined by $\alpha$ and $m_{\rm agb, l}$ for a fixed $m_{\rm agb, u}$
and a fixed star formation history.
These models can reproduce the observed Mg-Al anticorrelation at least 
qualitatively,
though they cannot reproduce the E populations with [Mg/Fe]$<0$ observed in some
GCs such as NGC 2808.
ME20 reported a turn over in the Mg-Al anticorrelation among stars in the most metal poor GCs, specifically 
M15 and M92, where Al abundances decreased rather than continue to rise at low [Mg/Fe]. This behavior is interpreted as 
evidence that HBB temperatures in the polluters of these GCs were sufficiently extreme to activate proton capture on 
$^{27}$Al to $^{28}$Si. This reflects the temperature sensitivity of the 
Mg-Al chain. The 8 ${\rm M}_{\odot}$ model begins to reach these extreme [Mg/Fe] 
abundances, but no model can replicate the [Al/Fe] turnover that was reported for the most metal poor GCs.

The two  models with different $\alpha$ yet the same $m_{\rm agb, l}$ have very similar
evolutionary loci on the [Mg/Fe]-[Al/Fe] plane, which implies that the IMF is not 
crucial in the distributions of GC stars on this plane.
The difference in [Mg/Fe] between 1P and 2P stars is more remarkable
in the models with different $m_{\rm agb, l}$,  which confirms that
$m_{\rm agb, l}$ is more important than $\alpha$ in determining chemical abundance
patterns of GCs with MPs.

Fig. 12 shows the distributions of [Al/Fe] (histograms) in 8 models with
different $F_{\rm dil}$ and [Fe/H]. 
Two peaks in [Al/Fe] distributions can be more clearly seen in the models
with lower $F_{\rm dil}$ (=0.3 and 1.0) due to more efficient chemical enrichment
by AGB stars. However, 
the predicted bimodality is less pronounced than observed in ME20, 
likely reflecting the simplified treatment of star formation history within GMCs in the present one-zone framework.
It is found that
1$\sigma$ dispersions in [Mg/Al] ($\sigma$( {\rm [Al/Mg]} )
are larger for smaller $F_{\rm dil}$:
$\sigma$([Al/Mg])=0.27 and 0.19 for 
$F_{\rm dil}=0.3$ and $10.0$, respectively. 
Pancino et al. (2017) investigated the Mg-Al anticorrelations
of GCs using  data from
{\it Gaia}-ESO Survey and found a positive
correlation between 
$\sigma$( {\rm [Al/Mg]} ) and $M_{\rm gc}$. 
This positive correlation is qualitatively
consistent with the above predictions 
because $F_{\rm dil}$ is smaller for larger  $M_{\rm gc}$ (due to larger $M_{\rm gmc}$)
in the present models,

Fig. 12 also demonstrates that (i) [Al/Fe] distributions are more extended for lower
[Fe/H] and (ii) the [Fe/H]=$-0.4$ model  does not clearly show a Mg-Al anticorrelation
due to almost no [Al/Fe] variation.
This [Fe/H] dependence of Mg-Al anticorrelations has been already discussed  by
Ventura et al. (2016). 
The metallicity dependence of $\sigma$([Al/Fe]) 
reflects the higher HBB temperatures reached in the AGB polluters at low [Fe/H], enabling more complete 
Mg-Al processing.
It is clear that 1$\sigma$ dispersions in [Al/Mg] are smaller 
for higher [Fe/H] in these models: $\sigma$([Al/Mg]=0.2 and 0.06 for 
[Fe/H]=$-1.8$ and $-0.4$, respectively.
Accordingly, the slope in this anticorrelation is;
\begin{equation}
\Delta \sigma( {\rm [Al/Mg]} )/\Delta {\rm [Fe/H]} \approx -0.1
\end{equation}
in these models. 
Pancino et al. (2017) 
found that this slope is $\approx -0.2$, which 
is approximately  twice as  
steep as  the predicted slope in  this study.
Their results imply that
although the present models can 
qualitatively reproduce the observed anticorrelation between $\sigma$( {\rm [Al/Mg]} ) and [Fe/H],
quantitative  reproduction of the anticorrelation 
would require inclusion of physical processes that are not implemented in
the present models.

\begin{figure}
\psfig{file=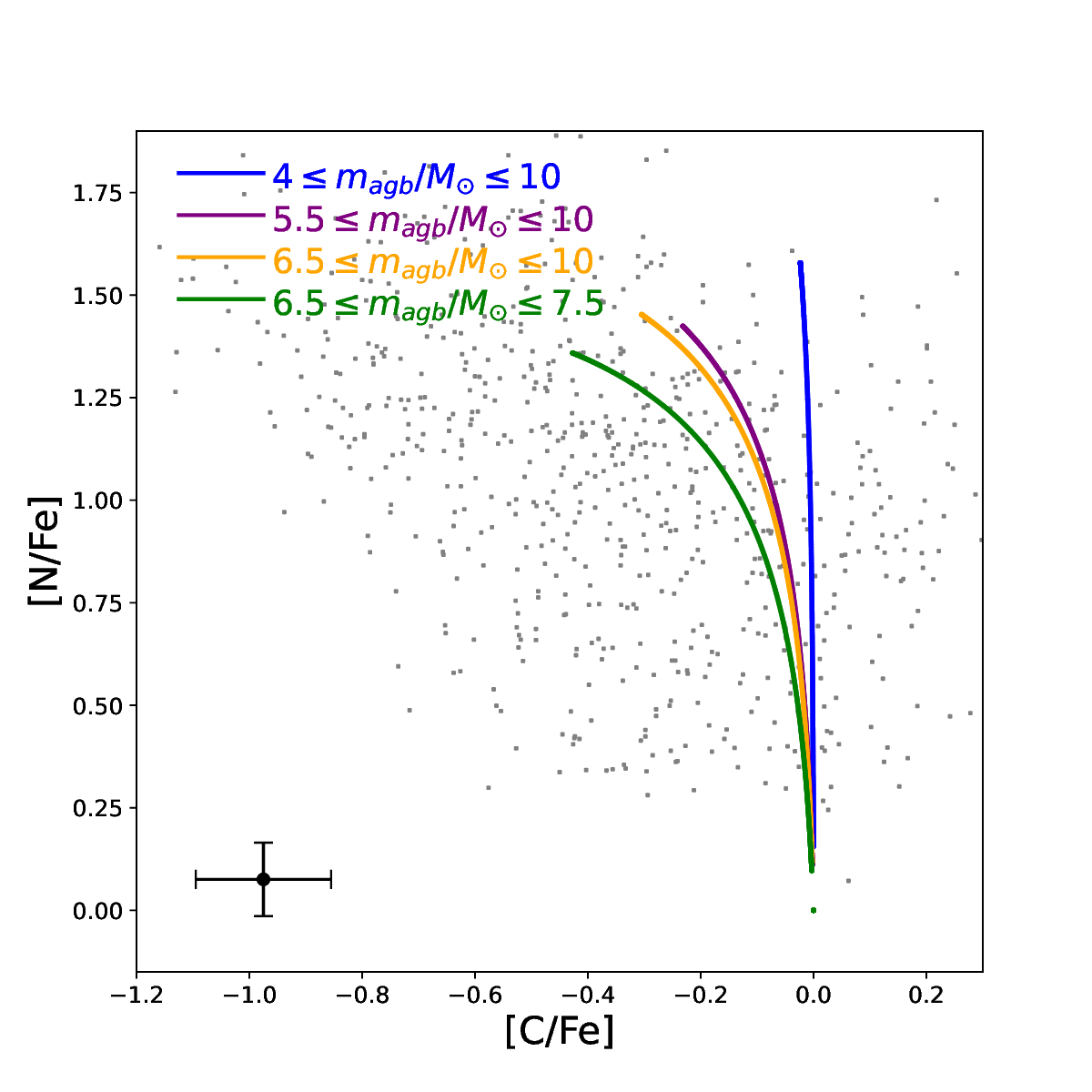,width=8.5cm} 
\caption{
Same as Fig. 9 but for C-N anticorrelations.
The AGB models by Ventura et al. (2013) are used to calculate the IMF-averaged AGB yields,
because D10 does not provide C and N yields.
Four models with $4 \le m_{\rm agb}/{\rm M}_{\odot} \le 10$ (blue),
$5.5 \le m_{\rm agb}/{\rm M}_{\odot} \le 10$ (purple),
$6.5 \le m_{\rm agb}/{\rm M}_{\odot} \le 10$ (orange),
and $6.5 \le m_{\rm agb}/{\rm M}_{\odot} \le 7.5$ (green) are plotted for $\alpha=2.3$.
Observational data from ME20 is used here.
}
\label{Figure. 13}
\end{figure} 

\subsection{C-N anticorrelation}

The C-N anticorrelation arises at the lowest temperatures of the 
proton-capture cycles relevant to GC chemical enrichment, with 
CN cycling converting $^{12}$C to $^{14}$N during HBB. Its near ubiquitous presence across GCs of different 
masses and metallicities reflects this comparatively low 
activation temperature, making it a less discriminating constraint on the 
precise masses of AGB polluters than the higher temperature cycles.
Fig. 13 shows (i) the distribution of GC stars from ME20 on the [C/Fe]-[N/Fe] plane
and (ii) the evolutionary loci of four models with
different mass ranges of AGB stars contributing chemical enrichment within GMCs 
on the plane.
Although C-N anticorrelations can be clearly seen in individual GCs with MPs (ME20),
such an anticorrelation is not so clearly apparent due to a large [C/Fe] dispersion
at a given [N/Fe] in this plot including all stars from different GCs.
This dispersion likely reflects a combination of genuine star-to-star abundance variations, measurement uncertainties 
arising from molecular CN and CO features, and evolutionary effects modifying initial C and N abundances.
Clearly,
the slopes of the simulated C-N  anticorrelations depend on the mass ranges
of AGB stars that can determine the IMF-averaged C and N yields.
There is almost no C-N anticorrelation (i.e., an almost vertical line) seen 
in the model with  $4 \le m_{\rm agb}/{\rm M}_{\odot} \le 10$, 
in which the IMF-averaged  [C/Fe] of AGB winds is almost identical to initial [C/Fe] due to
major contributions of lower mass AGB stars.
C-N anticorrelations can be more pronounced in
the models in which only high-mass stars with $m_{\rm agb}\ge 5.5 {\rm M}_{\odot}$ contribute
to chemical enrichment within GMCs. 
The model with
$5.5 \le m_{\rm agb}/{\rm M}_{\odot} \le 7.5$ shows the shallowest slope in
the C-N anticorrelation due to the lower IMF-average [C/Fe] and [N/Fe] of AGB winds.

The present models 
with AGB yields for [Fe/H]=$-1.6$ adopted from V13,  
has a  difficulty in reproducing 
[C/Fe]$<-0.6$ observed in a fraction of GC stars (ME20),
mainly because
the minimum possible [C/Fe] of AGB winds in the lowest metallicity model by V13 
is $-0.59$. This problem may not 
be critical, given that higher C yields would be possible in the updated
theoretical models for AGB stars.
V13 predicts that [C/Fe] of AGB winds can be as low as $-1$ in their higher metallicity
models ($Z=8 \times 10^{-3}$), which implies that the observed stars with [C/Fe]$\approx -1$
and high [N/Fe] $>1$ can be explained in the context of star formation from AGB winds
star with higher metallicity mixed with pristine GMC gas.
Although the present models predict that the slopes of C-N anticorrelations
could be diverse depending on the IMF and the mass ranges of polluting AGB stars,
the observed slopes are consistent across clusters  in ME20.
Also, an almost vertical C-N anticorrelation cannot be seen in GC samples by ME20,
which implies that AGB stars with $m_{\rm agb} \approx 4 {\rm M}_{\odot}$
are unlikely to significantly contribute to the chemical enrichment in GMCs. 
This is consistent with their marginal HBB temperatures with the potential to produce carbon enhanced, rather than nitrogen enhanced ejecta.

\begin{figure}
\psfig{file=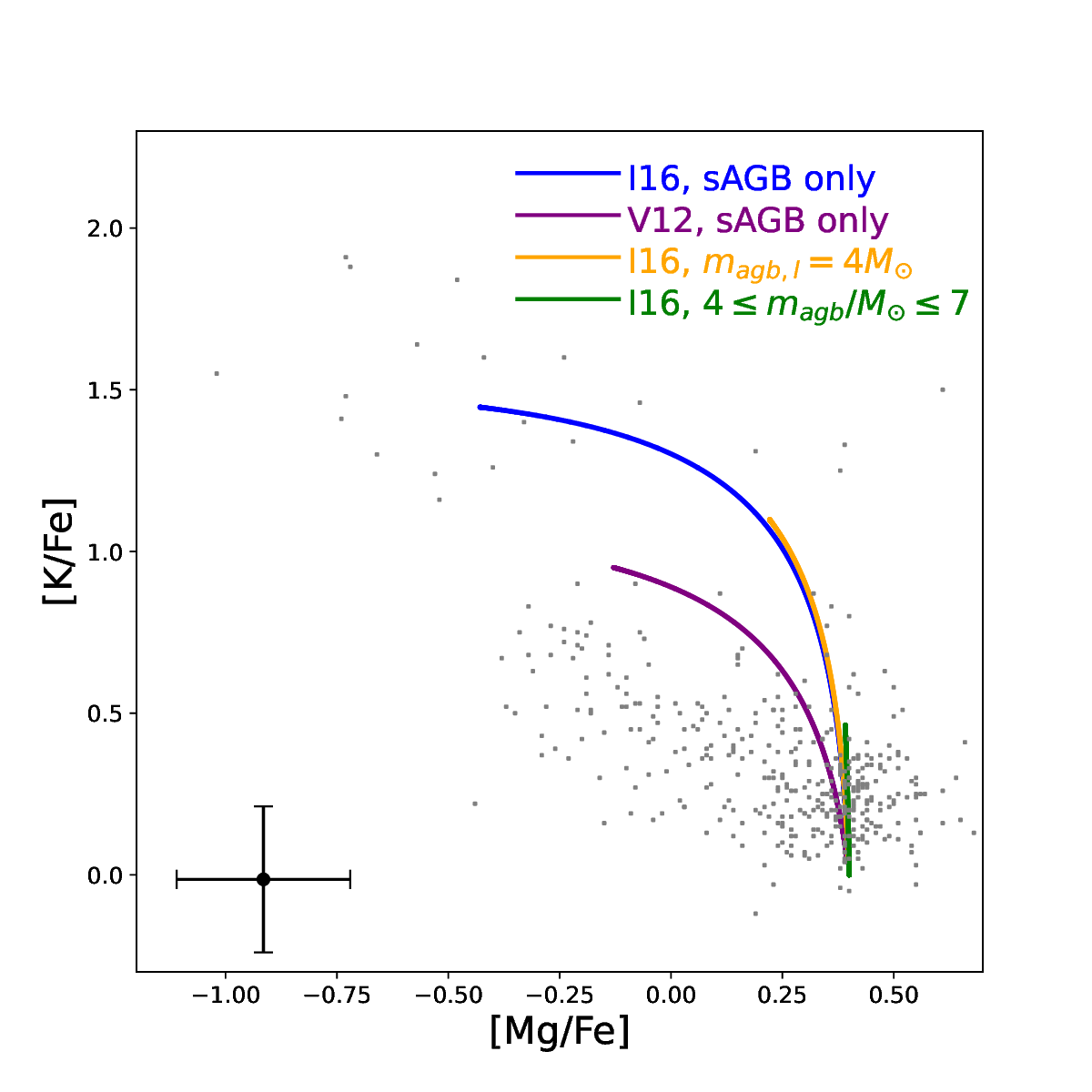,width=8.5cm} 
\caption{
Same as Fig. 9 but for Mg-K anticorrelations.
Different AGB yields are adopted in these four models. Only sAGB stars 
are assumed to contribute to chemical enrichment, and the  yields of sAGB stars 
from I16 (blue) and from V13 (purple) are adopted in the first two models.
The models with sAGB yields from I16 and IMF-averaged yields 
for $4 \le m_{\rm agb,l}/{\rm M}_{\odot} \le 10$ (orange)
and for $4 \le m_{\rm agb,l}/{\rm M}_{\odot} \le 7$ (green)
are also shown for comparison.
Observational data sets for $\omega$ Cen (Alvarez-Garay et al. 2024) and for NGC 2419 (Mucciarelli et al. 2010)
are used here.
}
\label{Figure. 14}
\end{figure}

\subsection{Mg-K}
The Mg-K anticorrelation represents the most extreme proton-capture signatures observed in GC stars, 
requiring HBB temperatures significantly higher than those needed to activate either the Ne-Na or Mg-Al cycles, 
through the Ar-K chain operating in the most massive sAGB polluters. Its detection in only a 
limited number of GCs is a direct consequence of this extreme temperature requirement.
Recent theoretical studies of stellar nucleosynthesis have demonstrated that
sAGB stars and nova are the best possible polluting stars responsible for the
observed Mg-K anticorrelation of NGC 2419 (V12;
I16; Fox et al. 2024). Clearly,
if only sAGB stars with $6.5 \le m_{\rm agb}/{\rm M}_{\odot} \le 10$
contribute to chemical enrichment in GC formation,
the observed Mg-K anticorrelation can be naturally reproduced for NGC 2419  (V12; I16).
However, both AGB and sAGB would  be able to
enrich GMCs through their stellar winds in real GC formation. 
We therefore consider the following two cases:
(i) only sAGB stars chemically enrich GMCs (``sAGB only'' model)
and (ii) both AGB and sAGB stars chemically enrich GMCs.
Different evolutionary loci of the models on the Mg-K plane
between the two cases 
can provide clues as to why the distributions of stars on
the Mg-K plane appear to be different between different GCs.
Since K yields are not available for AGB stars in D10 and V13,
we adopt an assumption of [K/Fe]=0 for all AGB stars just for simplicity.

Fig. 14 demonstrates that the sAGB only  model with I16 yields can show
a Mg-K anticorrelation, though it does not produce 2P stars with very low Mg
($\approx -1$) due to too much dilution of sAGB ejecta with pristine GMC gas.  
2P stars in the sAGB only model with V12 yields have systematically
lower [K/Fe] and higher [Mg/Fe], as expected for the chemical yields by V12.
A large fraction of stars have lower [K/Fe] for a given [Mg/Fe] compared
with the predictions of these two models, in particular,
for $-0.5 \le {\rm [Mg/Fe]} \le 0.25$. 
This apparent inconsistency will need to be reassessed  by our future  theoretical
models when sAGB yields for different $m_{\rm agb}$ and [Fe/H] are available.
It should be noted here that  2P stars with moderately
high [K/Fe] ($\approx 0.5$) and low [Mg/Fe] ($\approx -0.3$) observed in NGC 1786
(Alvarez-Garay  et al. 2025) cannot be well reproduced by any models in this study.
 We note that K abundance measurements in GCs carry additional systematic uncertainties. 
The primary K I diagnostic line at $\sim 7699$ \AA requires careful corrections for telluric contamination 
from the oxygen A band, while both KI resonance lines 
may be subject to significant departures from local thermodynamic equilibrium.

The model with $m_{\rm agb, l}=4 {\rm M}_{\odot}$, in which both AGB and sAGB
stars contribute to chemical enrichment within GMCs, shows a very steep
Mg-K anticorrelation with a narrow range of [Mg/Fe] in 1P and 2P stars. 
This result is due to  mixing of gaseous ejecta with high IMF-averaged [Mg/Fe] 
from AGB and sAGB stars with pristine GMC gas.
The model with $4 \le m_{\rm agb}/{\rm M}_{\odot} \le 7$ has a narrow mass range
of  sAGB stars with $6.5 \le m_{\rm agb} \le 7 {\rm M}_{\odot}$ 
so that the IMF-averaged [Mg/Fe] and [K/Fe] of AGB ejecta
can be very similar to [Mg/Fe] and [K/Fe] of pristine GMC gas.
Consequently, the 1P and 2P stars have similar [Mg/Fe] and [K/Fe] and no clear
Mg-Al anticorrelation can be seen in this model.
These results demonstrate that the mass range of AGB stars contributing
to chemical enrichment in GC formation is a crucial factor to determine
the distributions of 1P and 2P stars on the Mg-K plane.
Mg-K plane, and by extension, the HBB temperatures reached by the more massive sAGB stars, driving more 
extreme nucleosynthesis.

Based on these results, we suggest that the GMC forming NGC 2419
was polluted only by sAGB stars, possibly because the lifetime of the GMC
was not long enough so that  AGB stars with 
$m_{\rm agb}\le 6.5 {\rm M}_{\odot}$, which can eject gas later than sAGB stars,
can pollute the GMC.
It would be possible that a large number electron-capture supernovae from massive stars
with $8 \le m/{\rm M}_{\odot} \le 10$ expelled  the remaining pristine gas
of the GMC immediately from chemical enrichment by sAGB stars.
Such supernova explosions would be able to  truncate further chemical enrichment 
by AGB stars with $m_{\rm agb} \le 6.5 {\rm M}_{\odot}$  completely.
This explanation is highly speculative, thus its validity needs to be assessed by
numerical simulations of GC formation.

Alvarez Garay et al. (2026) has recently revealed no stars with [Mg/Fe]$<0$ in
M54, in which the majority of stars have [Fe/H] ranging from $-1.6$ to $-1.0$,
though M54 shows a clear Mg-K anticorrelation.
This observation cannot be simply explained by the present models with sAGB stars,
which implies that metallicities play a role in shaping the anticorrelation.
Dell'Agli et al. (2018) indeed demonstrated that 
[Mg/Fe] of gaseous ejecta from AGB stars with $Z=2.5 \times 10^{-3}$ and
$4.5 \le m_{\rm agb}/{\rm M}_{\odot} \le 6.5$ cannot be lower than 0.2,
though they did not show K yields of these stars.
It is therefore  possible that sAGB stars with higher [Fe/H] ($\approx -1$) cannot produce
gas with [Mg/Fe]$<0$.
Future models incorporating a broader grid of metallicity-dependent sAGB yields will enable a more complete 
exploration of how polluter metallicity shapes the Mg-K anticorrelation across the diversity of observed GCs.

Thus, both the mass ranges of polluting AGB stars and their metallicities
can determine the shapes of Mg-K anticorrelations in the new scenario.
The  mass ranges of polluting AGB stars (i.e., $m_{\rm agb, l}$ and $m_{\rm agb, u}$) 
in  GCs could be a complex function of star formation histories of their host galaxies,
growth histories of their parent GMCs,
and GMCs' lifetimes ($t_{\rm life}$).
Our future hydrodynamical simulations of GC formation therefore need to clarify 
what determines the mass ranges of AGB stars polluting GMCs in various galaxy
environments.

\begin{figure}
\psfig{file=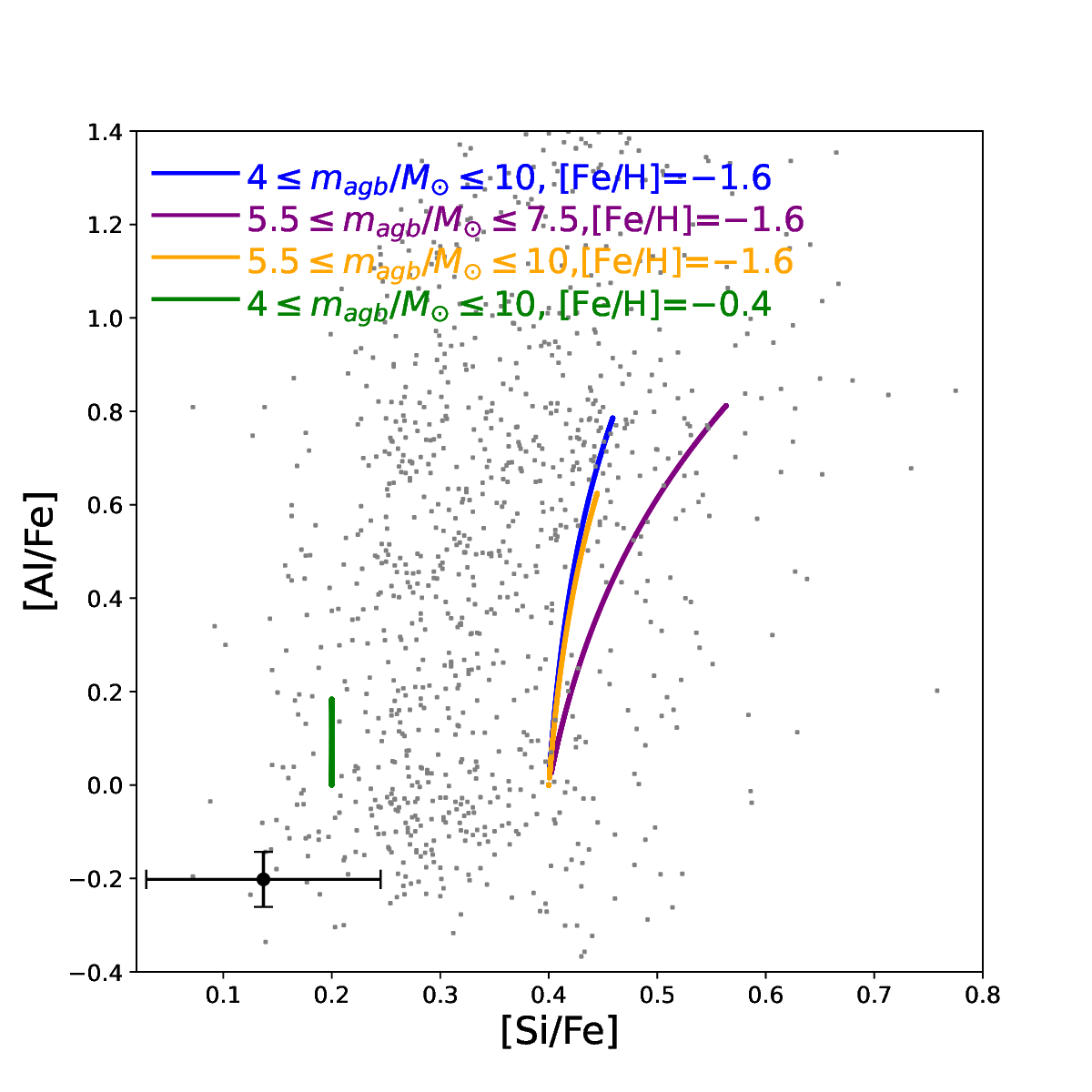,width=8.5cm} 
\caption{
Same as Fig. 9 but for Si-Al correlations.
The models 
with [Fe/H]=-1.6 and  $4 \le m_{\rm agb,l}/{\rm M}_{\odot} \le 10$ (blue),
with [Fe/H]=-1.6 and  $5.5 \le m_{\rm agb,l}/{\rm M}_{\odot} \le 7.5$ (purple),
with [Fe/H]=-1.6 and  $5.5 \le m_{\rm agb,l}/{\rm M}_{\odot} \le 10$ (orange),
and with [Fe/H]=-0.4 and  $4 \le m_{\rm agb,l}/{\rm M}_{\odot} \le 10$ (green),
are shown.  For the high-metallicity model, initial [Si/Fe]=0.2 instead of 0.4 is used
for consistency with model predictions by V13.
Data from ME20 is used here.
}
\label{Figure. 15}
\end{figure} 

\subsection{Si-Al correlation}

The Si-Al correlation arises when HBB temperatures are sufficient to drive proton capture beyond the 
Mg-Al chain, with $^{27}$Al serving as the seed nucleus for $^{28}$Si production via the 
$^{27}$Al(p,$\gamma$)$^{28}$Si reaction. 
The resulting positive correlation between Si and Al, rather than the anticorrelation seen in the 
Mg-Al plane, is therefore a signature of the most massive and metal-poor 
AGB polluters in which the full Mg-Al-Si nucleosynthetic pathway operates.
Fig. 15 shows the evolutionary loci of four models with
different AGB yields on the [Si/Fe]-[Al/Fe] plane and compares 
with the corresponding observations from ME20.
Clearly, steep  Si-Al correlations can be seen in 
the three low-metallicity models with [Fe/H]=-1.6, which is at least qualitatively
consistent with observations by ME20.
However, the adopted initial [Si/Fe] (=0.4), which is consistent with
the V13 model, appears to be significantly
higher than the observed mean [Si/Fe] of GC stars (ME20).
Consequently, the simulated Si-Al correlation deviates from the observed one,
which implies that lower initial [Si/Fe] ($\approx 0.3$)  needs to be adopted in the
three low-metallicity  models
to better reproduce the observed trend.
The higher metallicity  model with [Fe/H]=$-0.4$ shows an almost
vertical evolutionary locus and  a very narrow range
of [Al/Fe], which simply reflects the small Al yields of the AGB models
with [Fe/H]=$-0.4$  in V13. 
This result reflects the temperature requirements of Si production via HBB, that GCs with
unlikely to possess a clear Si-Al correlation.

The slopes of Si-Al correlations in the models with [Fe/H]=$-1.6$
depend on the ranges of AGB stars that
contribute to chemical enrichment in GCs, as seen in other elemental abundance patterns.
The slope is the steepest in the model in which
the IMF-averaged AGB yields for $4 \le m_{\rm agb}/{\rm M}_{\odot} \le 10$ 
are adopted, because the relative contributions of massive AGB stars,
which have high Si yields in V13,
are minor in the model.
If only massive AGB stars with 
$5.5 \le m_{\rm agb}/{\rm M}_{\odot} \le 7.5$  contribute to chemical enrichment
in GMCs, then the slope becomes flatter with a wider range of  [Si/Fe].
The model with 
$5.5 \le m_{\rm agb}/{\rm M}_{\odot} \le 10$ shows
narrower  ranges of [Si/Fe] and [Al/Fe] 
due to the lower Si ([Si/Fe]=0.44) and Al yields ([Al/Fe]=0.6) 
of  high-mass AGB stars with $m_{\rm agb} \ge 7.5 {\rm M}_{\odot}$
in V13.

The IMF-averaged yields of Si and Al are used in the present models.
Accordingly, even if AGB stars with a particular mass range 
(e.g., $5.5 \le m_{\rm agb}/{\rm M}_{\odot}  \le 6$) eject winds with
high [Al/Fe] ($>0.8$) and high [Si/Fe] ($>0.6$),  new stars formed in the models
do not show [Al/Fe]$>0.8$ and [Si/Fe]$>0.6$ due to mixing of ejecta from AGB stars
with different masses. 
Reproducing stars with [Al/Fe]$> 0.8$ and [Si/Fe]$ > 0.6$ simultaneously would require 
contributions from individual massive sAGBs operating at very high HBB temperatures 
where $^{27}$Al(p,$\gamma$)$^{28}$Si reaction proceeds efficiently enough to produce both 
extreme Al enrichment and Si enhancement before the ejecta are diluted by pristine GMC gas.
However, if gaseous ejecta
from AGB star with $m_{\rm agb}=5.5 {\rm M}_{\odot}$ is mixed with pristine
GMC gas to be converted into new stars, the observed high [Si/Fe]
and [Al/Fe] can be achieved in the present models. A clear goal of  our future study is
to understand how stellar winds from individual AGB stars with different masses
mix with pristine GMC gas and ejecta from other AGB stars.

\begin{figure*}
\psfig{file=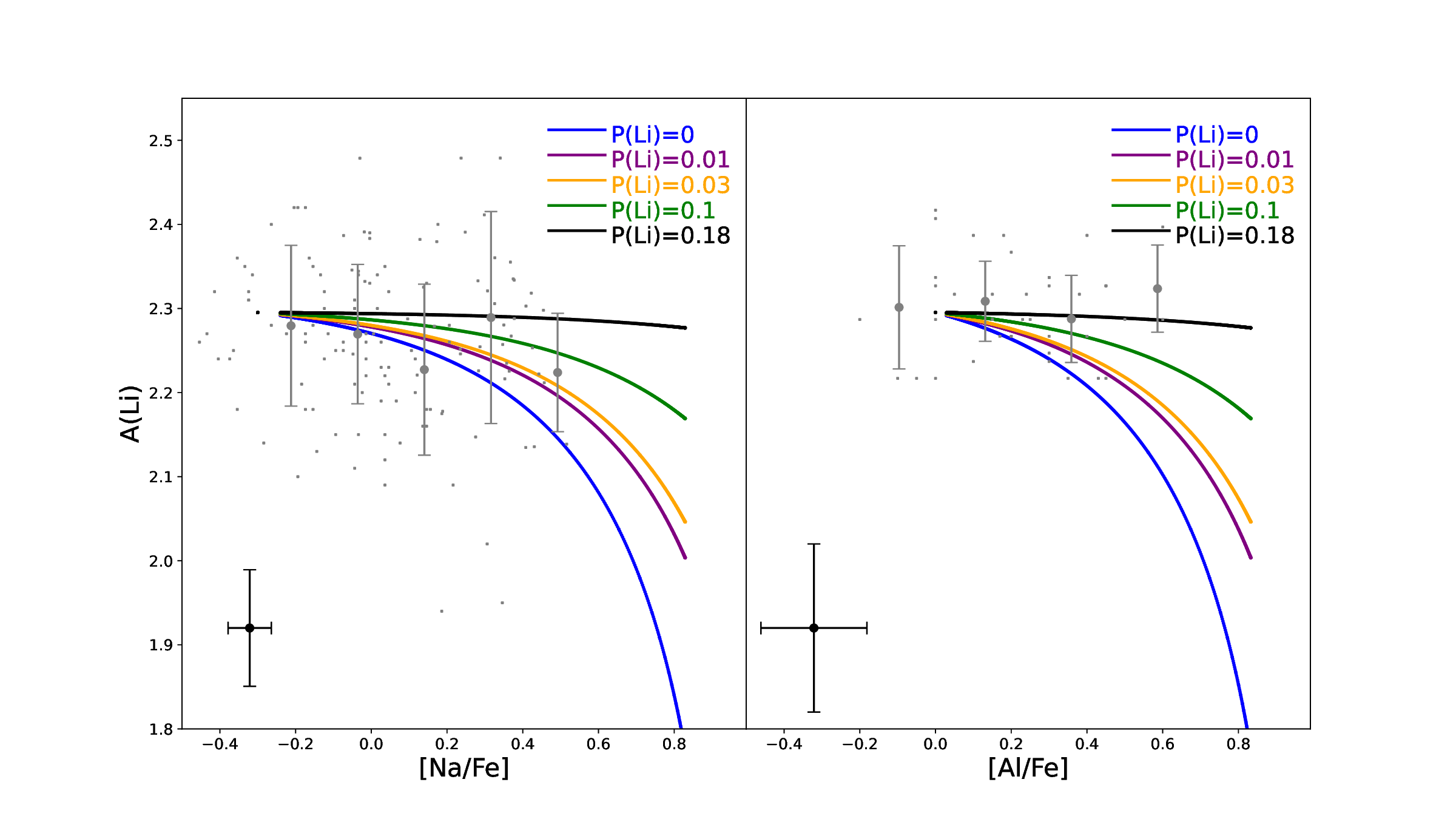,width=18cm} 
\caption{
A(Li) as a function of [Na/Fe] (left) and [Al/Fe] (right) 
in the five models with different fractions of Li-producing AGB stars 
(P(Li)): 0 (blue), 0.01 (purple), 0.03 (orange), 0.1 (green), and 0.18 (black).
Observational results  taken from Lind et al. (2009) and McKenzie et al. 2026 (MA26)  are plotted in the left panel,
and the mean and 1$\sigma$ dispersion for five [Na/Fe] bins are also  plotted with error bars.
Since A(Li) data from D'Orazi et al. (2014) is for evolved stars (i.e., Li-depleted),
we add 1.3 to A(Li) of these stars to reproduce the original A(Li) during unevolved phases in the right panel.
We thus make a physically meaningful comparison between these observations and the 
model predictions for the (anti)correlation between [Al/Fe] and A(Li).
}
\label{Figure. 16}
\end{figure*} 

\subsection{Li-Na and Li-Al relations}

Lithium represents a uniquely constraining diagnostic for GC formation scenarios because its surface abundance in 
giant stars is sensitive to both the original chemical enrichment by AGB 
polluters and the subsequent stellar evolution of the observed stars. In unevolved main sequence 
turn off stars, A(Li) reflects the original chemical composition of the gas from which they formed. However, as stars ascend the RGB, the first dredge up brings CN-cycle processed material to the surface, systemically 
depleting the original $^{7}$Li through convective dilution with the interior layers where it has been 
destroyed (e.g., Karakas \& Lattanzio 2014). For AGB polluters, $^{7}$Li can be transiently produced 
through the Cameron-Fowler mechanism, where $^{7}$Be 
synthesized at the base of the convective envelope during 
HBB is transported to cooler outer layers and converted to $^{7}$Li via electron capture before it can be destroyed 
(Cameron \& Fowler 1971; Sackmann \& Boothroyd 1992). 
Measuring A(Li) in the context of MPs therefor ideally requires unevolved turnoff or subgiant stars to 
avoid the complications of RGB depletion, however spectra of high enough quality for this analysis are observationally 
expensive.

We here assume that there are two types of polluting stars,  i.e.,  one ejecting Li-deficient gas with A(Li)=1.56
and the other ejecting Li-rich gas with A(Li)=2.8, in order to more clearly demonstrate whether Li-production
in polluting stars is required to reproduce the observed (anti)correlations of A(Li) with [Na/Fe] and [Al/Fe]. 
P(Li) is defined as  the mass fraction of polluters  producing Li (A(Li)=2.8)
among all stars with $4 \le m/{\rm M}_{\odot}\le  10$ 
and considered to be a free parameter  in the present study.
Fig. 16 describes  how the evolutionary loci of the present 
five  models on the A(Li)-[Na/Fe] and A(Li)-[Al/Fe] planes
depend on  Li production rates (P(Li)).
Steep A(Li)-[Na/Fe] and A(Li)-[Al/Fe] anticorrelations can be seen in
the model without Li production (P(Li)=0), which is inconsistent with
observational results (e.g., Lind et al. 2009; Monaco et al. 2010; D'Orazi et al. 2014). 
The models with larger P(Li) has flatter anticorrelations,
which result from  more A(Li) production  in these models.
A very flat A(Li)-[Na/Fe] relation can be seen only in the model with P(Li)=0.18,
which suggests that a significant fraction of AGB stars
need to produce Li-rich ejecta to reproduce
the observed very flat A(Li)-[Na/Fe] and A(Li)-[Al/Fe] relations in GCs.

Since not all of AGB stars can produce  Li with A(Li)$>2.3$ (e.g., D12; V13),
a crucial question  here is whether P(Li)$\approx 0.2$ is possible 
for a given  IMF.
Using the A(Li) table for AGB stars with different masses listed in D12, we find
that the mass fraction of Li-producing AGB stars is 0.29, 0.29 and 0.25 for
$\alpha=1.7$, 2.3, and 2.7, respectively.
Therefore, the required P(Li)$\approx 0.2$ for the flat A(Li)-[Na/Fe] and A(Li)-[Al/Fe]
relations is possible
in the present GC formation scenario. 
It is also found that the IMF-average A(Li) in AGB stars 
with $4 \le m_{\rm agb}/M_{\odot} \le 10$ is 2.4 for the Li yields
from D12. These results
imply that the present SCI scenario is quite promising in terms of 
explaining the observed flat relations.
Thus the  observed lack of an A(Li)-[Na/Fe] and A(Li)-[Al/Fe]  anticorrelations in GCs strongly
suggests that GC formation scenarios incapable of Li production
are disfavoured by the current observational evidence. The present scenario
predicts that some GCs with lower P(Li) should show very mild A(Li)-[Na/Fe] anticorrelations,
which can explain the observed A(Li)-[Na/Fe] in a few  GCs (e.g.,  Pasquini et al. 2005).

The 1$\sigma$ dispersions in A(Li) ($\sigma($A(Li))) and
differences in minimum and maximum A(Li) ($\delta_{\rm max}$(A(Li)) in simulated GCs
depend both on P(Li) and on
$F_{\rm dil}$, as summarized in Table 2. 
As expected, $\sigma($A(Li)) and $\delta_{\rm max}$A(Li)
are larger in the models with smaller P(Li) due to more efficient
formation of Li-deficient
stars from cold gas polluted by AGB ejecta with lower A(Li).
$\sigma$(A(Li)) and $\delta_{\rm max}$A(Li)
are larger in the models with smaller $F_{\rm dil}$,
which implies that more massive GCs are likely to have
large $\sigma$(A(Li)) and $\delta_{\rm max}$A(Li) due to 
smaller $F_{\rm dil}$ (caused by larger $R_{\rm s}$) in the present scenario.
It is currently unknown whether GCs have $\sigma$(A(Li)) and $\delta_{\rm max}$A(Li)
dependent on their total masses.

As pointed out by D12, 
2P stars can have very high A(Li) ($>2.9$),
if they are formed from gaseous ejecta from sAGB stars with 
$m_{\rm agb}=8 {\rm M}_{\odot}$ without mixing so much
with pristine gas.
D12 also suggested that a few  Li-rich stars observed in M4 and NGC 6397
could be intriguing  examples of 2P stars formed almost directly
from gaseous ejecta of sAGB stars.
If 2P stars form only after the ejecta of multiple AGB stars with different masses have 
mixed within the GMC, the resulting A(Li) will reflect an IMF-averaged value that is significantly 
lower than that of the most Li-rich individual AGB ejecta because of the contributions of 
Li-poor lower mass AGB stars that dominate by number.

It should be stressed here that 
the IMF-weighted Li yield (A(Li)) depends on the mass range of AGB stars,
[$m_{\rm agb,l}$,$m_{\rm agb, u}$]${\rm M}_{\odot}$, for a given IMF slope.
It is 2.39 for [3, 7.5],
2.20 for [3, 7.5],
2.20 for [4, 7.5],
2.25 for [5, 7.5],
3.19 for [3, 10.0],
and 3.46 for [5, 10.0] for the Salpeter IMF with $\alpha=2.35$.
Therefore, the average A(Li) depends strongly on how much sAGB stars 
with $m_{\rm agb}\ge 7.5 {\rm M}_{\odot}$ can continue to chemical enrichment
within GC-forming GMCs. 
The normal A(Li) observed in the Na-rich 2P stars in most GCs suggests that sAGB stars, 
despite their role in driving the Mg-K anticorrelation in extreme clusters such as NGC2419, 
do not dominate the bulk chemical enrichment responsible for the ubiquitous Na-O anticorrelation, 
consistent with their high predicted Li yields driving A(Li) far above observed values if they were the primary polluters.

\begin{table}
\centering
\begin{minipage}{85mm}
\caption{Li abundance properties of GCs with MPs predicted from 
 the new scenario. }
\begin{tabular}{lllll}
P(Li)  & $F_{\rm dil}$ &  ${\rm A}_{\rm m}$(Li) & $\sigma$(A(Li)) & $\delta_{\rm max}$A(Li) \\
0.0 & 3.0 &  2.22 & 0.13 & 0.52  \\
0.01 & 3.0 &  2.25 & 0.07 & 0.29  \\
0.03 & 3.0 &  2.25 & 0.06 & 0.25  \\
0.10 & 3.0 &  2.27 & 0.03 & 0.13  \\
0.18 & 3.0 &  2.25 & 0.00 & 0.02  \\
0.01 & 0.3 &  2.13 & 0.11 & 0.34 \\
0.01 & 1.0 &  2.20 & 0.10 & 0.33  \\
0.01 & 10.0 &  2.28 & 0.04 & 0.21  \\
\end{tabular}
\end{minipage}
\end{table}

\begin{figure*}
\psfig{file=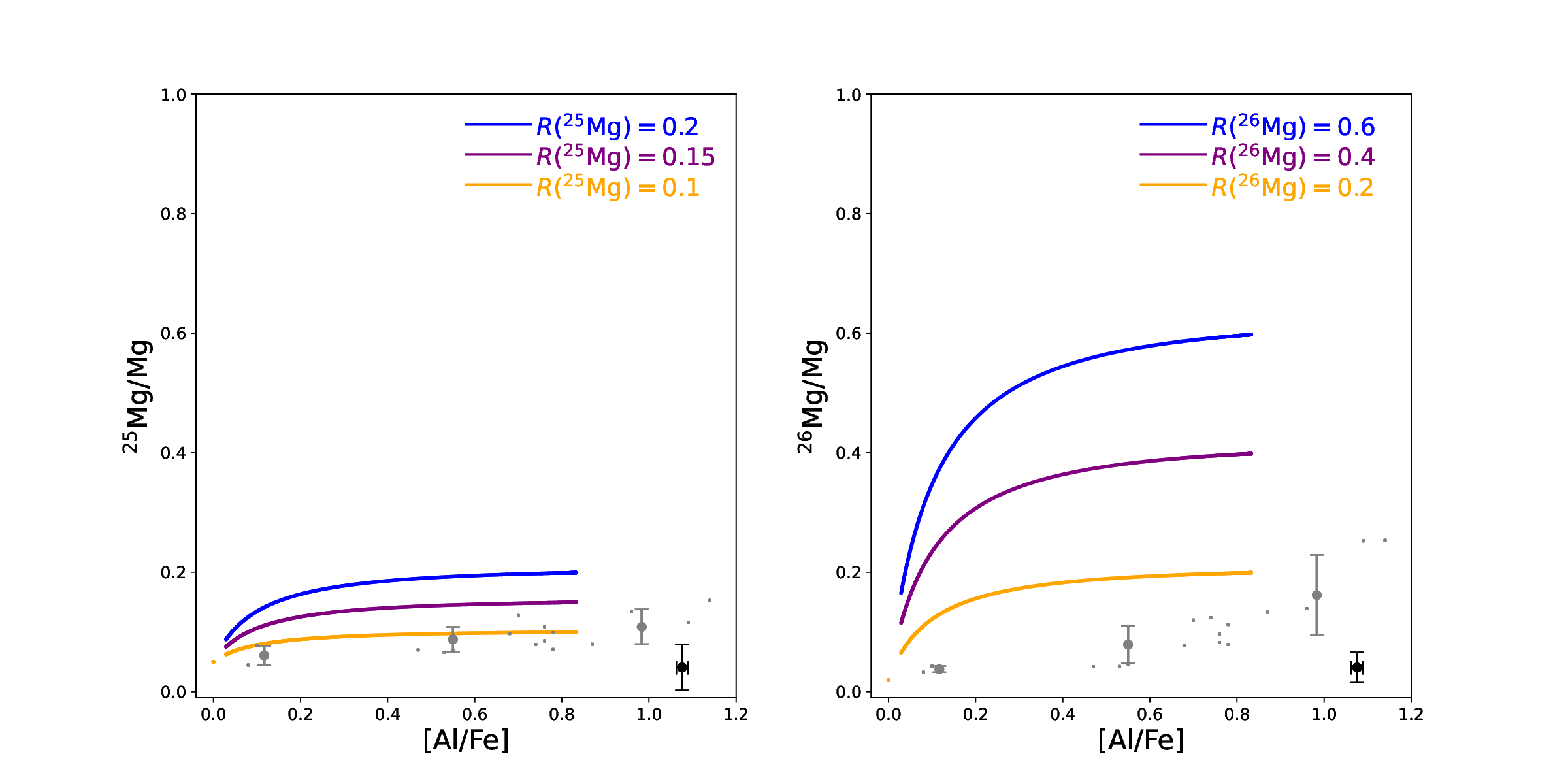,width=18cm} 
\caption{
${25}$Mg/Mg (left) and $^{26}$Mg/Mg ratios (right) as a function of [Al/Fe] for three models 
with different AGB ejecta compositions, denoted as R(${25}$Mg/Mg) and R(${26}$Mg/Mg), respectively. 
In the left panel, R(${25}$Mg/Mg) is set to 0.2 (blue), 0.15 (purple), and 0.1 (orange). In the right panel, R(${26}$Mg/Mg) is set to 0.6 (blue), 0.4 (purple), and 0.2 (orange). 
Reanalysis of  archival data from Yong et al. (2003) for NGC 6752  (McKenzie et al. in prep) is also plotted: 
the mean observational errors
are indicated by black error bars.
The mean and 1$\sigma$ dispersion for the [Al/Fe] bins are indicated by gray points and error bars, respectively. For all models, the initial $^{25}$Mg/Mg and $^{26}$Mg/Mg ratios of the pristine gas are assumed to be 0.05 and 0.02, respectively.
}
\label{Figure. 17}
\end{figure*}

\subsection{Mg isotope ratios}

Magnesium isotope ratios provide a uniquely sensitive probe of the temperatures and nuclear reaction conditions 
during HBB in AGB polluters, since $^{24}$Mg is destroyed by the Mg-Al chain 
while $^{25}$Mg and $^{26}$ Mg accumulate as intermediate products. 
The observed pattern of $^{26}$ Mg enhancement with increasing [Al/Fe] was first established in 
Yong et al. (2003), and through a reanalysis of the spectra by McKenzie et al. (in prep), the first evidence of a correlation between 25Mg and [Al/Fe] has also been detected.
Since the predicted  abundances of Mg isotopes (${}^{25}$Mg and ${}^{26}$Mg)
in stellar winds of AGB stars are given 
only for $m_{\rm agb}=5$ and $7 {\rm M}_{\odot}$ in V12, 
it is currently not possible with the available yields  for the present study
to derive the IMF-averaged yields of Mg isotopes.
We accordingly adopt a wide range of Mg isotope yields and thereby
investigate the time evolution of Mg isotope ratios in our GC formation models.
We here focus particularly on 
the model behaviour on the ${}^{25}$Mg/Mg-[Al/Fe] and ${}^{26}$Mg/Mg-[Al/Fe] planes to find
the best model that matches with observations.
Fig. 17 describes how the evolutionary loci of different models
on the ${}^{25}$Mg/Mg-[Al/Fe] 
and ${}^{26}$Mg/Mg-[Al/Fe] planes
depend on the adopted the mass ratios of ${}^{25}$Mg and  ${}^{26}$Mg  to Mg.

MA26 have recently discovered a weak positive correlation between [Al/Fe] 
and ${}^{25}$Mg/Mg extending to [Al/Fe] $\approx 1.1$ in NGC 6752.
However, this is the first cluster to see a trend in $^{25}$Mg with 
light element abundances (see Da Costa et al. 2013, Thygesen et al 2016).
Such a positive correlation can be reproduced well only by the model with 
$R({}^{25}{\rm Mg})=0.1$, though this ${}^{25}$Mg/Mg yield is lower than those predicted 
by Ventura et al. (2018) for $m=5 {\rm M}_{\odot}$ stars.
The apparent lack of stars between [Al/Fe] $=0.1$ and $0.4$ observed in NGC 6752 is not so well 
explained by the present models, which implies that the star formation histories of GCs in 
the models need to be revised. This could be also due to a selection effect.
Future work will explore the impacts of mass and metallicity as more 
Mg isotopic measurements are made available for a larger range of clusters.

Although the two models with $R({}^{26}$Mg)=0.4 and 0.6 shows clear positive
correlations between [Al/Fe] and ${}^{26}$Mg/Mg, the predicted non-linear
correlations (i.e., not straight lines) appear to be inconsistent with
observations. 
Furthermore, the models predict high  ${}^{26}$Mg/Mg ($>0.2$) already
at [Al/Fe]$<0.2$, which is not clearly seen in observations.
This high  ${}^{26}$Mg/Mg  is due largely to earlier chemical  enrichment of GMCs by
AGB stars in the models, which suggests that
time evolution of chemical enrichment by AGB stars needs to be fine-tuned
to explain the observations better in the present models.
The model with $R({}^{26}$Mg)=0.2 can better reproduce the observed correlation,
however, the model overpredicts ${}^{26}$Mg/Mg for [Al/Fe]$<0.8$.
These results imply that the observed distribution of GC stars
on the [Al/Fe] and ${}^{26}$Mg/Mg plane can contain both the Mg isotope
yields and the earlier chemical enrichment of GMCs by AGB stars.

\begin{figure}
\psfig{file=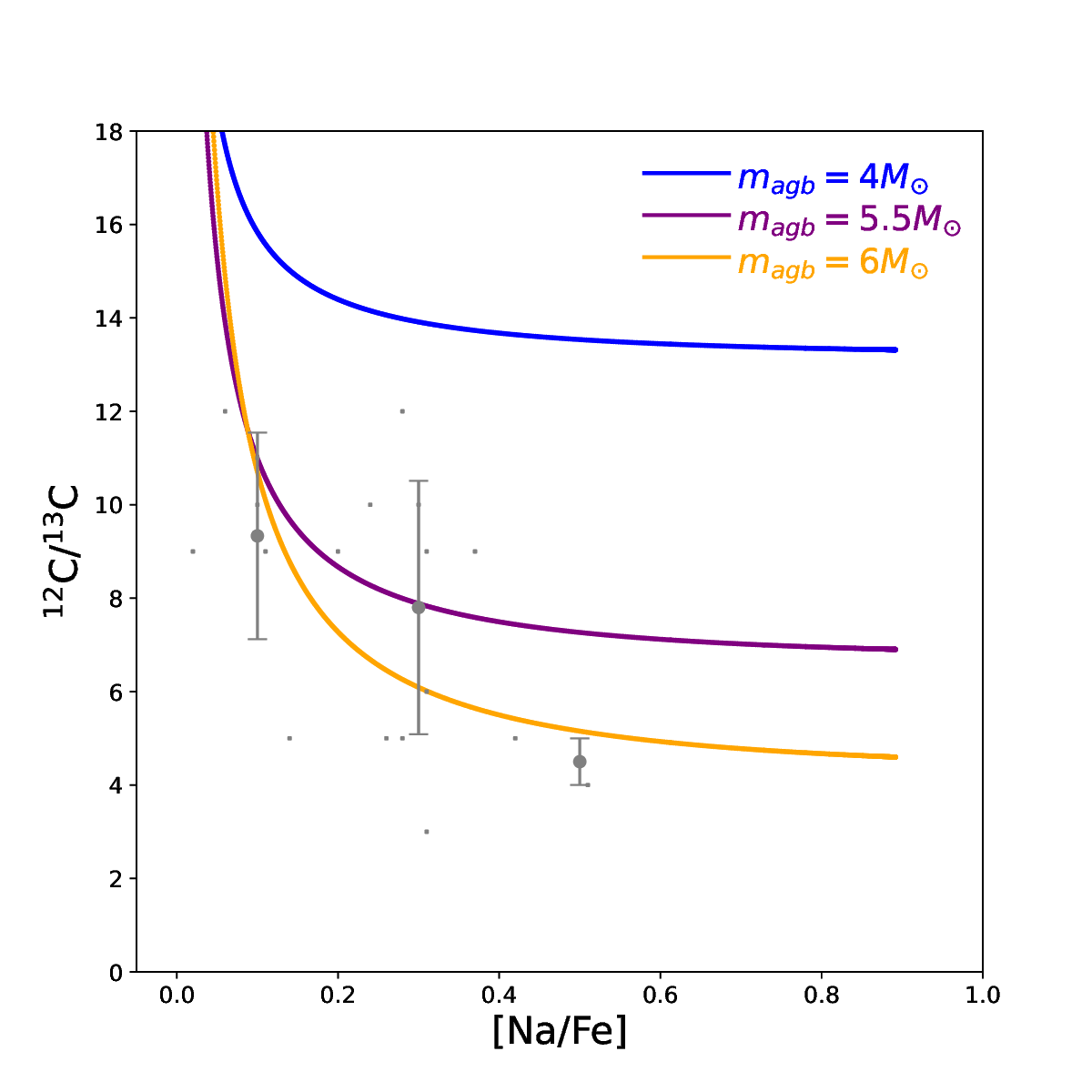,width=8.5cm} 
\caption{
$^{12}$C/$^{13}$C ratios as a function of [Na/Fe] in the three models with 
$m_{\rm agb}=4 {\rm M}_{\odot}$ (blue), 
$m_{\rm agb}=5.5 {\rm M}_{\odot}$ (purple), 
and $m_{\rm agb}=6 {\rm M}_{\odot}$ (orange). 
The observed data points from Carretta et al. (2015) are plotted by small gray circles,
and the mean and 1$\sigma$ dispersion  for [Na/Fe] bins are indicated by large gray points and error bars, respectively.
AGB yields from  a selected mass ($m_{\rm agb}$)
are used in each model (i.e., no IMF-averaged yields are used). 
}
\label{Figure. 18}
\end{figure} 

\subsection{C isotope ratios}

Carbon isotopic ratios provide a complementary diagnostic to lithium abundances in constraining GC formation, 
but share the same fundamental observational limitations. Both $^{12}$C/$^{13}$C
 and A(Li) are modified by stellar evolution as stars ascend the RGB. 
$^{12}$C is progressively converted to $^{13}$C 
through proton capture, producing lower ratios in the ejecta of more massive stars that reach higher temperatures. 
Therefore, meaningful comparisons between GC formation models and observations require measurements of unevolved main-sequence or subgiant stars.
The distributions of stars on the $^{12}$C/$^{13}$C-[Na/Fe] plane 
predicted from the present models  need to be
compared with those observed for unevolved stars in GCs.  Since
Carretta et al. (2005, C05) is the main study 
in which C isotope ratios are investigated for dwarfs and subgiants in a few
GCs,
we here use their observational results.
Since $^{12}$C and $^{13}$C yields are available for  a limited mass
range of AGB stars in K10, we are unable to calculate
the IMF-averaged yields. We accordingly use the yields  from a selected mass
$m_{\rm agb}$ (4.0, 5.5, and 6.0${\rm M}_{\odot}$) for $Z=0.0001$  from
K10. 
We here use  the yields from K10
and adopt the same model parameters (e.g., $F_{\rm dil}$) as those used
in the fiducial model.
Fig. 18 compares the predicted distributions of stars
on the [Na/Fe]-$^{12}$C/$^{13}$C plane
in three models with different $m_{\rm agb}$ 
with the corresponding observational results from C05.
Clearly, the models with $m_{\rm agb}$=5.5 and 6${\rm M}_{\odot}$ 
can reproduce the observed low $^{12}$C/$^{13}$C ratios of GC stars from C05,
which implies that 
the observed $^{12}$C/$^{13}$C ratios can provide some constraints on the mass range of AGB
stars polluting GMCs.

The observed anticorrelation between $^{12}$C/$^{13}$C and [Na/Fe]
is at least qualitatively reproduced by the present models with higher $m_{\rm agb}$.
However, 
it is observationally unclear whether or not [Na/Fe]-$^{12}$C/$^{13}$C anticorrelations
are as ubiquitous as O-Na anticorrelations  in GCs with different masses and metallicities
(i.e., not just observed in the three GCs by C05).
Given the number of unevolved stars with estimated $^{12}$C/$^{13}$C  in C05 
is only 18,
it would be too early for this study to discuss whether 
the shape of the observed [Na/Fe]-$^{12}$C/$^{13}$C  anticorrelation in C05
is consistent with the prediction.
Although Smith et al. (1995) and Maas et al. (2019) also derived 
$^{12}$C/$^{13}$C ratios in a few GCs, their sample stars are red giants 
(i.e., not unevolved stars) for which $^{12}$C/$^{13}$C ratios
have been much altered from their original values due to first dredge up, and mixing during the RGB bump.
Since the observed $^{12}$C/$^{13}$C ratios and [Na/Fe]-$^{12}$C/$^{13}$C
can give strong constraints on the theory of GC formation with MPs, as discussed later
in this paper,
it is crucial for future observational studies to derive
[Na/Fe] and $^{12}$C/$^{13}$C for a much larger number of unevolved stars in GCs.

\begin{table*}
\centering
\begin{minipage}{180mm}
\caption{ Comprehensive diagnosis of the SCI scenario.
If the predicted property for each diagnostic item 
is consistent (inconsistent)  with observations (1st column) at least qualitatively,
 \ding{51} (\ding{53})  is given in the second column.
It should be stressed here that  \ding{51} does not necessarily mean
"quantitatively consistent" with observations.
If the listed observation is yet to be explained by the scenario,
``?'' is given. The physical processes/conditions involved in reproducing
observations are briefly given in the 3rd column}
\begin{tabular}{lll}
Observation &  Consistency &  Required physics/conditions     \\
{\bf (A)} Correlation with GC mass  & &  \\
Anticorrelation between $F_{\rm 1P}$ and $M_{\rm gc}$  &  \ding{51}  &  
Higher  $R_{\rm s}$ for larger $M_{\rm gmc}$    \\
Correlation between $\delta  Y$ and $M_{\rm gc}$  &  \ding{51}  &  Higher  $R_{\rm s}$ for larger $M_{\rm gmc}$ 
    \\
{\bf (B)}  Abundance patterns  &   &  \\
Ubiquitous Na-O anticorrelations in the Galactic GCs  &  \ding{51}  &  Enrichment by AGB stars (IMF-averaged yields)    \\
Diversity in O-Na anticorrelations  &  \ding{51}  &  Diversity in IMF-averaged AGB yields     \\
Steep or no  O-Na anticorrelation in the bulge GCs  &  \ding{51}  &  Metallicity-dependent  AGB yields     \\
Metallicity-dependent Mg-Al anticorrelations &  \ding{51}  &  Metallicity-dependent AGB yields    \\
C-N anticorrelations  &  \ding{51}  &  Diversity in IMF-averaged AGB yields     \\
Mg-K anticorrelations  &  \ding{51}  &  Major contributions of sAGB stars     \\
Flat Na-Li and Al-Li relations  &  \ding{51}  &  Significant contributions of AGB stars producing Li     \\
Li variations  &  \ding{51}  &  Diversity in AGB fractions producing Li     \\
Positive $^{26}$Mg/Mg-Mg correlation  &  \ding{51}  &  Contributions of massive AGB stars      \\
Helium spreads between 1P and 2P  & \ding{53}  &  Helium yields need to be reduced ? \\
{\bf (C)} MPs in young clusters      & & \\
No MPs phenomena in young clusters  &  \ding{51}  &  Too low mass densities of AGB stars     \\
MPs phenomena in intermediate-age LMC clusters  &  \ding{51}  &  Higher mass densities of AGB stars due to starburst     \\
{\bf (D)} Type I-II GCs     &  &  \\
Type I and II GC dichotomy  & \ding{51}  &  Merging of GCs formed in gas-rich dwarf galaxies      \\
Two distinct O-Na anticorrelations in Type II & \ding{51}  &  Chemical enrichment in two different GMCs      \\
High fraction of s-rich stars in  Type II & \ding{51}  &  Chemical enrichment by field AGB stars      \\
Diverse fractions of  s-rich populations in  Type II & \ding{51}  &  Diversity in the mass-ratios of GC merging      \\
Chemical abundances of s-rich populations in  Type II & \ding{51}  &  Chemical enrichment by low-mass AGB stars      \\
Type II GC fraction in the Galaxy  & ?  &  Diverse GC merging probability in dwarfs ?      \\
{\bf (E)} 1P-2P difference     & &  \\
Structural differences between 1P and 2P  & ?  &  Efficient enrichment in inner GMCs ?     \\
Kinematic differences between 1P and 2P  & ?  & Radially dependent gas kinematics of GMCs ?      \\
Binary star fractions in 1P and 2P  & ?  &  Efficient binary formation in  inner GMCs ?      \\
Larger metallicity spreads in 1P  & \ding{51} &  Pre-existing 1P stars around GMCs      \\
{\bf (F)} Miscellaneous   & & \\
Possible dependence of $F_{\rm 2P}$ on GC ages and [Fe/H]  &  \ding{51}  &  
Higher  $R_{\rm s}$ for GC-host galaxies with lower [Fe/H] and higher $z$    \\
$F_{\rm 1P}$-$\delta_{\rm max} Y$ anticorrelation  &  \ding{51}  & Dilution factor  
dependent on GMC mass   \\  
Discrete MPs  & ?  &  Different levels of enrichment in smaller gas clouds ?     \\
Smaller abundance spread in light s-process elements  & \ding{51} &  Pollution by massive AGB stars      \\
P-rich stars in a few GCs  & \ding{51}  &  GMCs polluted by AGB stars and ONe nova      \\
Stellar halos around GCs  & \ding{51}  & Pre-existing field stars around GMCs    \\
Apparent lack of cold gas in young SCs  & \ding{51}  & All stars formed before GMC dispersal    \\
\end{tabular}
\end{minipage}
\end{table*}

\section{Discussion}

Based on these results, we here provide theoretical interpretations and implications for the various observed properties of GCs with MPs: 
Table~3 summarizes which observational properties can (and cannot) be reproduced well by the new GC formation scenario.
We do not discuss several key properties of GCs, such as the binary star fractions in 1P and 2P (e.g., Lucatello et al. 2015; Dalessandro et al. 2018) and the central concentrations of 2P populations (e.g., Leitinger et al. 2023), 
because this study does not utilize numerical simulations of GC formation.

\subsection{Surface densities of massive AGB stars ($\Sigma_{\rm agb}$) as a key to MP phenomena}

\subsubsection{$\Sigma_{\rm SFR}$–$F_{\rm 2P}$ correlation}

The interaction of GMCs with AGB stars is a physical process indispensable for the formation of GCs with 
MPs in the SCI scenario. If GCs all form within the gas disks of galaxies, then the surface number densities of 
AGB stars with ages of $0.02$--$0.1$~Gyr ($\Sigma_{\rm agb}$) within and around growing GMCs represent the key 
parameter controlling the GMC–AGB interaction process. $\Sigma_{\rm agb}$ could vary radially within a galaxy, and the 
mean $\Sigma_{\rm agb}$ could differ dramatically between galaxies of different masses and morphological types. 
Since $\Sigma_{\rm agb}$ is not a directly observable property of a galaxy, we have proposed that the star formation 
rate surface density ($\Sigma_{\rm SFR}$) can serve as an observable indicator to assess 
whether massive SCs in star-forming galaxies can develop MPs. Furthermore, the scenario predicts a 
positive correlation between $\Sigma_{\rm SFR}$ and $F_{\rm 2P}$, suggesting that the observed diversity in $F_{\rm 2P}$ can be discussed in the context of the baseline $\Sigma_{\rm SFR}$ in GC-forming galaxies.

For example, a rather high $\Sigma_{\rm SFR}$ ($\approx 3$--$10~{\rm M}_{\odot}~{\rm yr}^{-1}~{\rm kpc}^{-2}$) is 
observed in the gravitationally lensed Sunburst galaxy at $z = 2.37$ (Vanzella et al. 2022). Mowla et al. (2024) 
discovered 10 massive young SCs within a $1~{\rm kpc} \times 1~{\rm kpc}$ region of a star-forming, low-mass galaxy with a possible current SFR of $\approx 1~{\rm M}_{\odot}~{\rm yr}^{-1}$ at $z = 8.3$, implying that $\Sigma_{\rm SFR}$ in 
such systems is high enough to generate MPs within their constituent clusters. Conversely, young SCs observed in 
present-day galaxies with low $\Sigma_{\rm SFR} \le 1~{\rm M}_{\odot}~{\rm yr}^{-1}~{\rm kpc}^{-2}$ (e.g., Adamo et al. 2020) 
are highly unlikely to develop MPs. Main-sequence galaxies with baseline values of $\Sigma_{\rm SFR} \le 0.1~{\rm M}_{\odot}~{\rm yr}^{-1}~{\rm kpc}^{-2}$ are similarly highly unlikely to host young SCs with MPs, though they can still 
successfully form SCs without chemical anomalies. While the formation of young massive SCs is observed in 
interacting ultra-diffuse galaxies (UDGs) with low $\Sigma_{\rm SFR}$ (e.g., Buzzo et al. 2025), and 
recent numerical simulations have confirmed that massive SCs can form in interacting and merging gas-rich UDGs (Truman et al. 2026), we suggest that these SCs in UDGs are highly unlikely to host MPs due to their rather low ambient $\Sigma_{\rm SFR}$.

\subsubsection{Two types of young massive clusters with and without MPs}

The present study predicts that neither ages nor metallicities are fundamentally important parameters for the formation of GCs with MPs. It furthermore predicts that massive young clusters formed in gas-rich galaxies cannot possess MPs if the baseline $\Sigma_{\rm SFR}$ in their hosts is low. For example, the $\Sigma_{\rm SFR}$ of the present-day LMC is around $2 \times 10^{-3}~{\rm M}_{\odot}~{\rm yr}^{-1}~{\rm kpc}^{-2}$ (for an SFR of $0.15~{\rm M}_{\odot}~{\rm yr}^{-1}$ and $R = 5$~kpc), which is extremely low compared with the required threshold value of $\Sigma_{\rm SFR} \approx 1~{\rm M}_{\odot}~{\rm yr}^{-1}~{\rm kpc}^{-2}$. Accordingly, the present-day LMC is highly unlikely to form GCs with MPs, even if it can still form massive young star clusters via dynamical interactions with the SMC. Recent observations have shown that clusters with ages younger than $\approx 2$~Gyr do not display MPs, leading to the suggestion of an explicit ``age threshold'' for GC formation with MPs. However, the present study suggests that this observed apparent age threshold is caused not by a physical age constraint, but rather by the very low $\Sigma_{\rm SFR}$ characterizing the LMC over the last 2~Gyr.

The LMC is observed to have experienced a significant enhancement in its global SFR a few billion years ago, likely because it interacted strongly with both the MW and the SMC (e.g., Harris \& Zaritsky 2009). Such galaxy-scale interactions could dramatically increase the local $\Sigma_{\rm SFR}$ around star-forming GMCs, ultimately triggering the formation of GCs with MPs. Therefore, the observed existence of intermediate-age GCs hosting MPs (e.g., Martocchia et al. 2018; Li et al. 2021) could be due to a strong starburst event a few Gyr ago in the LMC rather than an intrinsic age effect. Similarly, the formation of intermediate-age GCs with MPs in the SMC (e.g., Niederhofer et al. 2017; Hollyhead et al. 2018) could be attributed to a strong starburst about 6~Gyr ago associated with the assembly of its spheroidal component through the major merger of two gas-rich dwarfs. 

The present scenario furthermore predicts that young massive SCs formed in luminous galaxy mergers (e.g., the Antennae) cannot possess MPs if the local $\Sigma_{\rm SFR}$ within their birthplace environments is too low. A prime example of such a young massive cluster is Cluster~1 in NGC~34, which features $M_{\rm gc} = 10^7~{\rm M}_{\odot}$ at an age of $\approx 100$~Myr (Cabrera-Ziri et al. 2014) yet lacks light-element variations. Thus, present-day galaxies with low global $\Sigma_{\rm SFR}$, such as the MW and the  LMC, are unlikely to host newly forming massive SCs with MPs.

Blue compact dwarf galaxies (BCDs) with highly compact sizes ($R \approx 1$~kpc) and high star formation rates ($> 1~{\rm M}_{\odot}~{\rm yr}^{-1}$) represent the most promising environments for hosting forming GCs with MPs in the present-day universe. However, it is currently impossible for spectroscopic studies of young clusters in BCDs to directly resolve star-to-star variations of chemical abundances. If the mean $[{\rm N}/{\rm Fe}]$ ratios of these unresolved SCs in BCDs are observed to be quite high ($\ge 0.5$), then these systems can be inferred to host a significant fraction of 2P stars with enhanced nitrogen. We predict a strong positive correlation between the mean $[{\rm N}/{\rm Fe}]$ of these SCs derived from integrated-light spectroscopy and the global $\Sigma_{\rm SFR}$ of their host star-forming dwarf galaxies.

\subsubsection{Can $F_{\rm 2P}$ depend on GC ages and metallicities?}

ME20 investigated how the 2P mass fraction ($F_{\rm 2P}$) of GCs correlates with cluster mass ($M_{\rm gc}$), age ($T_{\rm age}$), and metallicity ($[{\rm Fe}/{\rm H}]$) in Galactic GCs observed by APOGEE. Although they did not confirm the positive correlation between $F_{\rm 2P}$ and $M_{\rm gc}$ previously discovered by other studies (e.g., C09a; Milone \& Marino 2022), they revealed a positive correlation between $T_{\rm age}$ and $F_{\rm 2P}$ alongside an anticorrelation between $[{\rm Fe}/{\rm H}]$ and $F_{\rm 2P}$. At first glance, the new scenario appears unable to explain the origin of these relationships because the formation epochs and gaseous metallicities of GMCs are not the parameters that directly determine $F_{\rm 2P}$. However, if galaxies during their earlier assembly phases systematically experience higher $\Sigma_{\rm SFR}$, then the massive SCs formed during those epochs will evolve into the present-day  GCs characterized by older $T_{\rm age}$ and higher $F_{\rm 2P}$. Since these early-phase host galaxies also feature less enriched interstellar media, their constituent SCs will naturally evolve into modern GCs with lower $[{\rm Fe}/{\rm H}]$ and higher $F_{\rm 2P}$.

To quantitatively verify whether $F_{\rm 2P}$ can depend on $T_{\rm age}$ and $[{\rm Fe}/{\rm H}]$, we have run several one-zone chemical evolution models of galaxies using the numerical code developed by Bekki \& Tsujimoto (2012); 
Appendix~C briefly describes these models and their results. We find that $\Sigma_{\rm SFR}$ is systematically higher during the early epochs of galactic chemical evolution, which directly implies that GCs formed earlier (and thus possessing older $T_{\rm age}$) will exhibit higher $F_{\rm 2P}$ values. We also find that $\Sigma_{\rm SFR}$ is systematically higher at lower $[{\rm Fe}/{\rm H}]$, down to a threshold metallicity floor that depends on the gas infall timescale ($\tau_{\rm inf}$). This result similarly implies that GCs with lower $[{\rm Fe}/{\rm H}]$ are more likely to host larger 2P fractions. While the above arguments remain somewhat qualitative within these idealized one-zone models—given that the actual co-evolution of $[{\rm Fe}/{\rm H}]$ and $\Sigma_{\rm SFR}$ is determined by complex structural processes of galaxy formation—we will discuss the physical origins of these trends in a more quantitative manner in a forthcoming paper based on hydrodynamical simulations of GC assembly within evolving dark matter halos.

\subsubsection{Is $F_{\rm 2P}$ different between in situ and accreted GCs?}

Monty et al. (2024) recently demonstrated that the abundance ratios of europium to $\alpha$ elements ($[{\rm Eu}/\alpha]$) in Galactic GCs can be utilized to distinguish between in situ and accreted (ex situ) populations. One plausible reason for the elevated $[{\rm Eu}/\alpha]$ ratios in ex situ GCs is that their host dwarf galaxies possessed a top-light IMF characterized by either a steeper high-mass slope or a reduced upper-mass cutoff (e.g., Tsujimoto 2024). A top-light IMF has also been invoked to explain the detailed chemical abundance patterns of massive dwarf galaxies in the Local Group, such as the Fornax dwarf spheroidal (e.g., McWilliam et al. 2013). Such a top-light IMF is expected in star-forming galaxies with lower overall star formation rates according to the integrated galaxy-wide IMF theory (e.g., Weidner \& Kroupa 2005), implying that a top-light IMF is a natural consequence of lower $\Sigma_{\rm SFR}$.

If the defunct dwarf galaxies hosting ex situ GCs possessed lower baseline values of $\Sigma_{\rm SFR}$ (and thus a top-light IMF), then the SCI scenario suggests that ex situ GCs with elevated $[{\rm Eu}/\alpha]$ should systematically display lower $F_{\rm 2P}$ values for a given global cluster mass $M_{\rm gc}$. Because it is observationally established that $F_{\rm 2P}$ depends strongly on $M_{\rm gc}$ (e.g., C09a), discovering a clear, isolated signature in $F_{\rm 2P}$ between in situ and ex situ GCs may prove observationally challenging. Furthermore, this hypothesis assumes that a top-light IMF is the primary driver of elevated $[{\rm Eu}/\alpha]$ ratios in accreted clusters; if other environmental factors play a dominant role in determining $[{\rm Eu}/\alpha]$ in ex situ GCs, then the predicted difference in $F_{\rm 2P}$ between the two populations becomes less definitive. Notably, Lardo et al. (2026) have recently revealed a wide diversity of $F_{\rm 2P}$ values among different accreted GC streams (such as the Sequoia stream showing $F_{\rm 2P} \approx 0.45$). Within the framework of the new scenario, this behavior can be naturally explained by variations in the baseline $\Sigma_{\rm SFR}$ environments among the different defunct dwarf galaxies that originally hosted these clusters.

\begin{figure}
\psfig{file=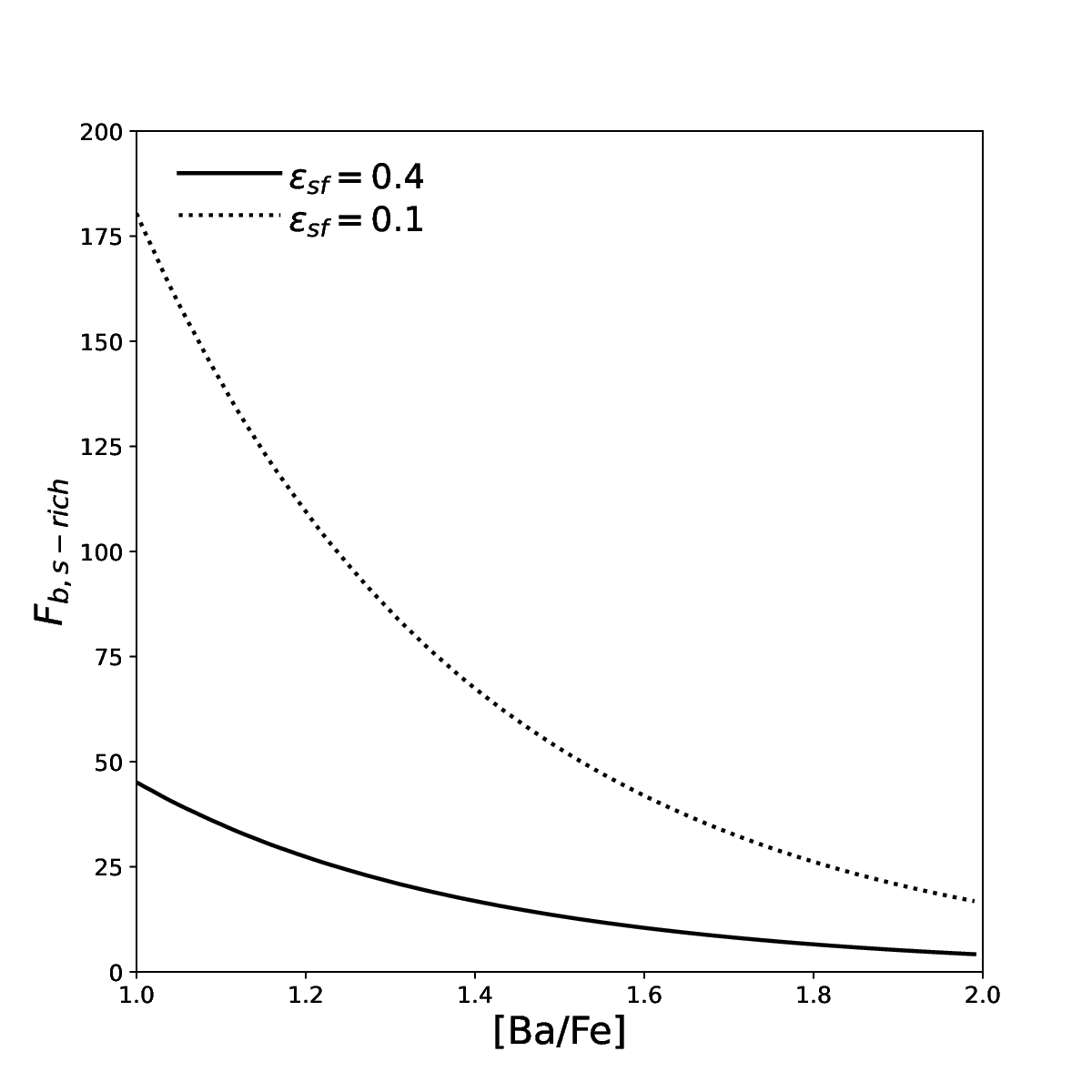,width=8.5cm} 
\caption{
Mass budget factors for $s$-rich stars ($F_{\rm b, s-rich}$) as a function of [Ba/Fe] of
polluting AGB stars for $\epsilon_{\rm sf}=0.4$ (solid) and 0.1 (dotted). The observed total stellar masses 
and  [Ba/Fe] in $s$-poor and $s$-rich stars in M22 are used to plot these.
$F_{\rm b, s-rich}>1$ means that the total mass required to explain
the observed [Ba/Fe] of $s$-rich stars is larger than the total-mass of $s$-poor stars.
The required large 
$F_{\rm b, s-rich}$ in this figure thus clearly demonstrates the mass budget problem for $s$-rich stars
in M22.
}
\label{Figure. 19}
\end{figure}

\subsection{Type II GC formation from GC merging}

The origin of metal-complex Type II GCs is yet to be fully understood. The observed significant [Fe/H] spreads can result either from (i) secondary star formation and the subsequent iron enrichment in existing GCs (e.g., D'Antona et al. 2016) or 
(ii) the merging of GCs with different [Fe/H] (BY12, BT16). 
The self-enrichment scenario of star-forming GMCs by multiple CCSNe 
(e.g., Bailin 2018; Wirth et al. 2021) might also be responsible for the formation of Type II GCs, 
although these studies did not discuss the origin of the $s$-poor and $s$-rich populations within the clusters.
We first point out a severe mass-budget problem regarding $s$-rich populations in Type II GCs based on the observed properties of M22, and then discuss how GC merging can better explain the physical properties of Type II GCs.

\subsubsection{The severe mass-budget problem of s-rich populations}

We here divide the stars of a Type II GC into two categories: an $s$-poor population with lower [Fe/H] and [Ba/Fe], and an $s$-rich one with higher [Fe/H] and [Ba/Fe]. The total GC mass is accordingly the sum of the total mass of the $s$-poor ($M_{\rm gc, s-poor}$) and $s$-rich populations ($M_{\rm gc, s-rich}$):
\begin{equation}
M_{\rm gc}=M_{\rm gc, s-poor}+M_{\rm gc, s-rich}.
\end{equation}
Using the observed properties of M22 (e.g., an $s$-rich fraction of $\approx 0.5$), we can estimate (i) the total mass of the initial gas ($M_{\rm o}$) a fraction of which can evolve into low-mass AGB stars and (ii) the mass-budget factor for the $s$-rich population:
\begin{equation}
F_{\rm b,s-rich}=\frac{ M_{\rm o} }{ M_{\rm gc, s-poor} }.
\end{equation}
If this factor is quite large ($>10$, i.e., $M_{\rm o} \gg M_{\rm gc, s-poor}$), then it suggests that the $s$-poor population is unlikely to be responsible for the formation of the $s$-rich population in M22. As discussed in \S 2.3, the present-day mass of the $s$-rich population should be smaller than its original mass because it can lose a fraction of its original mass through stellar evolution and internal/external dynamical processes:
\begin{equation}
M_{\rm gc,s-rich}=(1-f_{\rm lost})\epsilon_{\rm sf}M_{\rm o},
\end{equation}
where $\epsilon_{\rm sf}M_{\rm o}$ corresponds to the original mass of the $s$-rich population, and $f_{\rm lost}$ and $\epsilon_{\rm sf}$ represent the mass loss fraction and the star formation efficiency, respectively. We here adopt $f_{\rm lost}=0.4$, and $\epsilon_{\rm sf}=0.1$ and $0.4$.

We here assume that the $s$-rich population is formed from pristine gas mixed with stellar winds from the $s$-poor, low-mass AGB stars. The observed difference in [Ba/Fe] between the two distinct populations is defined as follows:
\begin{equation}
\Delta {\rm [Ba/Fe] } ={\rm [Ba/Fe]_{\rm s-rich} } -{\rm [Ba/Fe]_{\rm s-poor} }.
\end{equation}
Likewise, we use $\Delta {\rm [Fe/H] }$ and $\Delta t_{\rm age} 
$ for the [Fe/H] and age differences between the two populations, respectively. The Ba abundance of the $s$-rich population ($A_{\rm Ba, s-rich}$) can be calculated using the total mass of pristine gas ($M_{\rm g}$), the AGB ejecta mass ($M_{\rm ej}$), and the Ba yield of low-mass AGB stars ($f_{\rm Ba, ej}$):
\begin{equation}
A_{\rm Ba, s-rich } =\frac{ f_{\rm Ba, g} M_{\rm g} +  f_{\rm Ba,ej}M_{\rm ej}  } { M_{\rm g} + M_{\rm ej} },
\end{equation}
where $f_{\rm Ba, g}=f_{\rm Ba, s-poor}$ and $M_{\rm o}=M_{\rm g}+M_{\rm ej}$. 
Since [Ba/Fe]$){\rm s-rich}$ ($=0.55$), 
[Ba/Fe]$_{\rm s-poor}$ ($=0$), and the fraction of Ba-rich stars ($=0.5$) are observed, we can estimate $M_{\rm g}$ (and thus $M_{\rm o}$ and $F_{\rm b,s-rich}$) for a given $M_{\rm ej}$. Since a narrow mass range of AGB stars polluting the intracluster gas ($3 \le m_{\rm agb}/{\rm M}_{\odot} \le 4$) is required to produce Ba-rich ejecta and explain the observed possible age difference between the $s$-poor and $s$-rich populations (e.g., Marino et al. 2015; McKenzie et al. 2022), we adopt $f_{\rm agb}=0.03$ as a fiducial value. We investigate $F_{\rm b, s-rich}$ for [Ba/Fe] yields ranging from 1 to 2 for low-mass stars.

As shown in Fig. 19, $F_{\rm b, s-rich}$ is quite high, ranging from 20 to 180 for $\epsilon_{\rm sf}=0.1$, which means that $M_{\rm o}$ should be as large as $8.8 \times 10^6 {\rm M}_{\odot}$ for $\epsilon_{\rm sf}=0.1$. Given that the total mass of $s$-poor stars in M22 is only $\approx 2 \times 10^5 {\rm M}_{\odot}$, the required large $M_{\rm o}$ is clearly the manifestation of a ``mass-budget problem'' that is more severe than the mass-budget problem for GCs with high 2P fractions. $F_{\rm b, s-rich}$ ranges from 5 to 45 for $\epsilon_{\rm sf}=0.4$, which means that the required $M_{\rm o}$ is still much larger than the mass of $s$-poor stars. Thus, chemical enrichment by AGB stars from $s$-poor stellar populations cannot explain the observed total mass of $s$-rich stars in M22.

Chemical evolution models of dwarf galaxies predict that [Fe/H] and [Ba/Fe] can increase by $\approx 0.2$ dex and $\approx 0.5$ dex, respectively, within 300 Myr, depending on the adopted star formation histories and IMFs (BT16). 
Therefore, 
if two GCs form from separate GMCs at different epochs within their host dwarf galaxy
separated by $\approx 300$ Myr and then merge together to form a new single GC, 
the new GC can have both $s$-poor (formed earlier) and $s$-rich (formed later) populations. Since each of the merger progenitor GCs contains MPs, 
the new GC should have MPs in both its $s$-poor and $s$-rich populations. Accordingly, this GC merger scenario can naturally and self-consistently explain both (i) the observed high fraction of $s$-rich populations and (ii) the presence of MPs in both the $s$-poor and $s$-rich populations of M22. We thus suggest that GC merging is not limited to M22, but is an essential ingredient in the formation of Type II GCs.

\subsubsection{Diversity in Type II GC properties?}

 GC merging is more likely to occur among more massive GCs with higher internal 
stellar velocity dispersions ($\sigma_{\rm gc}$) within their host dwarf galaxies, because $\sigma_{\rm gc}$ should be comparable to the stellar velocity dispersion of the hosts. The mass ratio of two merging GCs can vary, which explains the observed diversity in the fractions of metal-rich populations. Using $N$-body simulations of GC merging, Gavagnin et al. (2016) demonstrated that the radial profiles of 1P-to-2P ratios and the rotational kinematics of the merger remnants (e.g., solid-body or differential rotation) depend on the mass and density ratios of the two progenitor clusters. Their results imply that there can be diversity in the radial profiles of $F_{\rm 1P}$ and rotational kinematics among Type II GCs.

Our recent hydrodynamical simulations of GC formation in gas-rich mergers have shown that multiple mergers among a number of massive young star clusters (SCs) lead to the formation of nuclear SCs with large [Fe/H] and age spreads (Matsui et al. 2025; Truman et al. 2026). The origin of MPs in Type II GCs like $\omega$ Cen and M54, which might have originated from the nuclei of defunct dwarf galaxies, can therefore be understood in the context of GC merging. Since age differences between stars in Type II GCs can be quite diverse due to the different formation epochs of the two merging GCs in this scenario, there could exist Type II GCs with large age spreads (e.g., 3 Gyr). However, such large age spreads have not been observationally confirmed in Type II GCs with two distinct populations. Possibly, two GCs forming at similar epochs can share similar 3D positions within their host galaxy (i.e., clustered GC formation), allowing them to merge together to form a new single Type II GC.

The observed fraction of Type II GCs ($\approx 20\%$) among Galactic GCs suggests that only 20\% 
of the initial GC populations formed in their host galaxies merged to form Type II GCs. 
It is not entirely clear why only $\approx 20\%$ of GCs experienced merger events within their host galaxies in the GC merging scenario. 
Our future studies need to clarify the origins of these observed diverse Type II GC properties 
based on more sophisticated simulations of GC formation in galaxies.

Terzan 5 is observed to host  two distinct populations with $\rm [Fe/H] = -0.8$ and $0.3$ at least 
(e.g., Ferraro et al. 2009; Zullo et al. 2026), 
featuring a very large age spread of $\approx 7$ Gyr between them (e.g., Zullo et al. 2026).
Although the origin of this GC or ``bulge fossil'' (Ferraro et al. 2009)
can also be discussed in the context of GC merging
within the Galactic bulge, it remains unclear how a GC merger could occur 7 Gyr
after the formation of the progenitor GC with $\rm [Fe/H] = -0.8$.
Furthermore, all delayed CCSNe and most SNe Ia would have already exploded
7 Gyr after the formation of the initial $\rm [Fe/H] = -0.8$ population.
Therefore, for this particular cluster, secondary star formation fueled by gas accretion onto the GC and
a subsequent starburst (e.g., McKenzie \& Bekki 2018; Romano et al. 2023)
provides a compelling alternative scenario for the formation of Terzan 5.

\subsection{Larger metallicity spreads in 1P stars?}

Our recent hydrodynamical simulations of GC formation in gas-rich dwarf galaxies have predicted that GC-forming GMCs can have large [Fe/H] spreads (typically 0.1 dex) due to (i) the merging of smaller clouds with different [Fe/H] formed at different locations within their host galaxies and (ii) self-enrichment by stellar feedback effects (McKenzie \& Bekki 2021b). Furthermore, they have also predicted that (i) GC-forming GMCs can gravitationally trap the surrounding field stars of their host dwarf galaxies and (ii) these trapped stars can eventually become a more metal-poor precursor population (``0P''). Since the 1P consists of stars formed earlier in GMCs and field stars trapped by GMCs in this new scenario, these simulation results imply that the 1P can have larger metallicity spreads. It is possible that GCs formed in galaxies with steeper metallicity gradients have larger metallicity spreads among their 1P stars in the new scenario.

The 2P stars can form from GMC gas mixed with AGB ejecta later than 1P stars in this scenario. Accordingly, they are highly likely to form from gas that has already experienced metal homogenization due to turbulent mixing within the GMCs. 2P stars can therefore have smaller metallicity spreads than 1P stars in this scenario. Since this explanation is speculative and not very quantitative, we will need to investigate whether and how metal diffusion processes within GC-forming massive GMCs can reduce the initial metallicity spreads of the GMCs using new hydrodynamical simulations of GC formation in galaxies.

Recent HST observations have shown that 1P stars display larger degrees of metallicity spreads compared with their 2P 
counterparts (e.g., Legnardi et al. 2022). However, Carretta \& Bragaglia (2025) were not able to confirm such large metallicity spreads in 1P stars. Lardo et al. (2022), on the other hand, confirmed large metallicity spreads among red giant branch stars from the 1P in M92, NGC 2808, and NGC 6362. Although ME20 found that the typical intrinsic [Fe/H] scatter in GCs observed by 
APOGEE is about 0.1 dex and does not depend on GC masses and ages, they did not discuss whether 1P stars have 
larger [Fe/H] spreads. 
The question of whether 1P stars have larger [Fe/H] spreads  remains actively debated, 
and disentangling intrinsic metallicity variations from systematic photometric and spectroscopic 
effects will require coordinated approaches to large homogeneous GC samples. 
The next generation of high resolution spectroscopic facilities will help to resolve these tensions.

\subsection{Li abundances as a crucial diagnosis for different GC formation scenarios}

Recent spectroscopic studies of Li abundances (A(Li)) and their correlations with light-element abundances in GCs with MPs have revealed that Li is not very depleted in 2P stars with rather high [Na/Fe] (e.g., McKenzie et al. in prep). 
These results disfavour
GC formation scenarios in which polluters destroy Li but cannot produce it at all. As shown in this paper, these observations are consistent with the SCI scenario, and they can be used as a new constraint on the fraction of AGB stars that can produce Li due to the Cameron-Fowler effect. 
G19 have already shown that the simple dilution of gaseous ejecta from polluters by pristine gas cannot 
explain the observed levels of Li depletion in GCs; we have shown for the first time that the observed levels can be naturally explained by our models, with $\approx 20$\% of AGB stars producing Li.

The observed Li depletion (A(Li) $\approx 0.2$ dex) 
in the extreme ``E'' populations of some GCs indicates that these stars were 
formed from pristine gas chemically enriched by polluters in which Li was destroyed (G19). This observation and the observed lack of an Li-Al anticorrelation combine to suggest that there should be polluters producing and destroying Li during GC formation: a single type of polluter cannot explain the observations self-consistently. 
D12 showed that (i) A(Li) depends strongly on the masses of AGB stars ($m_{\rm agb}$ in units of ${\rm M}_{\odot}$), (ii) A(Li) can be rather large, reaching A(Li)=2.75 and 
4.39 for $m_{\rm agb}=7.5$ and 8.0, respectively, and (iii) A(Li) 
values are less than 2.3 for $4 \le m_{\rm agb} \le 6.3$. If GMCs are polluted by larger fractions of 
AGB stars with $4 \le m_{\rm agb} \le 6.3$, GCs can have larger fractions of E populations. Given that IMF slopes determine the fractions of these intermediate-mass AGB stars, the observed diverse fractions of E populations might imply that IMF slopes vary across different GCs.

The observed anticorrelation between A(Li) and [Na/Fe] in NGC 6752 and NGC 6397 is suggested to be consistent with the pollution of pristine gas by intermediate-mass AGB stars (e.g., Pasquini et al. 2005; Lind et al. 2009). 
Shen et al. (2010) pointed out that the observed extended Li-O correlation in unevolved stars can be explained not by 
Li-poor ejecta mixed with (i.e., diluted by) 
pristine gas alone, but by Li-poor ejecta mixed with both pristine gas and Li-enriched gas. Although 
Bonifacio et al. (2007) found an Li-Na anticorrelation among 4 turn-off stars in 47 Tuc, Dobrovolskas et al. (2014) did not find convincing evidence for it among 110 turn-off stars. D'Orazi \& Marino (2010) 
discovered that both the 1P and 2P in M4 share the same Li abundance and thus suggested that Li production is required to reproduce the observed flat [Na/Fe]--A(Li) relation. Mucciarelli et al. (2011) found an almost flat [O/Fe]--A(Li) relation among stars with a wide range of [O/Fe] (i.e., both 1P and 2P) in M4, and Villanova \& Geisler (2011) also showed almost identical mean A(Li) values between N-poor (1P) and N-rich (2P) stars in M4.

D'Orazi et al. (2014) revealed that Li production is required to reproduce the observed [Al/Fe]-A(Li) relations in M5 and M12 and suggested that GC masses can possibly determine the extent to which Li production occurs. 
These existing observations, as well as observations from the Galactic Archaeology with HERMES 
(GALAH) survey (Buder et al. 2025), are difficult to reconcile with scenarios where the ejecta of the 
primary polluter is devoid of Li. If MP progenitors such as very 
massive stars, supermassive stars and massive interacting binary stars can not either retain, or 
regenerate Li, it may be challenging for them to reproduce these observations.
The SCI scenario predicts that (i) the fraction of AGB stars producing Li (i.e., $P{\rm (Li)}$) can determine the [Al/Fe]-A(Li)
 relations in GCs and (ii) $P{(\rm Li)}$ 
is a function of the IMF (e.g., $\alpha$). If less massive GCs have larger $P{\rm (Li)}$, 
this new scenario can explain the observed smaller scatter of A(Li) in less massive GCs (D'Orazi et al. 2014). 
If less massive GCs have less top-heavy IMFs, they could have larger $P{\rm (Li)}$ 
due to a smaller fraction of massive AGB stars with low A(Li) in their winds. 
It is theoretically unclear, however, how the IMF might depend on GC masses at their birth. It is thus the focus of our future study to investigate how the IMFs of GCs determine $P{\rm (Li)}$ and eventually cause the observed A(Li) dispersions that possibly depend on GC masses.

\subsection{$^{12}$C/$^{13}$C ratios in unevolved stars as a constraint on GC formation}

C05 first  derived $^{12}$C/$^{13}$C ratios for 21 dwarfs and subgiants in NGC 6752 and NGC 104 using the CH molecular band. One intriguing result is that the $^{12}$C/$^{13}$C ratios are quite low, ranging from 3 to 12 (see their Table 1), 
which led them to conclude that evolved intermediate-mass AGB stars are the most likely polluters for these GCs. 
If $^{12}$C/$^{13}$C ratios are investigated for \textit{unevolved stars} unaffected by stellar evolution in GCs with MPs, 
the observed ratios can provide crucial constraints on GC formation because the ratios are predicted to be 
significantly different between stellar winds from AGB stars and massive stars (e.g., Figs. 4 and 5 in Rizzuti et al. 2025).

The $^{12}$C/$^{13}$C ratio in stellar winds from AGB stars with $m=3 {\rm M}_{\odot}$ and $Z=0.0001$ 
is $\approx 50$ (K10), which suggests that chemical enrichment by low-mass AGB stars 
cannot explain the observed low $^{12}$C/$^{13}$C ratios ($<12$). These predictions will be tested against observational results when $^{12}$C/$^{13}$C ratios are estimated for many unevolved 1P and 2P stars in GCs. 
Rotating and non-rotating massive stars with $13 \le m/{\rm M}_{\odot} \le 120$ at [Fe/H]$=-2$ and $-1$ can eject stellar winds with rather large $^{12}$C/$^{13}$C ratios ranging from $\approx 10$ to $10^5$ (Rizzuti et al. 2025); 
the IMF-averaged $^{12}$C/$^{13}$C ratio is expected to be rather large ($>100$) for a Salpeter IMF. 
Accordingly, it is possible that the $^{12}$C/$^{13}$C ratio of 2P stars formed from these ejecta can be quite large too ($>100$). 
This means that the future observational derivation of $^{12}$C/$^{13}$C ratios in unevolved stars can provide a new constraint on GC formation scenarios. 
It is not clear whether the $^{12}$C/$^{13}$C ratios of stellar winds from very massive stars, supermassive stars, and massive interacting binaries are as large as those of the massive stars mentioned above. 
It is thus safe to say that the low $^{12}$C/$^{13}$C ratios observed in C05 are consistent with the predictions from the new scenario with $4 \le m_{\rm agb}/{\rm M}_{\odot} \le 10$.

\begin{figure}
\psfig{file=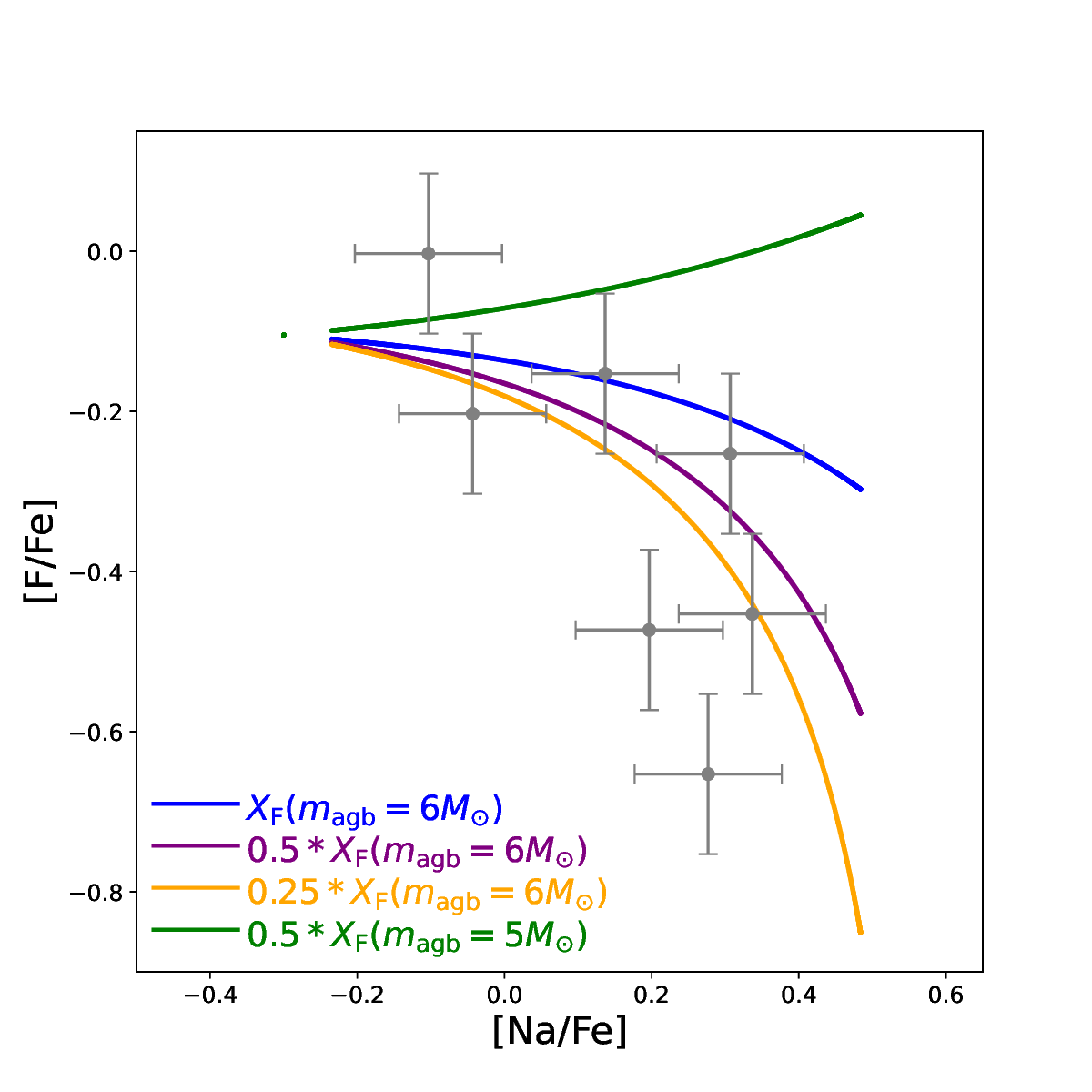,width=8.5cm} 
\caption{
Evolution of the four models with $X_{\rm F}=X_{\rm F}(m_{\rm agb}=6 {\rm M}_{\odot}) $ (blue),
$0.5 \times X_{\rm F}(m_{\rm agb}=6 {\rm M}_{\odot}) $ (purple),
$0.25 \times X_{\rm F}(m_{\rm agb}=6 {\rm M}_{\odot}) $ (orange),
and $0.5 \times X_{\rm F}(m_{\rm agb}=5 {\rm M}_{\odot}) $ (green) on 
the [Na/Fe]-[F/Fe] plane. The observed abundances taken from
Smith et al. (2005) for M4 are shown by gray circles with observational error bars.
The observed Na-F anticorrelation can be better reproduced  by the models with reduced F yields.
}
\label{Figure. 20}
\end{figure} 

\subsection{[Na/Fe]-[F/Fe] anticorrelations ?}

Fluorine occupies a unique nucleosynthetic position in the context of GC chemical enrichment, 
being produced and destroyed at the interface between the CNO and Ne-Na proton capture cycles. 
[F/Fe] is sensitive to both the depth of the third dredge up, and the efficiency of HBB, making the 
Na-F anticorrelation a probe of the balance between these two competing processes in AGBs 
(Forestini et al. 1992, Lugaro et al. 2004, Karakas \& Lattanzio 2014).
Previous observations revealed (anti)correlations of [F/Fe] with
[O/Fe] and [Na/Fe] in several Galactic GCs, which provided 
a constraint on the nucleosynthesis site of polluting stars in GC formation
(e.g., Cunha et al. 2003; Smith et al. 2005; Yong et al. 2008
Alves-Brito et al. 2012; D'Orazi et al. 2013).
It is, however, observationally unclear whether these correlations and anticorrelations
are ubiquitous in the vast majority of GCs (e.g.,  de Laverny \&  Recio-Blanco 2013).
Nevertheless it is  worthwhile for this study to investigate whether or not
these observations  can be well reproduced by 
the present scenario. We here particularly
investigate the [Na/Fe]-[F/Fe] anticorrelation observed in M4
(Smith et al. 2005) by adopting the F yields from K10 for different 
$m_{\rm agb}$ at $Z=10^{-4}$ in the fiducial model.
Since the mass fraction of F  ($X_{\rm F}$) is available only for 
AGB stars with $m_{\rm agb} \le 6{\rm M}_{\odot}$ (i.e., unable to derive
an IMF-averaged F yield),
we here demonstrate what $X_{\rm F}$ (thus what particular $m_{\rm agb}$)  can
better reproduce the observed Na-F anticorrelation.
We adopt the initial [Na/Fe] of $-0.3$ and  [Na/Fe]=0.6 in AGB ejecta
in all four  models.  

Fig.  20 shows the evolutionary loci 
of the four models with different $X_{\rm F}$
on the [Na/Fe]-[F/Fe] plane.
For example, the model
with 0.5 $X_{\rm F}(m_{\rm agb}=6 {\rm M}_{\odot}$) means that
the F yield is the half of $X_{\rm F}$ for $m_{\rm agb}=6 {\rm M}_{\odot}$ adopted
from K10.
The model with $X_{\rm F}(m_{\rm agb}=6 {\rm M}_{\odot}$ shows a
[Na/Fe]-[F/Fe] anticorrelation that is significantly flatter than the observed one.
As shown in the models 
with $X_{\rm F}=0.25 X_{\rm F}(m_{\rm agb}=6 {\rm M}_{\odot}$ 
and 0.5 $X_{\rm F}(m_{\rm agb}=6 {\rm M}_{\odot}$,
the F yields for $im_{\rm agb}=6 {\rm M}_{\odot}$ 
need to be reduced to match well with the observed Na-F anticorrelation.
These results suggest that  (i) $m_{\rm agb}=6 {\rm M}_{\odot}$  AGB models 
by K10 appear to 
overproduce $F$ and (ii) other models with smaller  $X_{\rm F}$ 
for massive AGB stars
(e.g., Siess 2010) can possibly explain the Na-F anticorrelation better.
The model with  $X_{\rm F}=0.5 X_{\rm F}(m_{\rm agb}=5 {\rm M}_{\odot}$)
shows a Na-F {\it correlation},
which implies that AGB stars with $m_{\rm agb} \le 5 {\rm M}_{\odot}$ cannot
contribute greatly to chemical enrichment in  forming GCs.

Doherty et al. (2017) and Siess (2010)
predicted [F/Fe]$\approx 0$ and $-2.4$, respectively
for sAGB stars with $m_{\rm agb}=7.5 {\rm M}_{\odot}$
at $Z=10^{-4}$. These vastly different F yields from the two groups
make it difficult to assess the possible role of sAGB stars
in the formation of the Na-F anticorrelation.
If the F yields of sAGB stars are indeed very low ([F/Fe]$<-1$),
mixing of pristine GMC gas with sAGB stars can end up with
the formation of stars with [F/Fe]$<-0.6$.
So far, other theoretical models based on other polluters (e.g., supermassive stars etc)
are yet to comprehensively explore the constraints placed on the relationship between Na and F.
It would be thus too early to make a robust conclusion on whether the Na-F anticorrelation
is due to chemical enrichment by AGB and sAGB stars in forming GCs.

\begin{figure}
\psfig{file=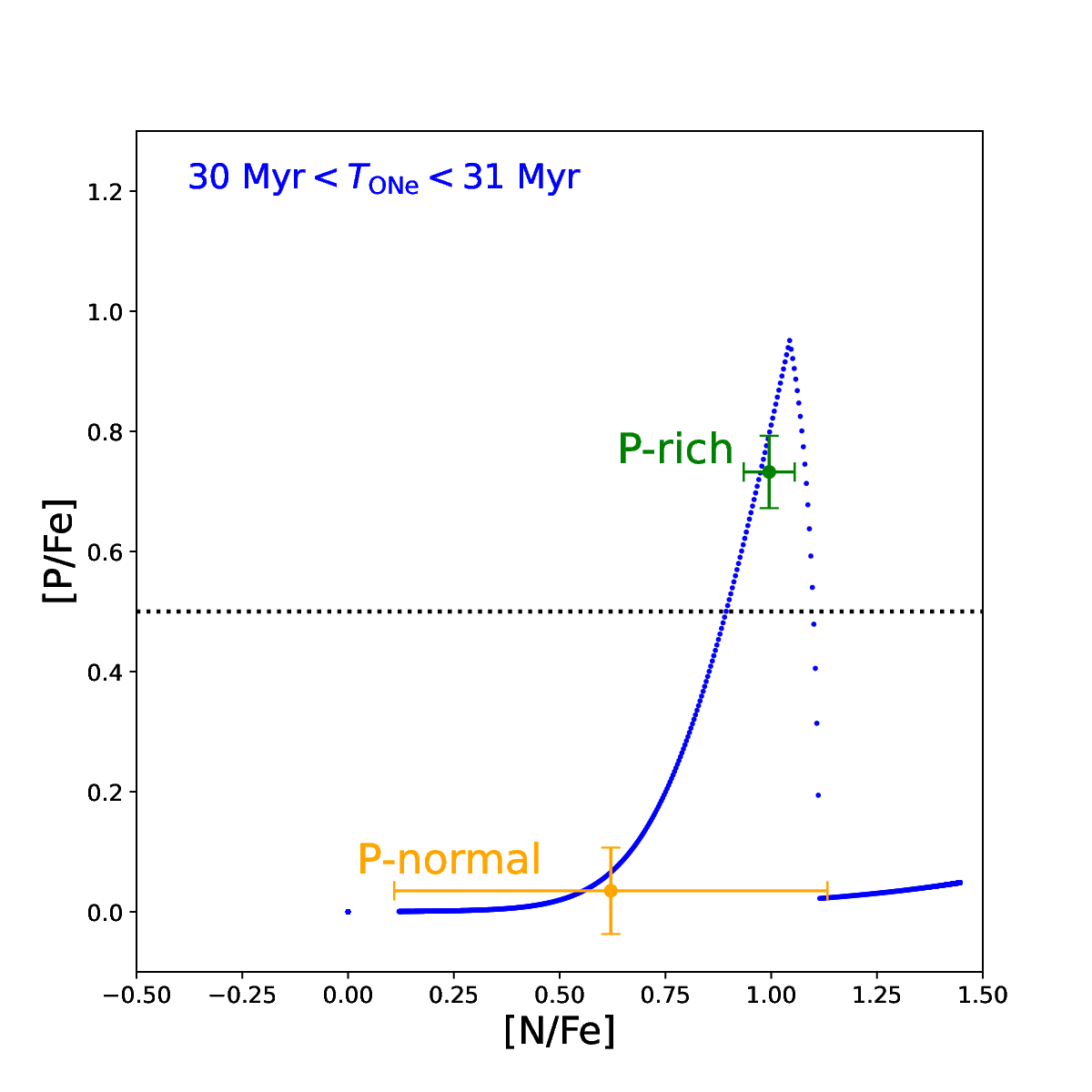,width=8.5cm} 
\caption{
Evolution of the fiducial model with P-enrichment by ONe novae during GC formation
within a GMC (blue dots). It is assumed in this model that ONe nova explosions can enrich cold gas of the GMC
at time steps ($T_{\rm ONe}$)  between 30 and 31 Myr. 
The mean (filled circles) and 1$\sigma$ dispersion of [P/Fe] (error bars) are shown separately
for ``P-normal'' ([P/Fe]$<0.5$; orange) and ``P-rich'' stars ([P/Fe]$\ge 0.5$; green). 
}
\label{Figure. 21}
\end{figure} 

\subsection{P-rich stars as a new clue to MPs phenomena}

Recent observations have discovered ``P-rich stars''  in some Galactic
GCs, such as M4 and NGC 6316 (e.g., Brauner et al. 2024; Barbuy et al. 2025).
Although the nucleosynthesis sites for P production are yet to be fully understood 
(e.g., Masseron et al. 2020),  one of the sites can be ONe novae formed from binary
stellar systems with massive stars with $m > 7 {\rm M}_{\odot}$ (e.g., Bekki \& Tsujimoto 2024).
If multiple ONe nova explosions occur within and around star-forming GMCs,
then the new stars formed from the gas mixed with ONe nova ejecta
should be able to have high [P/Fe] depending
on the dilution levels.
Given that ONe nova explosions start to occur $\sim 10^8$ yr after
star formation events (e.g., Kemp et al. 2022),
chemical enrichment of GMCs by ONe novae (from pre-existing 0P populations)
can be discussed in the context of the SCI scenario.
The SCI scenario accordingly needs to explain
why star-forming GMCs can be chemically enriched by ONe novae so efficiently
and why only some GCs have such P-rich stars.

In order to discuss P-enrichment in forming GCs in a more quantitative manner,
we here investigate the fiducial model with $t_{\rm life}=5 \times 10^7$ yr and $F_{\rm dil}=1$,
in which
the ONe nova model adopted in Bekki \& Tsujimoto (2023) is implemented.
Fig. 21 shows the evolutionary locus of the model on the [N/Fe]-[P/Fe] plane
in the model in which the mass of P per unit star-forming mass, initial [P/Fe],
and [N/Fe] of AGB ejecta are assumed to be 
$1.69 \times 10^{-7}$, 0, and 1.5, respectively.
Although the epoch of ONe nova explosions ($T_{\rm ONe}$) determines the
locus of a model on the [N/Fe]-[P/Fe] plane,
ONe novae are assumed to occur at 30 Myr $\le T_{\rm ONe} \le $ 31 Myr in this model: It should 
be stressed here that this is only a possible locus on the [N/Fe]-[P/Fe] plane
among many possible loci.
Clearly, P-rich stars with [P/Fe]$>0.5$ can be formed only after ONe nova explosions
in this model, which demonstrates that ONe novae can be a possible candidate for polluting
stars responsible for the origin of P-rich stars.

The predicted fraction of P-rich ([P/Fe])  stars is 0.05
in this particular model with a short timescale of P-enrichment by ONe novae,
which will be able to be compared with future observational studies on the fraction of 
P-rich stars in GCs.
Obviously, the P-rich star fractions in GCs strongly depend on
the P yields and 
the epoch and the duration of ONe novae.
If ONe novae occur later in GC formation
(i.e., later in GMC evolution),
as in this model, P-rich stars can also be N-rich ($>0.5$) and have
a smaller dispersion in [N/Fe] compared with P-normal stars with
a larger [N/Fe] dispersion.
Indeed, Barbuy et al. (2025) have found P-rich stars with high [N/Fe] in the Galactic bulge GCs,
which is consistent with the model with 30 Myr $\le T_{\rm ONe} \le $ 31 Myr.
On the other hand,
if ONe novae can occur only in the early GC formation phases,
P-rich stars can have lower [N/Fe], because field AGB stars have not yet chemically enriched the pristine GMC gas.
Therefore, whether P-rich stars are also N-rich is determined by when ONe novae explode.

It is also likely that ONe novae cannot always occur within and around GC-forming GMCs, 
because binary fractions of massive stars could be quite small in some GMCs, 
which means that only some GCs can have P-rich stars.
Since the total number of ONe novae can be small due to the short lifetime of GMCs,
the level of P-enrichment is likely to show a large cluster-to-cluster variation.
This stochastic ONe nova enrichment scenario can explain why Barbuy et al. (2025) 
found P-rich stars with [P/Fe]$>0.5$ only in a few GCs.
This observed presence and absence of P-rich stars in GCs can 
be nicely explained by the P-enrichment of GMCs by sporadic ONe nova events.

CCSNe originating from massive stars with pre-supernova carbon-oxygen (C-O) shell merging
are demonstrated to produce a large amount of P and are thus a major driver for the chemical
evolution of P in the Galaxy (e.g., Ritter et al. 2018).
However, such CCSNe with O-C shell merging cannot be responsible for the formation of P-rich
stars in GCs with small [Fe/H] spreads, because
CCSNe can introduce large [Fe/H] spreads in GCs with MPs.
Therefore, ONe novae originating from binary stars
appear to be the only polluter that is responsible for
the formation of P-rich stars in GCs. If this is the case,
then it would be reasonable to propose that P-rich stars discovered in the Galactic field
(e.g., Masseron et al. 2020)
can also be formed from gas enriched by ONe novae. It is, however, unclear whether
these P-rich field stars were formed in GC-forming GMCs or in the ISM enriched 
by ONe novae. It is likely that the high-density cold gas of GMCs can dramatically decelerate
the high-speed ($1600-6000$ km s$^{-1}$) ONe nova ejecta to finally trap it for further
star formation.
The present study accordingly suggests that the chemical properties of P-rich stars
in GCs with small [Fe/H] spreads can also provide a clue to the origin of the Galactic P-rich 
field stars.

\subsection{Abundance spreads in light $s$-process elements ?}

Internal abundance spreads in $s$-process elements have been so far revealed
in a number of the Galactic GCs with MPs
such as NGC 1851, and  M22 (Yong \& Grundahl 2008, Yong et al. 2009, 2014;
Marino et al. 2015; McKenzie et al. 2022).
The detailed abundance patterns of
GC stars with abundance enhancement in $s$-process elements
can not only determine the nucleosynthesis site (e.g., AGB stars or
massive stars) but also 
constrain the mass range of the polluters
(e.g., Shingles et al. 2014; McKenzie et al. 2022).
For example, Yong et al. (2014) suggested that 
AGB stars with  3-5${\rm M}_{\odot}$ are the major polluters
responsible for the formation of MPs in M2
based on  the observed  ratios of light ($ls$) and heavy ($hs$) $s$-process
abundances ($[ls/hs]$) as well as [Pb/La].
Yong et al. (2012) also found a possible  correlation
of [Y/Fe] and [Zr/Fe] (light $s$-process elements ) with [Na/Fe]
in M2, which would support chemical pollution by AGB stars in forming GCs.

The present study did not investigate the time evolution of light $s$-process elements
in forming GCs, though recent AGB and sAGB models 
have provided theoretical predictions on these chemical yields.
For example, Doherty et al. (2017) predicted [X/Fe] for light $s$-process elements
can be as high as 0.5 for sAGB stars with $m=7 {\rm M}_{\odot}$
and $Z=10^{-3}$ (see Lugaro et al. 2012 for the yields of AGB stars with  [Fe/H]$=-2.3$).
This result implies that even 2P stars in Type I GCs can have 
significant abundance spreads ($>0.05$) 
in light $s$-process elements depending on $F_{\rm dil}$
in the SCI scenario. 
Doherty et al. (2017) also predicted that [X/Fe] 
of heavy $s$-process elements (e.g., Ba) can be significantly
smaller than those of light $s$-process
elements in low-metallicity sAGB stars.

Schiappacasse-Ulloa et al. (2025) have derived [Y/Fe] and [Zr/Fe] of stars in
14 GCs and thereby investigated the spreads of these abundances separately
for 1P and 2P stars. 2P stars in some GCs appear to show $\Delta$[Y/Fe] and $\Delta$[Zr/Fe] 
larger than the typical uncertainty in the abundance measurements (0.15 dex),
which implies that massive AGB and sAGB stars polluted GC-forming GMCs.
However, it is not so clear why only some GCs show such larger abundance spreads in the SCI 
scenario: Possibly, only GCs with small $F_{\rm dil}$ can have observable
spreads in these elements.
There appears to be no evidence supporting that $\Delta$[Y/Fe] and $\Delta$[Zr/Fe]
are systematically larger than $\Delta$[Ba/Fe] and $\Delta$[La/Fe] in
Schiappacasse-Ulloa et al. (2025).
Future spectroscopic observations with more accurate abundance measurements (errors of $<0.05$ dex)
will be able to reveal  abundances spreads of  light 
and heavy $s$-process elements in 2P stars
caused by chemical enrichment by massive AGB and sAGB stars in the new scenario.

\subsection{R-process enrichment}

Internal abundances spreads for $r$-process elements have been so far confirmed
only in a small number of  Galactic GCs such as M15 and M92 
(e.g., Sneden et al. 1997; Roederer 2011; Worley et al. 2013; Kirby et al. 2023).
Recent theoretical models and numerical simulations of GCs (e.g., M5)
with abundance spreads in $r$-process elements (e.g., [Eu/Fe])
have shown that the origin of the spreads
can results from chemical enrichment by neutron star (NS) merging  in the early
formation histories of GCs (e.g., Bekki \& Tsujimoto 2017; Tarumi et al. 2021).
NSs originating from 1P stars explode during 2G formation so that gaseous
ejecta from NS merging  can enrich intracluster gas in the classic AGB scenario
(Bekki \& Tsujimoto 2017). As a result of this, 2P stars can have
higher abundances of $r$-process elements and 
a larger abundance spread in $r$-process elements (e.g., [Eu/Fe]) compared with
1P stars. Recent observations have shown that the abundance spread in  $r$-process 
elements is larger in 1P than in 2P for M15 (Henderson et al. 2025).
If this trend is observationally confirmed in other GCs for a large number of stars,
chemical enrichment by merging of NSs from 1P can be ruled out.

In the SCI scenario,  AGB stars polluting star-forming GMCs in a host galaxy
can initially have
a large abundance spread in their $r$-process element due to the possible
large abundance spreads in field stars of  the host.
If metal diffusion processes in high-density environments of GMCs
can significantly reduce the initial abundance spreads in $r$-process elements within
a few $10^7$ yr, then 2P stars are likely to  have the smaller  abundance spread
compared with 1P stars.
The present study is unable to provide predictions on how much  initial abundance spreads
in GMCs can be reduced during GMC growth due to metal diffusion.
It is thus our future study to understand the origin of the observed possible
large abundance spreads of $r$-process elements in 1P stars of GCs
using numerical simulations of GC formation
based on the SCI scenario.

\subsection{Discrete MPs of GCs  from GMC growth  merging}

Although the distributions of stars along the anticorrelations were assumed to be continuous in previous observational and theoretical studies of GCs, 
Recent precise spectroscopic measurements of [Mg/Fe] and [Al/Fe] 
for GC stars in NGC 2808 have revealed
that at least three distinct groups of stars exist
in the Mg-Al diagram  (Carretta 2014). Such discrete stellar populations 
have been found in other GCs like 
NGC 6752 and M22 (Marino et al. 2011;
Carretta et al. 2012; Milone et al. 2013), which implies 
that the distributions of stars along the anticorrelations of light
elements could not be continuous. The origin of this discreetness is yet to be
fully understood, though our previous theoretical models of GC formation
suggested that the observed discrete distributions are due to
sudden truncation of star formation by supernova feedback effects (Bekki et al. 2017).

If a GMC grows through merging of smaller molecular clouds (e.g., Kobayashi et al. 2017)
and if the smaller clouds are enriched by AGB stars to different degrees,
a group of stars formed from a cloud should be able to have
chemical abundance pattern that are distinct from those of any other groups of stars
formed in other clouds.
Accordingly, the young massive SC  formed from this growing  GMC
through molecular cloud merging should be able
to have  discrete MPs.
The establishment of discrete MPs within the building blocks of GC-hosting GMCs
has been already suggested by Elmegreen (2017), though the chemical enrichment
processes are quite different between his model and the  present SCI scenario.
It is our future study to demonstrate that the formation of discrete MPs
is indeed possible in GC formation from growing GMCs
based on our new hydrodynamical simulations of GC formation.

\begin{figure}
\psfig{file=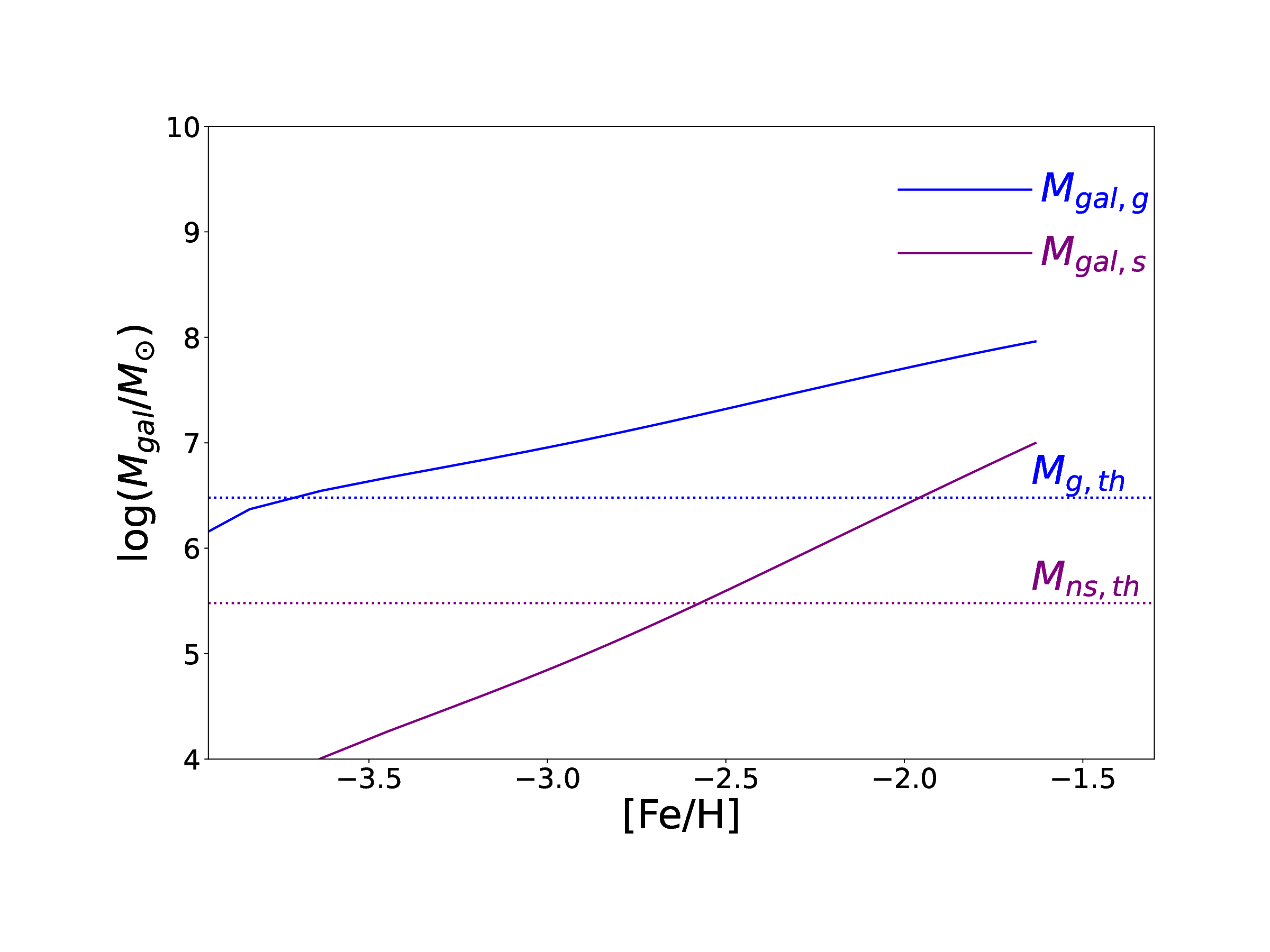,width=8.5cm} 
\caption{
Total gaseous (blue) and stellar masses (purple)  of
a gas-rich dwarf 
galaxy  as a function of [Fe/H] in the one-zone chemical evolution
model of the  galaxy for the first 1 Gyr evolution.
 The blue and purple dotted lines indicate the threshold 
gaseous ($M_{\rm g, th}$) and stellar masses ($M_{\rm ns, th}$) 
for GC formation with MPs, respectively.
The crosspoint between the solid and dotted lines
can mark the threshold [Fe/H] 
([Fe/H]$_{\rm th}$;  ``metallicity floor'')
above which GC formation with MPs is possible
in the SCI scenario.
Since the threshold [Fe/H] is higher for new stars (i.e., the crosspoint
between the two purple lines), the metallicity floor in this particular
model is around $-2.5$.
}
\label{Figure. 22}
\end{figure} 

\subsection{A possible minimum [Fe/H]  for MPs formation in  the Galactic and extragalactic 
GC systems (GCSs)}

The present new scenario predicts that only massive GMCs can become GCs with
MPs if they are chemically polluted by a sufficient number of AGB stars.
This means that there is  a threshold total gas mass of GMCs ($M_{\rm g, th}$)
over which GCs with MPs can be formed from GMCs.
The required larger ratio of new stars interacting with a GMC to the 
GMC mass ($R_{\rm s}=M_{\rm ns}/M_{\rm gmc}>0.1$) 
also means that there is a threshold total mass of new stars ($M_{\rm ns, th}$)
over which the GMC can be converted into a GC with MPs.
In order for a galaxy to form a GC with MPs,  its total gaseous and 
stellar masses ($M_{\rm gal,s}$ and $M_{\rm gal, g}$, respectively) should 
meet the following conditions at least:
\begin{equation}
M_{\rm gal, g} \ge M_{\rm g, th}
\end{equation}
and 
\begin{equation}
M_{\rm gal, s} \ge M_{\rm ns, th}.
\end{equation}
Not all  GMCs can form GCs in galaxies, and not all stars are new stars
with ages of $0.02-0.1$ Gyr, Therefore, the above required $M_{\rm gal, g}$ and 
$M_{\rm gal, s}$ are regarded as lower limits.
Since $M_{\rm gal, g}$ and $M_{\rm gal, s}$ can be related to [Fe/H] through
galactic chemical evolution,
we can  derive the minimum [Fe/H] required for the formation of GCs with MPs
([Fe/H]$_{\rm min}$) using our chemical evolution models (Bekki \& Tsujimoto 2012, BT12). 
Here we investigate the time evolution
of [Fe/H], $M_{\rm gal, g}$, and $M_{\rm gal s}$ for the first 1 Gyr
based on  our chemical evolution models
of dwarf galaxies with the final total masses of $M_{\rm gal}=10^8 {\rm M}_{\odot}$ at $T=1$ Gyr.
The model parameters such as gas the IMF and chemical yields
in these models
are exactly the same as those adopted in BT12.
We particularly discuss the results of the model with the Salpeter IMF ($\alpha=2.35$) 
in which the gas infall timescale ($\tau_{\rm inf}$) and the star formation efficiency ($C_{\rm SF}$)
are assumed to be 3 Gyr and 0.02, respectively.
Based on these results, we derive 
[Fe/H]$_{\rm min}$ for $M_{\rm gc}=3 \times 10^5 {\rm M}_{\odot}$,
$F_{\rm b}=10$, and $R_{\rm s}=0.1$).

Fig. 22 shows that the blue solid line for $M_{\rm gal, g}$ 
evolution crosses with the dotted line for $M_{\rm th, g}$
at [Fe/H]$ \approx -3.7$, which means that [Fe/H] should be higher than -3 for
the formation of massive GMCs: The gas might not
be in the form of molecular hydrogen due to such low dust abundances though.
On the other hand, the cross point is around [Fe/H]$\sim -2.6$ for $M_{\rm gal, s}$,
 which
is higher than [Fe/H]$_{\rm min}$ estimated from $M_{\rm gal, g}$ and $M_{\rm g, th}$.
This means that when a dwarf galaxy becomes as metal-rich as  [Fe/H]$\sim -2.6$, then
it can start to form  a GC with MPs in this particular model.

It should be noted here that [Fe/H]$_{\rm min}$ can become lower or higher
depending on the model parameters of chemical evolution models. 
For example,  [Fe/H]$_{\rm min}$ can be higher in dwarf galaxies with
lower $M_{\rm gal}$ for a given set of model parameters for star formation
and gas infall. It is beyond the scope of this paper to describe
how [Fe/H]$_{\rm min}$ depends on galaxy masses, star formation histories,
and gas infall rates etc in detail.
However, the present study indicates that there should be
[Fe/H]$_{\rm min}$ for the formation of GCs with MPs
in each of galactic building blocks.
The minimum of [Fe/H]$_{\rm min}$  among all building blocks of the Galaxy
 corresponds to
the observed ``metallicity floor'' (e.g., Beasley et al. 2019).

We note that the metallicity floor predicted by the SCI scenario arises from a distinct physical mechanism compared to that 
proposed by Kruijssen (2019), who argued that the observed paucity of GCs with [Fe/H] $\le  -2.5 $
reflects the galaxy mass-metallicity relation at high redshift. 
In the SCI scenario, an additional floor specific to MP formation 
arises from the requirement for sufficient AGB star 
density around GC forming GMCs, which is only met above a threshold galactic gaseous and stellar mass that 
corresponds to ${\rm [Fe/H]_min} \approx   -2.5$ 
 in the models. Both mechanisms may operate simultaneously, with the 
Kruijssen (2019) floor setting an absolute lower limit on GC survival and the SCI floor setting an additional 
constraint on which surviving GCs can develop MPs.

The metallicity floors of extragalactic GCSs
could be quite different from that of the Galactic GCS due to different formation
histories of 
galactic building blocks.
For example,  if the building blocks of a galaxy have systematically longer star-formation
timescales (i.e., slower chemical enrichment), then [Fe/H]$_{\rm min}$ of the GCS
can be higher. It is currently very difficult to observationally investigate
the presence or absence of MPs in all GCs of a galaxy outside the Local Group and thereby infer
[Fe/H]$_{\rm min}$ of the GCS, because it is exceptionally challenging  to
find observational evidence of MPs in integrated spectrum of such extragalactic GCs.
Nevertheless it is an intriguing observational question whether
GCSs have different [Fe/H]$_{\rm min}$ and how their [Fe/H]$_{\rm min}$ depend on
the global properties of their host galaxies.

The chemical evolution models described above can also predict the total mass of
intermediate-mass stars ($4 \le m/{\rm M}_{\odot} \le 10$), denoted as $M_{\rm TO}$, that 
have just left the main sequence at any given time step (every Myr) in a galaxy. This quantity, $M_{\rm TO}(4 \le m_{\rm TO}/{\rm M}_{\odot} \le 10)$, serves as a useful gauge for determining when (or at what $[{\rm Fe/H}]$) a galaxy
contains a sufficient amount of gas ejected from AGB stars to form a  GC with MPs within the SCI scenario.
Fig. 23 illustrates how the time evolution of $M_{\rm TO}$ depends on the star formation efficiency or rapidity ($C_{\rm SF}$) and the IMF slope ($\alpha$)
across six different models. Here, $M_{\rm TO}$ is normalized by the total mass of gas ($M_{\rm gas}$) that
infalls onto the galaxy over its entire evolution.
For example, if $M_{\rm TO}(4 \le m_{\rm TO}/{\rm M}_{\odot} \le 10) = 10^{-5}$,
the total mass of stars that have entered the AGB phase
over the last $10^6$~yr is $10^4~{\rm M}_{\odot}$ for a galaxy with $M_{\rm gas} = 10^9~{\rm M}_{\odot}$. This implies that the total gas mass ejected from massive AGB stars over the last $10^8$~yr is $\approx 9 \times 10^5~{\rm M}_{\odot}$, assuming an ejection fraction of $f_{\rm ej} = 0.9$.

Fig. 23 clearly demonstrates that at a given time $T$, $M_{\rm TO}$ for massive AGB stars is larger in models with a higher $C_{\rm SF}$ for a fixed $\alpha$,
which is driven by more rapid star formation. Regardless of the
chosen $C_{\rm SF}$ and $\alpha$, $M_{\rm TO}$ remains very low  during the initial
epoch ($T = 0$--$30$~Myr). This implies that the formation of GCs with MPs is
less likely during the very early phases of galaxy evolution. Given that stars formed earlier typically possess lower $[{\rm Fe/H}]$, this result
supports the existence of a metallicity floor for the formation of GCs with MPs.
Furthermore, models featuring a top-heavy IMF ($\alpha = 1.5$) systematically exhibit larger $M_{\rm TO}$ values across all three $C_{\rm SF}$, suggesting that the IMF is also a key determinant of the 2P fraction ($F_{\rm 2P}$) in GCs.
It should be emphasized that within the SCI scenario, both the total mass of the AGB ejecta and the mass
function of the AGB stars polluting GMCs are crucially important. Appendix D describes how $M_{\rm TO}$ evolves as a function of the turn-off mass ($m_{\rm TO}$) over time ($T$),
thereby demonstrating when GMCs can be exclusively polluted by massive AGB stars.

\begin{figure}
\psfig{file=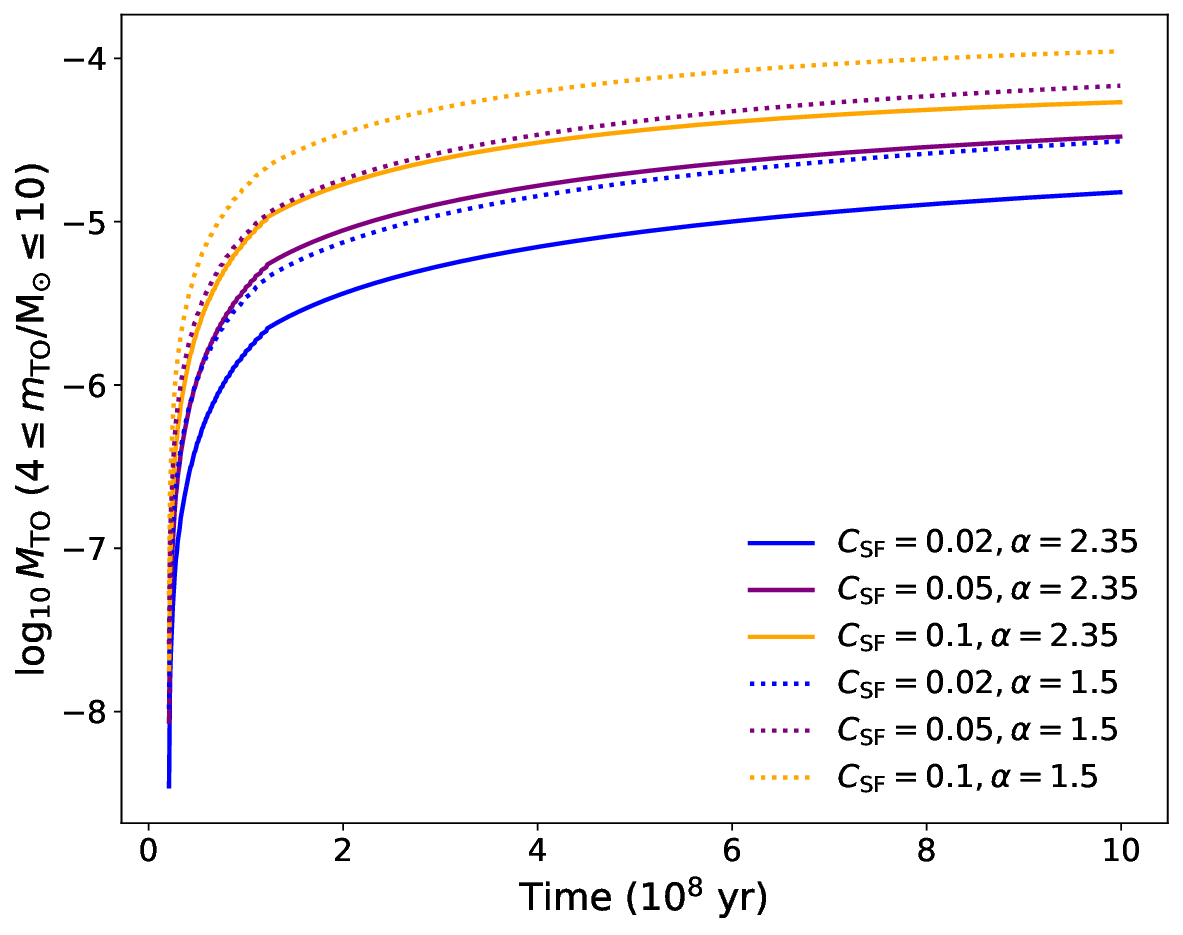,width=8.5cm} 
\caption{
$M_{\rm TO}$ of intermediate-mass stars as a function of time ($T$) in six models with $C_{\rm SF}=0.02$ (blue),
0.05 (purple), and 0.1 (orange) for the IMF slope $\alpha=2.35$ (solid lines) and 
1.5 (dotted lines).
The total mass of intermediate-mass stars ($M_{\rm TO}(4 \le m_{\rm TO}/M_{\odot} \le 10$)) that have 
just left the main-sequence at a give time step is calculated in these models.
$M_{\rm TO}$ normalized by the total mass of in falling gas is plotted every Myr in this figure.
Larger $C_{\rm SF}$ means that gas is more rapidly consumed by star formation.
}
\label{Figure. 23}
\end{figure} 

\begin{table*}
\centering
\begin{minipage}{180mm}
\caption{Predicted properties of GCs with MPs from
 the new scenario to be compared with
future observations.}
\begin{tabular}{lll}
Properties  &  Predictions & Comments \\
$F_{\rm 2P}-\Sigma_{\rm SFR}$ relation & 
Higher $F_{\rm 2P}$ in GCs forming  with higher $\Sigma_{\rm SFR}$  &
Higher $F_{\rm 2P}$ is likely for GCs formed at higher $z$.  \\
$F_{\rm 2P}-$[Eu/Mg] relation  & 
Lower $F_{\rm 2P}$ in GCs with higher [Eu/Fe] &
This can be clearly seen for a given $M_{\rm gc}$.   \\
Li-Al (Li-Na) anticorrelations  & 
Diverse slopes (from flat to shallow)  &
The polluter's mass range  determines the slopes. \\
$\delta$A(Li)$-M_{\rm gc}$  relation  & 
Larger A(Li) spreads in larger $M_{\rm gc}$   & 
This is due to $F_{\rm dil}$ dependent on $M_{\rm gc}$.\\
Nitrogen of P-rich stars  & 
High [N/Fe]  ($>0.5$)   & 
P-rich stare can be  preferentially found in 2P. \\
$^{12}$C/$^{13}$C ratios  in unevolved  stars  &
Very low ratios ($\approx 5$) in 2P depending on $F_{\rm dil}$ &
1P-2P difference in this ratio can be quite large. \\
$^{12}$C/$^{13}$C$-$[Na/Fe] anticorrelation   & 
Lower $^{12}$C/$^{13}$C ratios for higher [Na/Fe] &
This is expected for other light elements.\\
Precursor (0P) stars in  GCs' halos   & 
Peculiar chemical abundance patterns   &
They originate from  field stars of GC-hosts. 
\end{tabular}
\end{minipage}
\end{table*}

\subsection{Formation of N-rich stars from disintegration of low-mass
clusters}

The inner regions of the Galaxy are observed to have field stars with
[N/Fe]$>0.5$, the origin of which is yet to be fully understood 
(e.g.,  Fern\'andez-Trincado et al. 2019; Spite et al. 2022; Kane et al. 2026).
Although these N-rich stars might  have originated from disintegration of 
GCs with MPs  (e.g., Martell et al. 2010; Fern\'andez-Trincado et al. 2019) 
or from disruption of the Galactic building blocks
containing N-rich stars (e.g., Bekki 2019c),
we here suggest an alternative idea about their origins.
Although low-mass  clusters in the present-day Galaxy
do not have MPs with  anticorrelations between light elements
(e.g., Pancino et al. 2011),
the present model predicts that even low-mass SCs can 
possess MPs if their progenitor GMCs are chemically enriched by AGB stars to
a large extent  due to
rather high $\Sigma_{\rm SFR}$ of their host galaxies.
NGC 6535 with $M_{\rm gc}\approx 10^{3.53} {\rm M}_{\odot}$ and
a O-Na anticorrelation
(e.g., Bragaglia et al. 2017)
is an example of such low-mass SCs with MPs.
These low-mass SCs
can be quickly disintegrated  by their host galaxies' tidal fields and two-body
dynamical relaxation effects to finally become field stars.
These field stars can have high [N/Fe] 
and thus be observed as N-rich stars in the inner region 
of the Galaxy. The total number and mass of these low-mass SCs with N-rich
stars are much larger than those of the Galactic GCs, which implies that
the observed large fraction ($\sim 0.15$) of N-rich stars can be
naturally explained.
Therefore  the present study suggests that 
the stripping of 2P stars  from the Galactic 
GCs with MPs is not the only process for the
formation of the observed N-rich stars in the Galaxy:
the vast majority of N-rich stars in the Galaxy might originate
from disintegrated young low-mass SCs with N-rich stars.

Low-mass SCs with N-rich stars could be the fundamental building blocks
of a galaxy, if the building blocks (i.e., dwarf galaxies) of the
galaxy have rather high $\Sigma_{\rm SFR}$.
There are a number of spectroscopic observations that support the presence
of N-rich stellar populations in giant elliptical galaxies
(e.g., Schiavon 2007) and ultra-compact dwarf galaxies 
(e.g., Strader et al. 2013). It is our future work 
to discuss whether the building blocks of these galaxies could have
rather high $\Sigma_{\rm SFR}$ for longer timescales based on
our new hydrodynamical simulations of these galaxies.
The disintegration of such low-mass SCs with N-rich stars could be related
to the long-standing problem of the UV upturn phenomenon  in giant elliptical
galaxies (Bekki 2012; Goudfrooij 2018; Goudfrooij et al. 2026), 
because N-rich stars are also He-rich in the SCI scenario.

\subsection{Observational evidence of AGB-GMC interaction in our backyards}

The chemical abundance patterns of presolar grains have long been
considered to contain fossil records of interactions between
the parent GMC of our  solar system ("Sun-hosting MC")
and various types of stars (e.g., CCSNe, AGB stars, novas)
encountering with the GMC (e.g., Lugaro 2005). 
For example, the elevated  ${}^{26}$Al abundances
of presolar grains were suggested to result from mixing of gaseous ejecta
from massive AGB stars with original gas of the Sun-hosting GMC
(e.g., Kastner \& Myer 1994; Wasserburg et al. 2006; Trigo-Rodriguez et al. 2009;
Parker \& Schoettler 2003).

Trigo-Rodriguez et al. (2009) showed that if 
a massive AGB star with $m=6.5 {\rm M}_{\odot}$ polluted the Sun-hosting
MC with a dilution factor ($M_{\rm g}/M_{\rm ej}$)  of $\approx 300$, then
the observed abundances of 
${}^{26}$Al, 
${}^{41}$Ca, 
${}^{60}$Fe, 
and ${}^{107}$Pd 
in presolar grains can be self-consistently
reproduced. The large dilution factor implies that the initial mass of the GMC ($M_{\rm gmc}$)
is at least $\approx 2000 {\rm M}_{\odot}$ (for $f_{\rm ej}=0.9$).
If the entire GMC was polluted by multiple AGB stars, 
and if the initial mass of the SC ($M_{\rm sc}$)
hosting the Sun is  $10^3 {\rm M}_{\odot}$ (Adams 2010),
then the total number of AGB stars ($N_{\rm agb}$) polluting the MC  is $\approx 6$
for $\epsilon_{\rm sf}=0.1$. Therefore, it is likely that
the Sun-hosting MC might have been chemically enriched by multiple  massive
AGB stars.

If the Sun-hosting GMC was polluted by AGB stars with low $N_{\rm agb}$
due to  low $\Sigma_{\rm SFR}$ of the Galaxy about 4.6 Gyr ago,
it would be quite reasonable that GMCs within compact dwarf galaxies
with rather high $\Sigma_{\rm sf}$ at high-$z$ are polluted by a 
large number of AGB stars.
The present study predicts that low-mass SCs can have star-to-star abundance inhomogeneities
to much smaller degrees due to small numbers of AGB stars interacting with their parent GMCs.
However,
it is very difficult for observational studies of low-mass SCs
to reveal subtle differences in chemical abundances between their member  stars.
The observed extremely accurate  chemical abundance patterns of presolar grains are
the only fossil records that can be used to discuss
whether GMC-AGB interactions occurred in the early stages of low-mass SC  formation from GMCs.
Thus, AGB-GMC interactions are relevant to not only GC formation with MPs but also to the origin of presolar 
grains in our own solar system - representing an independent but cohesive line of evidence for the physical 
processes central to the SCI scenario.

\subsection{Comments on other scenarios}

Recent hydrodynamical simulations of star formation within GMCs have demonstrated that pre-SN feedback effects of massive stars, such as radiation pressure and photoevaporation, can  dramatically  decrease the
 SFEs (e.g., 2--10\%; Kim et al. 2021). Given that stellar feedback effects of massive stars with $m=[10-100] {\rm M}_{\odot}$ are incorporated in these simulations, the feedback effects of very massive and supermassive stars with $m >1000 {\rm M}_{\odot}$ could be by several orders of magnitude stronger than those predicted in these simulations (owing to $L \propto M^{3.5}$, where $L$ and $M$ are stellar luminosity and mass, respectively). Accordingly, although such very massive stars 
may form in young SCs (e.g., Lahań et al. 2025), such stars are highly likely to quench star formation rather than promote it. This enormously strong feedback effect of very massive and supermassive stars on star formation needs to be quantitatively investigated in GC formation scenarios with these stars being major polluters.

As previously discussed in preceding sections of this paper, 
the observed slopes in the anticorrelations (correlations) of A(Li) with [Al/Fe] ([O/Fe]) rule out GC formation scenarios in which intracluster gas is chemically enriched {\it only by a polluter} that cannot produce Li. Accordingly, the observed slopes imply that stellar sources ejecting Li-free or extremely Li-depleted gas, such as very massive and supermassive stars, fast-rotating massive stars, and massive interacting binaries, cannot be a major polluter responsible for the formation of MPs in GCs. However, if such Li-free gas is mixed with both pristine gas and Li-enhanced ejecta from other stellar sources (e.g., AGB stars) to form 2P stars, the observed slopes in Li-Al anticorrelations can be well explained. Therefore, a hybrid scenario in which both AGB stars and massive stars chemically enrich pristine GMC gas to form 2P stars can be explored for a better formation scenario of GCs with MPs.

Cluster environments soon after the removal of all remaining pristine gas of GMCs by the 
feedback effects of CCSNe might not be as hospitable for secondary star formation as the classic AGB scenario envisaged. If 
15\% of all CCSNe are delayed ones that can explode 50--200 Myr after 
cluster formation (Zapartas et al. 2017), then the time 
interval ($\delta t_{\rm dCCSN}$) between two consecutive explosions of delayed CCSNe can be as 
short as $\approx 3 \times 10^4$ yr for a total CCSN number of $\approx 10^4$, 
which is reasonable for GCs with $M_{\rm gc,0}=10^6 {\rm M}_{\odot}$ with the Salpeter IMF. 
If this $\delta t_{\rm dCCSN}$ is shorter than the typical timescale of star formation ($t_{\rm sf}$), 
the classic AGB scenario has a serious problem. Likewise, if the timescale of disruptive interactions between proto-stellar cores and already existing (1P) stars ($t_{\rm int}= 10^3-10^4$ yr; Bobrick et al. 2025) 
is shorter than $t_{\rm sf}$, the scenario also has a serious problem. 
Thus, GC formation models based on the classic AGB scenario will need to investigate whether and how $t_{\rm sf}$ can be shorter than $\delta t_{\rm dCCSN}$ and $t_{\rm int}$ in order to assess the validity of the scenario.

Almost all models for GC formation with MPs assume that star formation from gaseous ejecta of polluters mixed with pristine gas is possible even after the formation of clusters of massive stars ($m>10 {\rm M}_{\odot}$). These models have not yet assessed the validity of the assumption using hydrodynamical simulations of GC formation with various stellar feedback effects; the assumption could be unrealistic and unreasonable. The present study has shown that chemical enrichment of GMCs by existing AGB stars before the formation of clusters of massive stars ($m>10 {\rm M}_{\odot}$) is closely associated with the formation of GCs with MPs in galaxies with high $\Sigma_{\rm SFR}$. Given that AGB stars are not the only stellar objects that can influence the chemical and dynamical evolution of GMCs, other sources such as CO and ONe novae, CCSNe, delayed CCSNe, and SNe Ia surrounding GMCs can also play some roles in GMC evolution within galaxies with high $\Sigma_{\rm SFR}$. We will need to investigate the roles of these various star-cloud interactions in GMC evolution, because such interactions might be imprinted on the chemical abundance patterns of GCs.

\section{Conclusions}

We have presented a scenario in which globular clusters (GCs) with multiple stellar populations (MPs) 
can be formed from massive star-forming gas clouds being chemically polluted by pre-existing Asymptotic Giant Branch (AGB) stars in gas-rich dwarf galaxies. In this star-cloud-interaction (``SCI'') scenario, it is assumed that the GC-forming gas clouds have physical properties similar to present-day giant molecular clouds (GMCs), though the mass fractions of molecular hydrogen gas could be lower than those of typical GMCs due to low metallicities. This new scenario differs from the classic AGB scenario in that AGB stars formed before GMC formation can enrich star-forming GMCs through their stellar winds. Accordingly, the new scenario does not suffer from the potentially serious problems inherent to the classic one, such as the mass-budget and the dilution timing problems.

Using idealized analytic models for GC formation with MPs based on the new scenario, we have investigated various physical properties of GCs—such as correlations between first-population (1P) fractions and global GC masses, as well as chemical abundance patterns—and thereby compared the results with corresponding observations. We have found that most of the observed GC properties, such as chemical abundances, can be well reproduced by the new scenario with a particular set of model parameters, though a number of properties are yet to be fully reproduced (see Table~3). We have also provided key theoretical predictions of the scenario to be compared with future observational studies of GCs; Table~4 summarizes these predictions. 
The principal results are as follows: \\

(1) The surface number densities of intermediate-mass stars with ages ranging from 0.02 to 0.1~Gyr ($\Sigma_{\rm agb}$) around and within star-forming GMCs can determine whether GCs formed within the GMCs can develop MPs. 
GCs can host MPs if they are formed from GMCs that are chemically polluted by AGB stars due to a high $\Sigma_{\rm agb}$. It is proposed that the observable surface densities of star formation rates ($\Sigma_{\rm SFR}$) can be used 
as a proxy for $\Sigma_{\rm agb}$ in discussing whether GCs with MPs can be 
formed in galaxies. A value of $\Sigma_{\rm SFR, th} \approx 1~{\rm M}_{\odot}~{\rm kpc}^{-2}~{\rm yr}^{-1}$ is 
suggested as a threshold SFR density above which GCs with MPs can form within galaxies.

(2) More massive GCs can have smaller 1P fractions ($F_{\rm 1P}$), mainly because more massive 
GC-forming GMCs can interact with larger numbers of AGB stars (per unit mass) within and around the GMCs 
to be chemically enriched by their stellar winds to a greater extent (i.e., 
larger $R_{\rm s}$ for larger $M_{\rm gmc}$). The predicted $M_{\rm gc}$–$F_{\rm 1P}$ relation is consistent with the observed one, though the simulated dispersions of $F_{\rm 1P}$ for a given $M_{\rm gc}$ are smaller than observed. The present model predicts that low-mass SCs ($< 10^4~{\rm M}_{\odot}$) can initially host a small fraction of second-population (2P) stars in galaxies with high $\Sigma_{\rm SFR}$, though they can be completely disintegrated over 13~Gyr.

(3) More massive GCs can also have larger spreads in helium abundance ($\delta Y$) 
between 1P and 2P stars, alongside larger differences between the minimum and 
maximum $Y$ ($\delta_{\rm max} Y$). 
Therefore, $F_{\rm 1P}$ is anticorrelated with $\delta Y$ and $\delta_{\rm max}Y$ in simulated GCs. 
The simulated $1\sigma$ dispersion of $\delta Y$ for each $M_{\rm gc}$ bin is
significant ($\approx 0.01$), resulting primarily from the initial $R_{\rm s}$
dispersion for a given GMC mass. Notably, this simulated $1\sigma$ dispersion
is consistent with the corresponding observational data.

(4) The predicted $M_{\rm gc}$–$\delta Y$ relation is significantly steeper 
than the observed one, 
which means either that the $Y$ yields adopted in the present study are too large or that the SCI scenario has a potentially serious problem in explaining these helium observations.
It should be noted here that
the $\delta Y$  values must be derived indirectly 
from multiband photometric comparisons of population sequences using synthetic spectra and isochrone fitting (M18).

(5) The SCI scenario can successfully reproduce the observed O–Na and C–N anticorrelations of GCs, 
consistent with the established results of  the classic AGB scenario. 
The evolutionary loci of the present models on the $[{\rm O}/{\rm Fe}]$–$[{\rm Na}/{\rm Fe}]$ plane depend on the adopted model parameters, such as the stellar initial mass function (IMF), mass ranges of polluting AGB stars, and metallicities. This result implies that the observed diverse O–Na anticorrelations can be caused by variations of these parameters during GC formation. The high-metallicity models ($[{\rm Fe}/{\rm H}] = -0.4$) show significantly steeper O–Na anticorrelations compared with the low-metallicity ones ($[{\rm Fe}/{\rm H}] = -1.8$), which appears to be consistent with the latest observations.

(6) Both the Mg–Al anticorrelation and the Si–Al correlation observed in low-metallicity GCs can be qualitatively reproduced by the new scenario. There is expected to be no such correlation or anticorrelation in high-metallicity GCs under the new scenario, because gaseous ejecta from massive AGB stars at high metallicities does not exhibit Mg-depletion and Si-enhancement. The observed Si–Al relation shows a larger spread in $[{\rm Si}/{\rm Fe}]$ compared with the model predictions, which suggests that further improvement of the models and AGB yields is required to better match the observations.

(7) The Mg–K anticorrelation observed in NGC~2419 can be reproduced well by the models in which: (i) super-AGB (sAGB) stars are the major sources polluting star-forming GMCs, and (ii) sAGB yields from Iliadis et al. (2016) are adopted. 
However, these models do not reproduce the Mg-K anticorrelations observed in $\omega$  Centuari and NGC1786 with the same fidelity.
This inconsistency has the following two implications. One is that the Mg and K yields of sAGB stars contributing to the chemical enrichment of forming GCs could differ due to metallicity or mass differences among the sAGB stars. If the (IMF-averaged) Mg and K yields of sAGB stars are indeed different, diverse Mg–K anticorrelations observed in different GCs 
could be explained within the SCI scenario. 
The other implication is that sAGB stars are not responsible for the formation of Mg–K anticorrelations, meaning the 
SCI scenario has a serious problem. Since Mg and K yields dependent on the physical parameters of sAGB stars are currently lacking, we are unable to definitively conclude whether the SCI scenario has a serious problem in reproducing the observed Mg–K anticorrelations in GCs.

(8) The models with lithium production by massive AGB stars can much better reproduce the observed, remarkably flat relation between $A({\rm Li})$ and $[{\rm Al}/{\rm Fe}]$ compared with those without Li production. In particular, the observed 2P stars with $[{\rm Al}/{\rm Fe}] \approx 0.5$ and $A({\rm Li}) \approx 2.3$ can be best explained by the models in which the fraction of Li-producing AGB stars among all AGB stars ($P({\rm Li})$) is as high as 20\%. These results suggest that GC formation models with polluters incapable of producing Li are disfavoured  as a viable framework for the origin of MPs in GCs. The models with lower $P({\rm Li})$ show weak anticorrelations of $A({\rm Li})$ with $[{\rm Na}/{\rm Fe}]$ and $[{\rm Al}/{\rm Fe}]$, which can explain the origin of the trends observed in NGC~6752. Detailed comparison with the observed $A({\rm Li})$–$[{\rm Al}/{\rm Fe}]$ relation allows us to tightly constrain the fraction of AGB stars that need to produce Li in the new scenario.

(9) The observed relations between $[{\rm Al}/{\rm Fe}]$ and magnesium isotopes (${}^{25}{\rm Mg}/{\rm Mg}$ and ${}^{26}{\rm Mg}/{\rm Mg}$ ratios) can be well reproduced by the new scenario, but only if we adopt abundance ratios of ${}^{25}{\rm Mg}/{\rm Mg}$ and ${}^{26}{\rm Mg}/{\rm Mg}$ in the AGB ejecta that are a factor of 2–3 lower than the values predicted in the literature. This discrepancy suggests that current stellar models of AGB stars over-predict the abundances of ${}^{25}{\rm Mg}$ and ${}^{26}{\rm Mg}$ isotopes in their stellar winds. It remains possible that the required smaller abundances of these Mg isotopes could represent a potential problem for the SCI scenario.

(10) The new scenario predicts that the metallicity spreads in the 1P stars of GCs are inherited from those in their parent GMCs and from the field stars trapped by the GMCs (``0P''). It also predicts that the metallicity spreads in the 2P population can be smaller than those in the 1P population, because 2P stars form later from gas that has experienced substantial metal homogenization due to turbulent diffusion within the forming GMCs.

(11) The origin of the observed Type-I versus Type-II GC dichotomy can be well understood in the context of the SCI scenario. First, two distinct GCs form from separate GMCs polluted by AGB stars within their host dwarf galaxy at different epochs. The later-formed GC can have a higher $[{\rm Fe}/{\rm H}]$ and higher abundances of $s$-process elements (e.g., $[{\rm Ba}/{\rm Fe}]$) due to background chemical enrichment by core-collapse supernovae (CCSNe) and lower-mass AGB stars ($m < 4~{\rm M}_{\odot}$) within the host galaxy, provided it forms more than $\approx 100$~Myr after the earlier cluster. The two clusters then merge together to form a single, complex GC that displays two distinct populations with different $[{\rm Fe}/{\rm H}]$ (hence ``iron-complex'') and $[{\rm Ba}/{\rm Fe}]$ profiles ($s$-poor and $s$-rich), while each population internally maintains the characteristic light-element anticorrelations. This GC merging mechanism can naturally solve the severe mass-budget problem for $s$-rich stars in Type-II GCs (e.g., M22) and firmly links internal $[{\rm Fe}/{\rm H}]$ and $[{\rm Ba}/{\rm Fe}]$ spreads to galaxy-scale chemical evolution in the early universe.

(12) The observed fraction of Type-II GCs among Galactic clusters ($\sim 20\%$) can be understood in the context of GC merging probabilities within the Galactic building blocks (i.e., now-defunct dwarf galaxies). The differences in the mass fractions of $s$-rich stars among different Type-II GCs can be attributed to variations in the mass ratios of the two merging clusters in their host dwarf galaxies. Furthermore, the chemical abundance patterns of $s$-process elements in Type-II GCs are better explained by chemical pollution from low-mass AGB stars rather than the weak $s$-process in massive stars.

(13) The chemical abundances of phosphorus-rich stars with $[{\rm P}/{\rm Fe}] > 0.5$ 
recently discovered in a few Galactic GCs with MPs (e.g., M4) provide unique new constraints on the theory of GC formation. The chemical abundance patterns of such P-rich stars can be explained if they formed from P-rich gaseous ejecta from oxygen-neon (ONe) novae that mixed thoroughly with N-rich AGB ejecta within the cluster-forming GMCs. High-speed ($\approx 1000~{\rm km~s}^{-1}$) ejecta from ONe novae can be efficiently trapped by GMCs through hydrodynamical interactions between the fast ejecta and the cold, dense cloud gas, enabling it to mix with the ambient GMC material and AGB stellar winds for star formation. Thus, the SCI scenario offers a plausible explanation for the origin of P-rich stars, although the fine details of P-rich star formation must be investigated in depth by our future hydrodynamical simulations of GC formation with ONe novae.

(14) There should be a strict ``metallicity floor'' that defines the minimum metallicity ($[{\rm Fe}/{\rm H}]_{\rm min}$) required for GC formation with MPs. The new scenario predicts that GCs with MPs can form: (i) if the gas mass $M_{\rm g}$ is larger than a threshold cloud mass ($M_{\rm g, th}$), and (ii) if the pre-existing stellar mass $M_{\rm ns}$ is larger than a threshold mass of intermediate-mass stars with ages of 0.02–0.1~Gyr. Therefore, GC-host dwarf galaxies are required to have global gaseous and stellar masses ($M_{\rm gal, g}$ and $M_{\rm gal, s}$) that are larger than $M_{\rm g, th}$ and $M_{\rm ns, th}$, respectively. This lower limit on $M_{\rm gal, g}$ and $M_{\rm gal, s}$ directly implies the presence of a metallicity floor ($[{\rm Fe}/{\rm H}]_{\rm min}$), because $[{\rm Fe}/{\rm H}]$ strongly correlates with galactic gas and stellar masses through global star formation and chemical evolution in dwarf galaxies. The present scenario predicts that $[{\rm Fe}/{\rm H}]_{\rm min}$ can vary between different galactic GC systems because the masses and star formation histories of their primordial building blocks were likely quite diverse.

(15) Since $\Sigma_{\rm SFR}$ must be exceptionally high for the formation of GCs with MPs in the new scenario, it is highly unlikely that the present-day LMC can form GCs with MPs due to its low global $\Sigma_{\rm SFR}$. The LMC may last have formed GCs with MPs about a few billion years ago, when it experienced a massive starburst triggered by tidal interactions with the MW
 and the Small Magellanic Cloud (SMC). Therefore, the observed apparent age threshold above which LMC clusters exhibit MPs results not from a physical age constraint, but from the lack of intense starburst activity over the last few Gyr (leading to a lower $\Sigma_{\rm SFR}$ and thus an insufficient $\Sigma_{\rm agb}$). Under the new scenario, starbursting blue compact dwarf galaxies (BCDs) with compact sizes can successfully produce GCs with MPs, whereas star-forming, gas-rich ultra-diffuse galaxies (UDGs) are highly unlikely to form them.

(16) The present scenario predicts that the mass fraction of 2P stars ($F_{\rm 2P}$) in GCs with MPs formed in galaxies with higher $\Sigma_{\rm SFR}$ will be systematically larger due to more efficient chemical enrichment of the parent GMCs by AGB stars. The scenario also predicts that GCs with older ages and lower metallicities tend to have higher $F_{\rm 2P}$ values, because their primordial host galaxies were likely to feature much higher baseline values of $\Sigma_{\rm SFR}$. Therefore, the observed relationship between redshift evolution ($z$), $\Sigma_{\rm SFR}$, and metallicity in high-redshift star-forming galaxies can provide crucial hints regarding the origin of the observed dependence of $F_{\rm 2P}$ on cluster ages and metallicities.

(17) A top-light IMF is expected for lower $\Sigma_{\rm SFR}$ environments within the framework of the integrated galaxy-wide IMF theory (Weidner \& Kroupa 2005), and it represents a key ingredient in chemical evolution models that reproduce the lower $[\alpha/{\rm Fe}]$ and 
elavated  $[{\rm Eu}/\alpha]$ ratios observed in massive satellite galaxies in the Local Group (e.g., Tsujimoto 2024). It is thus possible that accreted (ex-situ) GCs with elevated $[{\rm Eu}/\alpha]$ ratios that formed in defunct dwarf galaxies characterized by top-light IMFs and lower $\Sigma_{\rm SFR}$ will exhibit lower $F_{\rm 2P}$ values for a given global GC mass.

(18) Future observational studies of ${}^{12}{\rm C}/{}^{13}{\rm C}$ ratios in unevolved main-sequence stars and their correlations with light-element abundances (e.g., $[{\rm Na}/{\rm Fe}]$) and $M_{\rm gc}$ are crucial, because these observations can provide completely independent constraints on GC formation scenarios. The new scenario explicitly predicts low ${}^{12}{\rm C}/{}^{13}{\rm C}$ ratios ($\approx 5$) for the enriched population and large differences in this ratio between 1P and 2P stars ($\approx 90$). It is thus important for alternative GC formation frameworks to predict their own expected ${}^{12}{\rm C}/{}^{13}{\rm C}$ ratios for 2P stars to allow for clean observational tests.

(19) The classic AGB scenario would still be promising if both the feedback effects of delayed CCSNe and the disruptive encounters of existing (1P) stars with protostellar cores are shown to be unable to severely suppress secondary star (2P) formation within dense stellar environments. Conversely, GC formation scenarios invoking polluters unable to produce lithium (e.g., massive stars and massive interacting binaries) face a serious problem in reproducing the observed, remarkably flat 
correlation of $A({\rm Li})$ with $[{\rm Al}/{\rm Fe}]$. Other scenarios 
based on gas accretion onto existing 1P stars, stellar collisions, or stellar mergers
have not yet been demonstrated to reproduce the full range of chemical abundance 
patterns characteristic of GCs, though further development of these frameworks may yet provide viable alternatives.
Now that recent observational studies of GCs with MPs have revealed a diverse array of physical properties and clear correlations with global cluster parameters, predictions from all these candidate formation scenarios must be tested against these rich empirical datasets in a comprehensive manner.

(20) Since the present study is based on idealized analytic models, our future studies will need to demonstrate that GC formation proceeds along the pathways outlined here using more sophisticated hydrodynamical simulations of GC formation from GMCs polluted by AGB stars. We will also need to compare the observed structural and kinematic differences between 1P and 2P stars within GCs hosting MPs against the corresponding multi-dimensional predictions from our forthcoming numerical simulations. Ultimately, future observational searches for young massive star clusters with MPs across a wide range of host galaxy environments will validate or disprove this new scenario.

\section{DATA AVAILABILITY}
The data used in this paper (outputs from computer simulations)
will be shared on reasonable request
to the corresponding author.

\section{Acknowledgment}
The hydrodynamical simulations of gas-rich disks galaxies
were  performed on the OzSTAR national facility at Swinburne University of Technology.
The OzSTAR program receives funding in part from the Astronomy National
Collaborative Research Infrastructure Strategy (NCRIS) allocation provided by the
Australian Government, and from the Victorian Higher Education State Investment
Fund (VHESIF) provided by the Victorian Government.
Support for this work was provided by NASA through the NASA Hubble Fellowship 
grant \#HST-HF2-51560 awarded by the Space Telescope Science Institute, which is operated by the Association of Universities 
for Research in Astronomy, Inc., for NASA, under contract NAS5-26555.

{}

\appendix

\section{Fractions of AGB stars within and around GMCs}

\begin{figure*}
\psfig{file=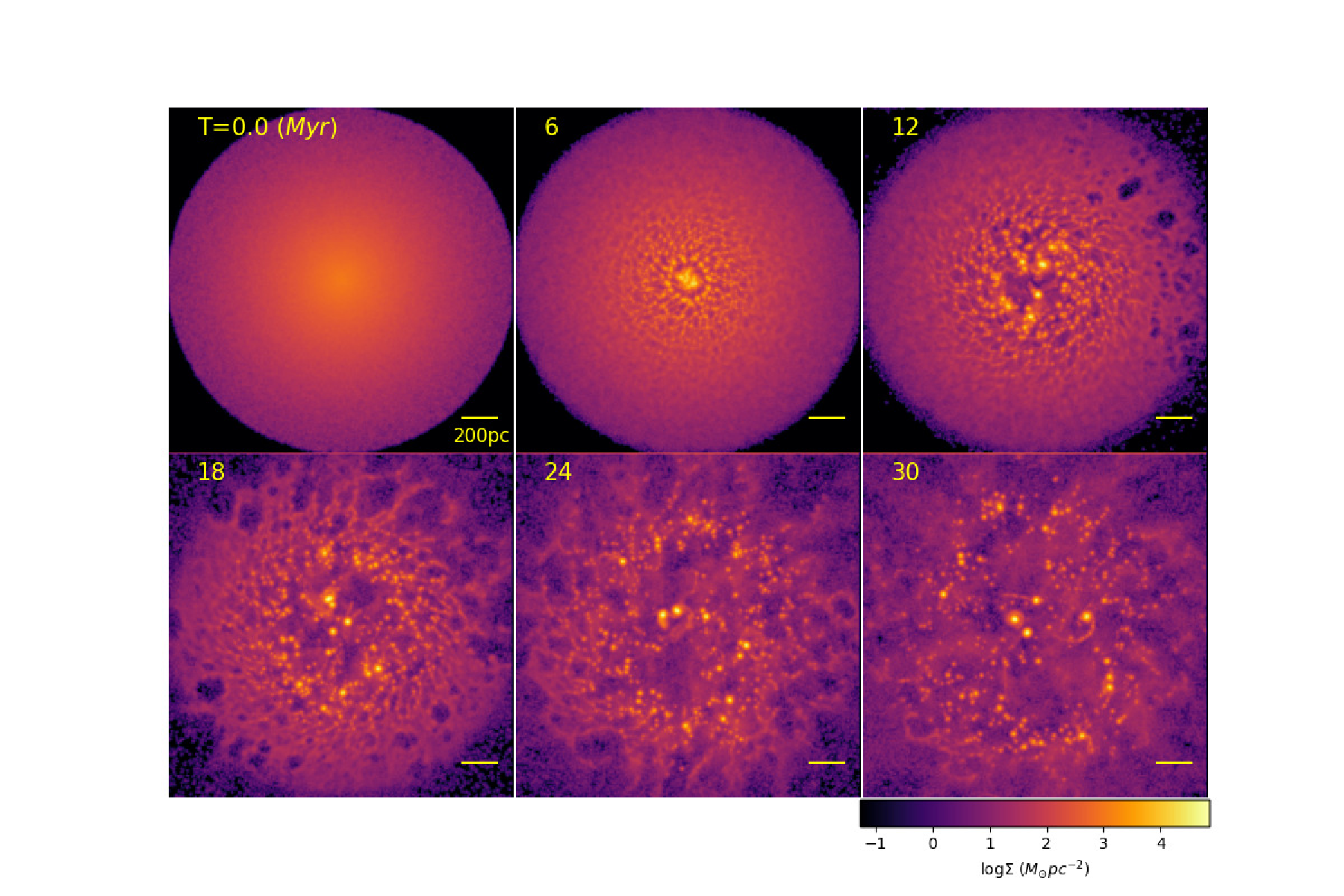,width=18.5cm} 
\caption{
Time evolution of the surface gas densities ($\Sigma_{\rm g}$) projected onto the $x$-$y$
plane for a gas-rich compact dwarf galaxy model in which massive GMCs can be formed
during the evolution of the gas disk.
The time $T$ that has elapsed since the start of this simulation
is shown in the upper left corner at each panel, and the bar in
the lower right corner measures 200 pc.
}
\label{Figure. 24}
\end{figure*} 

\begin{figure}
\psfig{file=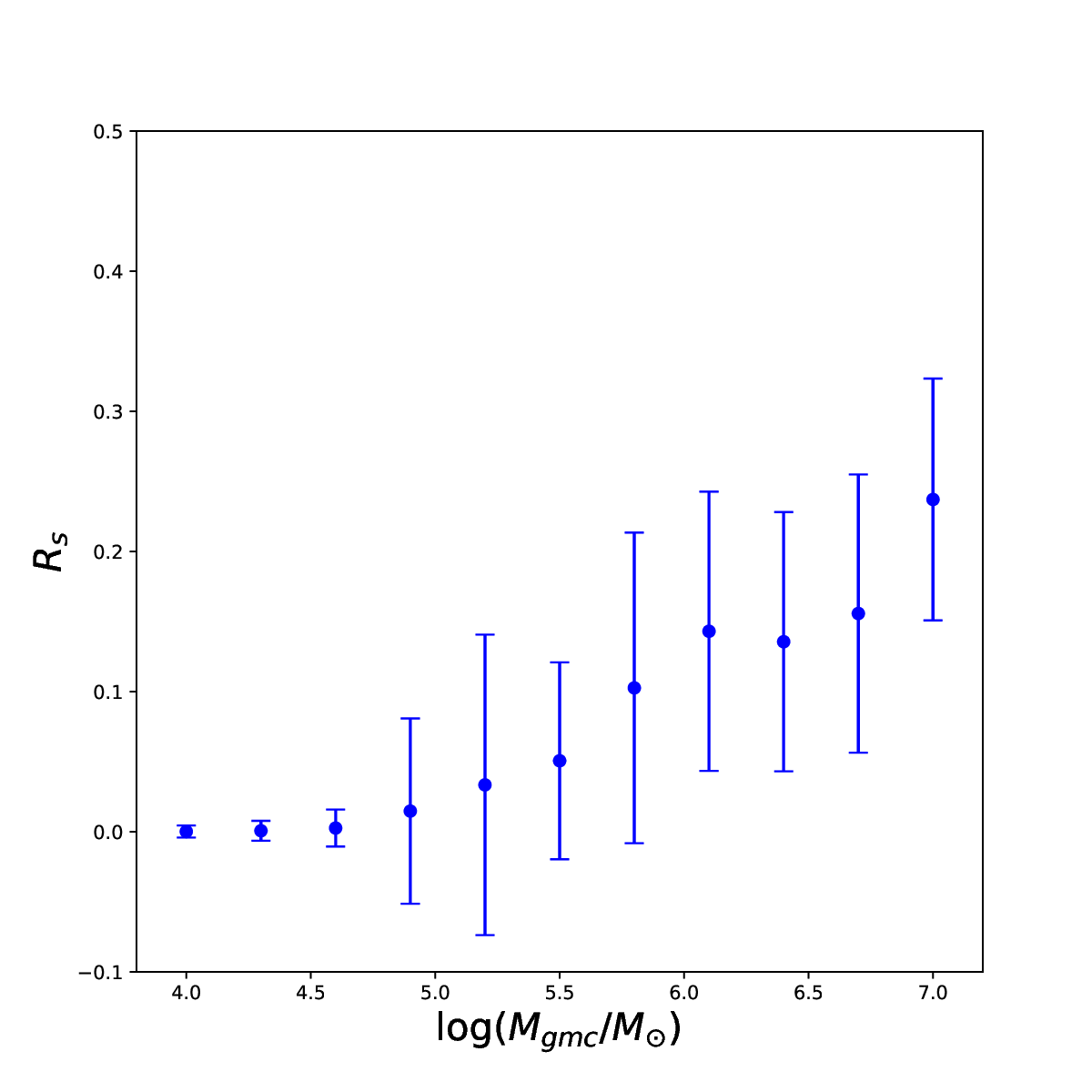,width=8.5cm} 
\caption{
$R_{\rm s}$ ($M_{\rm ns}/M_{\rm gmc}$)  as a function
of $M_{\rm gmc}$) in the selected five models.
The error bar indicates the 1$\sigma$ dispersion of $R_{\rm s}$ in each $M_{\rm gmc}$ bin.
This plot is based on $R_{\rm s}$ for GMCs selected from the models at different time
steps.
}
\label{Figure. 25}
\end{figure}

The mass ratios of AGB stars within and around GMCs to total GMC masses ($R_{\rm s}$) is a crucial parameter for GC formation with MPs in the SCI scenario.
Although simple analytical models can be used to estimate $R_{\rm s}$, as done in this study, numerical simulations of galaxy evolution with ${\rm H_2}$ formation on dust grains and new star formation within GMCs enable us to more accurately estimate $R_{\rm s}$ in galaxies of different masses and various types.
Using our original hydrodynamical simulations with ${\rm H_2}$ formation on dust grains, dust and metal enrichment, and star formation (Bekki 2015, 2025), we here investigate how $R_{\rm s}$ depends on $M_{\rm gmc}$ in simulated galaxies.
To do so, we first identify GMCs in each simulated galaxy at each time step, and then investigate $M_{\rm gmc}$ and the total mass of intermediate-mass stars ($M_{\rm ns}$) with $4 \le m/{\rm M}_{\odot} \le 10$ and ages ranging from 0.02 to 0.1 Gyr within $2R_{\rm gmc}$ from the center of each GMC, where $R_{\rm gmc}$ is the size of the GMC.
Since the total mass of young intermediate-mass stars can be a good proxy for $M_{\rm agb}$, we investigate $R_{\rm s}$ for each GMC in simulated galaxies.

The total masses of dark matter, stars, and gas in an initial gas-rich dwarf disk galaxy are denoted as $M_{\rm gal,dm}$, $M_{\rm gal, s}$, and $M_{\rm gal, g}$, respectively.
The density profile of the dark matter halo is represented by the NFW model with a central concentration of 12.0.
The stellar and gaseous disks are assumed to have exponential profiles, with the size being $R_{\rm gal,d}$ for both components.
The disk galaxy is assumed to have an initial gaseous metallicity of [Fe/H]=-1.6 and no radial gradient of [Fe/H].
The standard (fixed) Kroupa IMF with the three IMF slopes, $\alpha_1$, $\alpha_2$, and $\alpha_3$ being 1.3, 2.3, and 2.3, respectively, is adopted in estimating the total mass of intermediate-mass stars from the mass of each new stellar particle.
The size-mass relation of GMCs ($R_{\rm gmc} \propto M_{\rm gmc}^{0.5}$) adopted 
by Bekki \& Mackey (2009) is also used in selecting intermediate-mass stars around each GMC.
It should be stressed here that $M_{\rm ns}$ is a value estimated at a given time step (i.e., an instantaneous value), whereas $M_{\rm agb}$ is the total mass of AGB stars that have interacted with a GMC over the GMC's lifetime (i.e., an integrated value).
Therefore, $M_{\rm agb}$ is not exactly the same as $M_{\rm ns}$.

We mainly describe the results of the model in which $M_{\rm gal,dm}=10^{10} {\rm M}_{\odot}$, $M_{\rm gal,s}=1 \times 10^{8} {\rm M}_{\odot}$, $M_{\rm gal,g}=9 \times 10^{8} {\rm M}_{\odot}$, $c=12$, and $R_{\rm gal,d}=1$ kpc, and where the gas mass and size resolutions are $100 {\rm M}_{\odot}$ and 3 pc, respectively.
$M_{\rm gmc}$ and $M_{\rm ns}$ are estimated for each GMC at 6 selected time steps.
Fig. A1 shows the time evolution of the projected gas density in this gas-rich dwarf disk model for 0.1 Gyr.
Clearly, multiple high-density gaseous clumps are formed within the gas disk, where massive star clusters can form from the clumps.

Fig. A2 describes how $R_{\rm s}=M_{\rm ns}/M_{\rm gmc}$ depends on $M_{\rm gmc}$ in this model, in which 
many GMCs are identified at the selected time steps.
Here, $R_{\rm s}$ can be used to estimate the total mass of AGB stars that can contribute to the chemical enrichment of GMCs (see the main text for details).
Clearly, $R_{\rm s}$ is systematically higher in GMCs with larger $M_{\rm gmc}$, though the dispersion in $R_{\rm s}$ at a given $M_{\rm gmc}$ is not small.
This result can thus justify the adopted positive $M_{\rm gmc}$-$R_{\rm s}$ correlation in the SCI scenario (see Eqn. (22) in the main text).
It should be stressed here, however, that the slope of the $M_{\rm gmc}$-$R_{\rm s}$ correlation depends on the model parameters of the simulations.
For example, the slope becomes flatter in models with lower initial mean gas densities (e.g., due to lower gas mass fractions).
This means that the coefficients $a_{\rm s}$ and $b_{\rm s}$ in Eqn. (22) could be different in different galaxies.
Also, $R_{\rm s}$ becomes rather small ($<0.01$) in models with larger gas disk sizes due to lower gas densities, which means GCs cannot have MPs.
We will extensively discuss how $R_{\rm s}$ depends on model parameters for gaseous disks in dwarf galaxies in our forthcoming works based on numerical simulations of dwarf galaxies.

\section{Relations between $F_{\rm dil}$, $F_{\rm 1P}$,
and $\delta_{\rm max}Y$}

\begin{figure}
\psfig{file=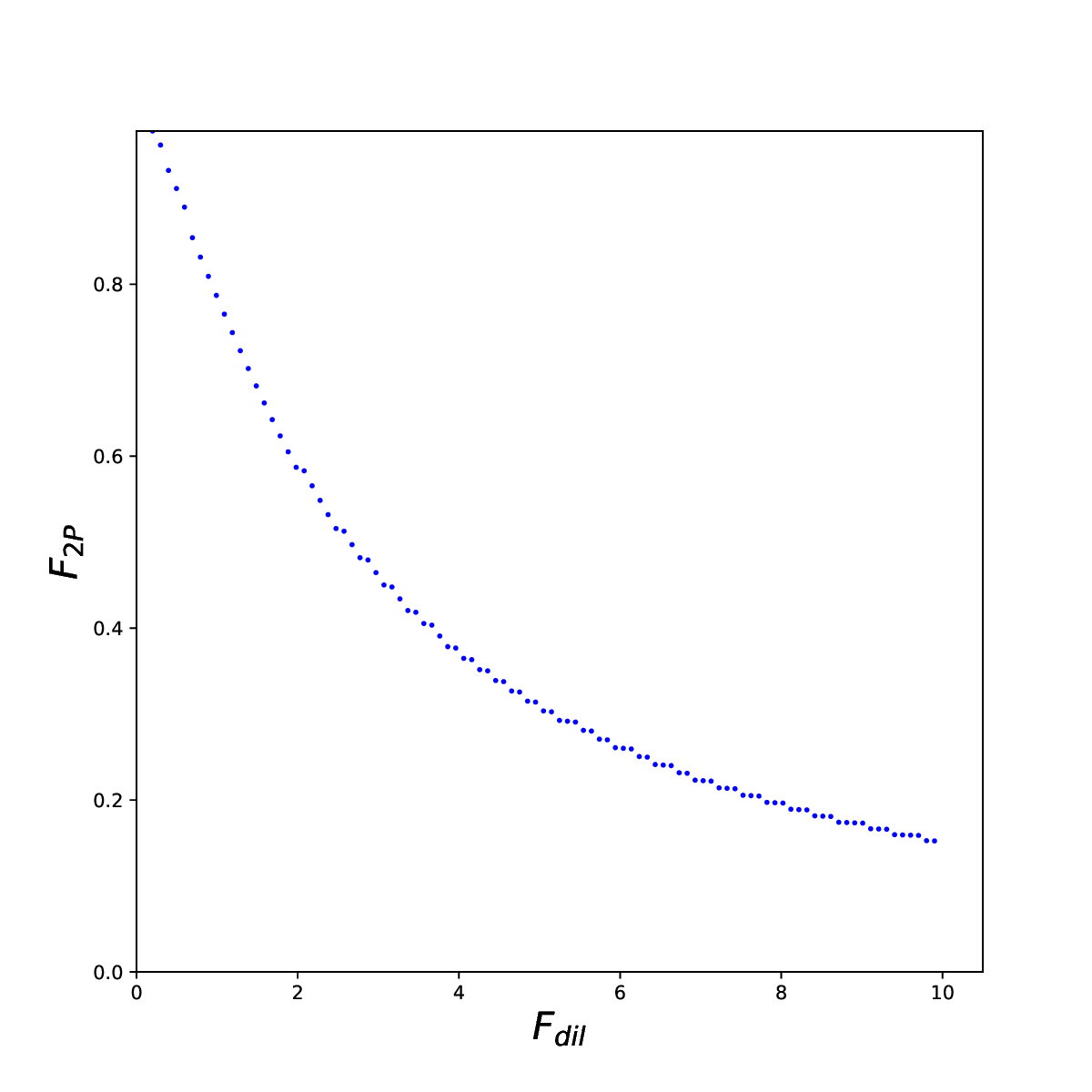,width=8.5cm} 
\caption{
$F_{\rm 2P}$ as a function of $F_{\rm dil}$ in a simple one-zone model
of chemical enrichment in GC formation. Data points at different 100 time steps
are plotted in this figure.
The anticorrelation between $F_{\rm dil}$ and $F_{\rm 2P}$ is a universal
trend in the adopted one-zone models, which enables this study
to develop a mathematical  formula for $F_{\rm dil}$-$F_{\rm 2P}$ relations adopted 
in the models for the observed $M_{\rm gc}$-$F_{\rm 1P}$ relation (see the main text).
}
\label{Figure. 26}
\end{figure} 

\begin{figure}
\psfig{file=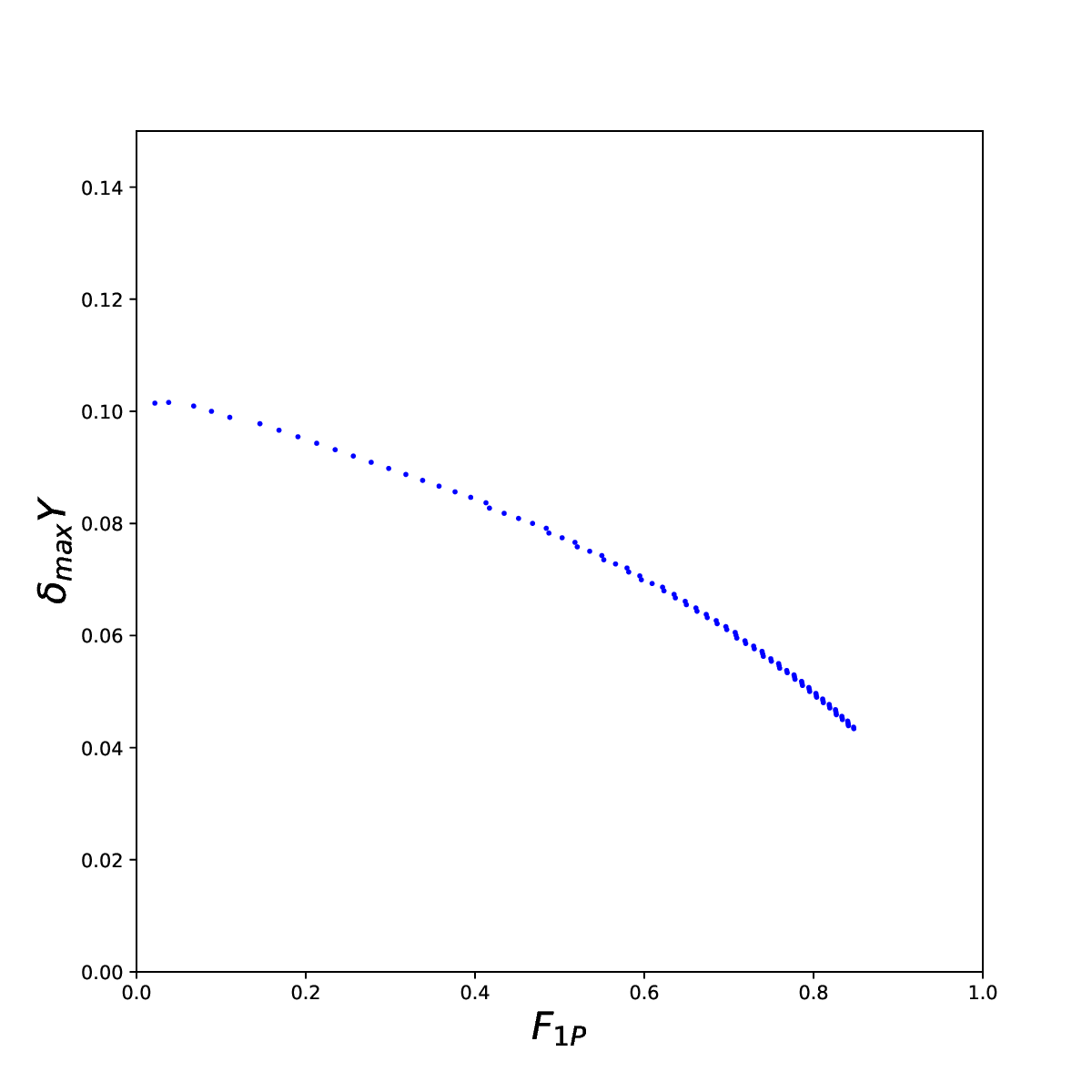,width=8.5cm} 
\caption{
Differences in $Y$ between new stars with minimum and maximum $Y$ ($\delta_{\rm max} Y$)
as a function of $F_{\rm 1P}$ in the one-zone chemical enrichment model for
GCs with MPs.
This $F_{\rm 1P}$-$\delta_{\rm max}Y$ relation is expected, because
$F_{\rm 1P}$ and $\delta_{\rm max}Y$ correlates and anticorrelated with 
$F_{\rm dil}$, respectively.
}
\label{Figure. 27}
\end{figure} 

Using an idealized dilution model for AGB ejecta, we can investigate
how the dilution factors ($F_{\rm dil}$) determine the 
fractions of 1P stars ($F_{\rm 1P}$), 
the differences between minimum and maximum
$Y$ ($\delta_{\rm max}Y$), and 
those between minimum and maximum
[Na/Fe] ($\delta_{\rm max}$[Na/Fe]=[Na/Fe]$_{\rm max}$-[Na/Fe]$_{\rm min}$) 
in GCs.
Here we assume that (i) the initial total mass of pristine GMC gas to be 
mixed with AGB ejecta is $M_{\rm g}$, (ii) the total mass of AGB ejecta
within a GC-forming GMC ($M_{\rm ej}(t)$) increases 
steadily with time, and (iii) the mixed gas
is converted into new stars with the star formation efficiency 
of $\epsilon_{\rm sf}$. 
We consider that the increase rate of $M_{\rm ej}(t)$ 
is simply fixed at  $M_{\rm ej, t}/T_{\rm sf}$,
where $M_{\rm ej, t}$ is the total mass of AGB ejecta and 
$T_{\rm sf}$ is the duration of star formation in GMCs:
$M_{\rm ej}(t)=M_{\rm ej,t}t/T_{\rm sf}$.
Therefore, the total mass of new stars formed from the mixed gas ($M_{\rm ns}$)
at each time step is as follows:
\begin{equation}
M_{\rm  ns}(t) = \epsilon_{\rm sf}(M_{\rm g} + M_{\rm ej}(t)).
\end{equation}
At each time step in each model, we calculate the total mass, $Y$, and [Na/Fe] of
new stars. We run the models with $0 < F_{\rm dil} < 1$ in order to find
the correlations of $F_{\rm 1P}$, 
$\delta_{\rm max}Y$, and 
$\delta_{\rm max}$[Na/Fe] with $F_{\rm dil}$.
Following C09, we consider that stars with [Na/Fe] less than
[Na/Fe]$_{\rm min}$+0.3 are classified as 1P.

Fig. B1 clearly demonstrates that $F_{\rm dil}$ is a crucial parameter
that determines $F_{\rm 1P}$ for 
$\epsilon_{\rm sf}$=0.03. As expected,
$F_{\rm 1P}$ is larger for larger $F_{\rm dil}$, because most of new stars
formed from AGB ejecta mixed with pristine GMC gas  have lower [Na/Fe]
to be classified as 1P. It is also found that
both $\delta_{\rm max}$Y and $\delta_{\rm max}$[Na/Fe] are larger for smaller $F_{\rm dil}$.
Fig. A2 shows that 
$\delta_{\rm max}Y$  
is larger for smaller $F_{\rm dil}$,
because the later formed 2P stars can have larger $Y$
in the models with smaller $F_{\rm dil}$.
These clear correlations and anticorrelations of 
$F_{\rm 1P}$
and $\delta_{\rm max}Y$
with $F_{\rm dil}$
can be used to discuss how the total masses of
GMCs,  which determine $F_{\rm dil}$, can influence 
$F_{\rm 1P}$ and
$\delta_{\rm max}Y$) 
in the SCI scenario.
These results also justify the adopted relation between $F_{\rm dil}$
and $F_{\rm 1P}$ in the preset models (see \S 2).
 
\section{Possible correlations of $F_{\rm 2P}$  with ages and metallicities
of GCs}

The SCI scenario predicts that $F_{\rm 2P}$ can be larger in GCs forming in
galaxies with larger $\rho_{\rm agb}$ (and thus larger $\Sigma_{\rm SFR}$). Therefore,
if $\Sigma_{\rm SFR}$ is larger for galaxies with older ages and lower metallicities,
then the observed relations of $F_{\rm 2P}$ with ages and metallicities (ME20)
can be qualitatively explained by the scenario.
In order to briefly discuss whether $F_{\rm 2P}$ can depend on
the ages and metallicities of GCs, we here investigate how $\Sigma_{\rm SFR}$ depends on
the ages and metallicities of GC-forming galaxies 
by using our one-zone chemical evolution models of galaxies (BT12).
We adopt the ``wind model`` of BT12, in which gas can be ejected from galaxies by supernova feedback effects
through galactic winds, and about 30\% of the gas can be expelled from galaxies: the details
of the wind model are given in B12.
We assume that (i) the initial [Fe/H] of the gas is -4,
(ii) the gas infall time scale ($\tau_{\rm inf}$) is a free parameter,
and (iii) the half-mass radius of a gas-rich dwarf galaxy, $R_{\rm h}$, is 200 pc.
We investigate the time evolution of the SFR, [Fe/H], and $\Sigma_{\rm SFR}$
over 3 Gyr (from $T=0$ to 3 Gyr) in models with different $\tau_{\rm inf}$
in which the final stellar masses are [3-6]$\times 10^8 {\rm M}_{\odot}$, corresponding
to gas-rich dwarf galaxies.

Fig. C1 describes how $\Sigma_{\rm SFR}$ evolves with [Fe/H]
in the model with $\tau_{\rm inf}=10^8$ yr.
As the total gas mass increases with time through gas infall, both [Fe/H]
and $\Sigma_{\rm SFR}$ increase with time, which results in a positive
correlation between $\Sigma_{\rm SFR}$ and [Fe/H].
Then $\Sigma_{\rm SFR}$ reaches its peak at [Fe/H]$\approx-2.1$,
when the total gas mass starts to decrease due to gas consumption by star formation.
After its peak, $\Sigma_{\rm SFR}$ decreases with increasing [Fe/H] almost linearly.
Since a lower $\Sigma_{\rm SFR}$ means a lower $F_{\rm 2P}$, 
this result implies that $F_{\rm 2P}$ can be lower for higher [Fe/H] 
in GCs with [Fe/H]$>-2.1$.
It is confirmed that this anticorrelation between [Fe/H] and $\Sigma_{\rm SFR}$
can be seen in other models with different $\tau_{\rm inf}$, though the
[Fe/H] values at the $\Sigma_{\rm SFR}$ peak are different between these models.

Fig. C2 shows the time ($T$) evolution of $\Sigma_{\rm SFR}$ in the model
with $\tau_{\rm inf}=10^8$ yr. After reaching its peak at $T=0.3$ Gyr,
$\Sigma_{\rm SFR}$ starts to decrease with increasing $T$. 
This means that there should be a positive correlation between ages and $F_{\rm 2P}$ of GCs formed
after $T \approx 0.2$ Gyr in this model.
It is also clear that this age-$F_{\rm 2P}$ correlation after the $\Sigma_{\rm SFR}$
peak can be seen in other models with different $\tau_{\rm inf}$.
These results imply
that younger GCs are likely to have smaller $F_{\rm 2P}$,
because $F_{\rm 2P}$ can correlate with $\Sigma_{\rm SFR}$.
It is possible that GCs cannot be formed due to much smaller gas masses and densities
in the very early phases of galaxy formation before $\Sigma_{\rm SFR}$ peaks.
Although we adopt idealized one-zone models in these discussions, the models
can grasp some essential ingredients
of the physical processes that can possibly determine $F_{\rm 2P}$ in GC formation within galaxies.
It is doubtlessly our future work to investigate the origins of the possible correlation and anticorrelation
of $F_{\rm 2P}$ with the ages and metallicities of GCs using more sophisticated models of GC formation,
ideally  numerical simulations of GCs within high-$z$ galaxies.

\begin{figure}
\psfig{file=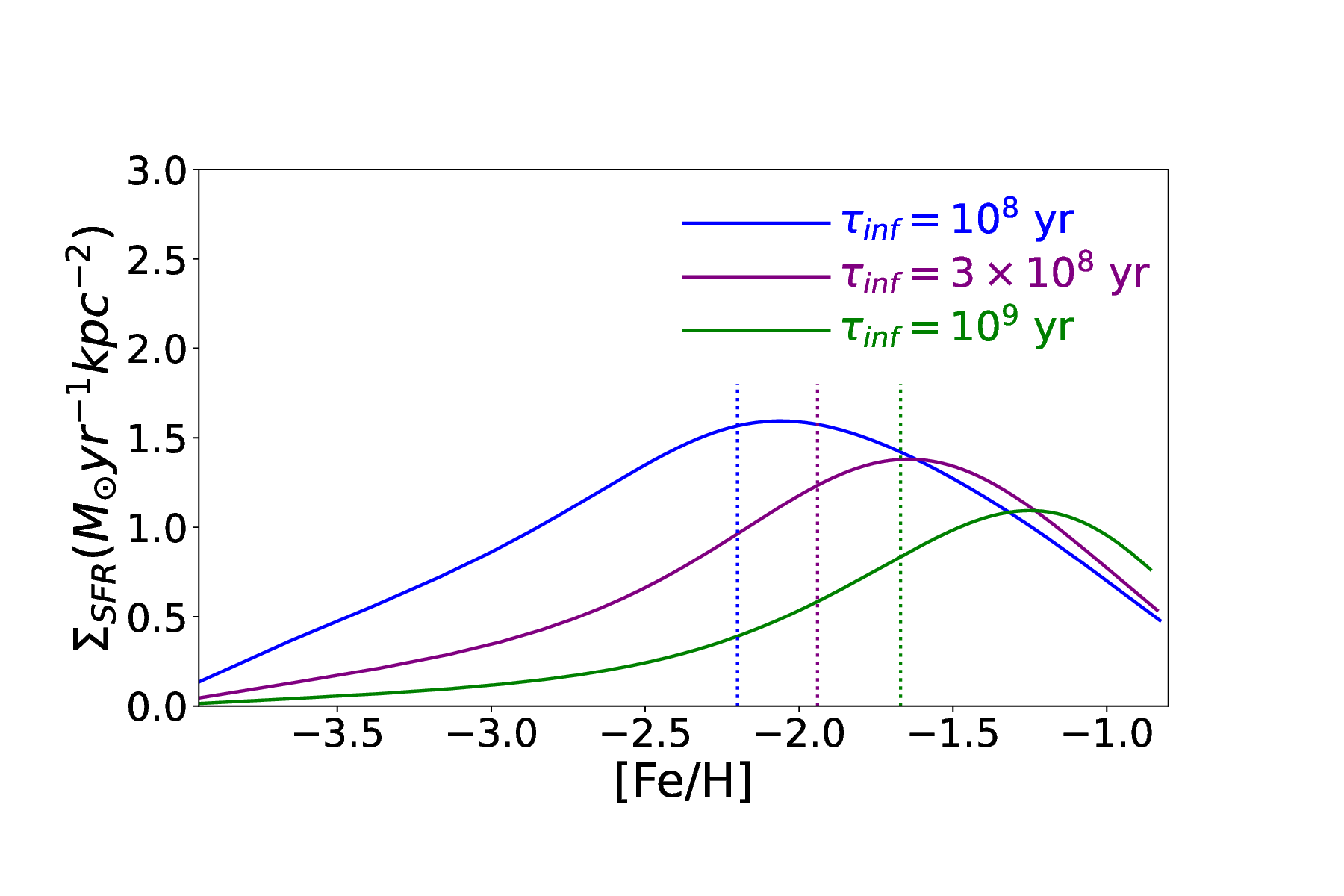,width=8.5cm} 
\caption{
$\Sigma_{\rm SFR}$ as a function of  [Fe/H] in one-zone chemical evolution models
for a gas-rich dwarf with $\tau_{\rm inf}=10^8$ yr (blue),
$\tau_{\rm inf}=3 \times 10^8$ yr (purple),
and $10^9$ yr (green).
A fixed half-mass radius (200pc) is assumed to calculate $\Sigma_{\rm SFR}$ at
different time steps in
these models.
The vertical dotted line for each model  indicates [Fe/H] when the total mass of new
stars ($M_{\rm ns}$) becomes $3 \times 10^7 {\rm M}_{\odot}$.
The result for 
the  dwarf model with $\tau_{\rm inf}=10^8$ yr accordingly implies that
if GCs are formed only after $M_{\rm ns} \ge 3 \times 10^7 {\rm M}_{\odot}$,
then the GCs with higher [Fe/H] can have lower $F_{\rm 2P}$ due to lower
$\Sigma_{\rm SFR}$. 
These results of the adopted idealized models can be used in the physical
interpretations of the observed relation between [Fe/H] and $F_{\rm 2P}$ and by ME20.
}
\label{Figure. 28}
\end{figure}

\begin{figure}
\psfig{file=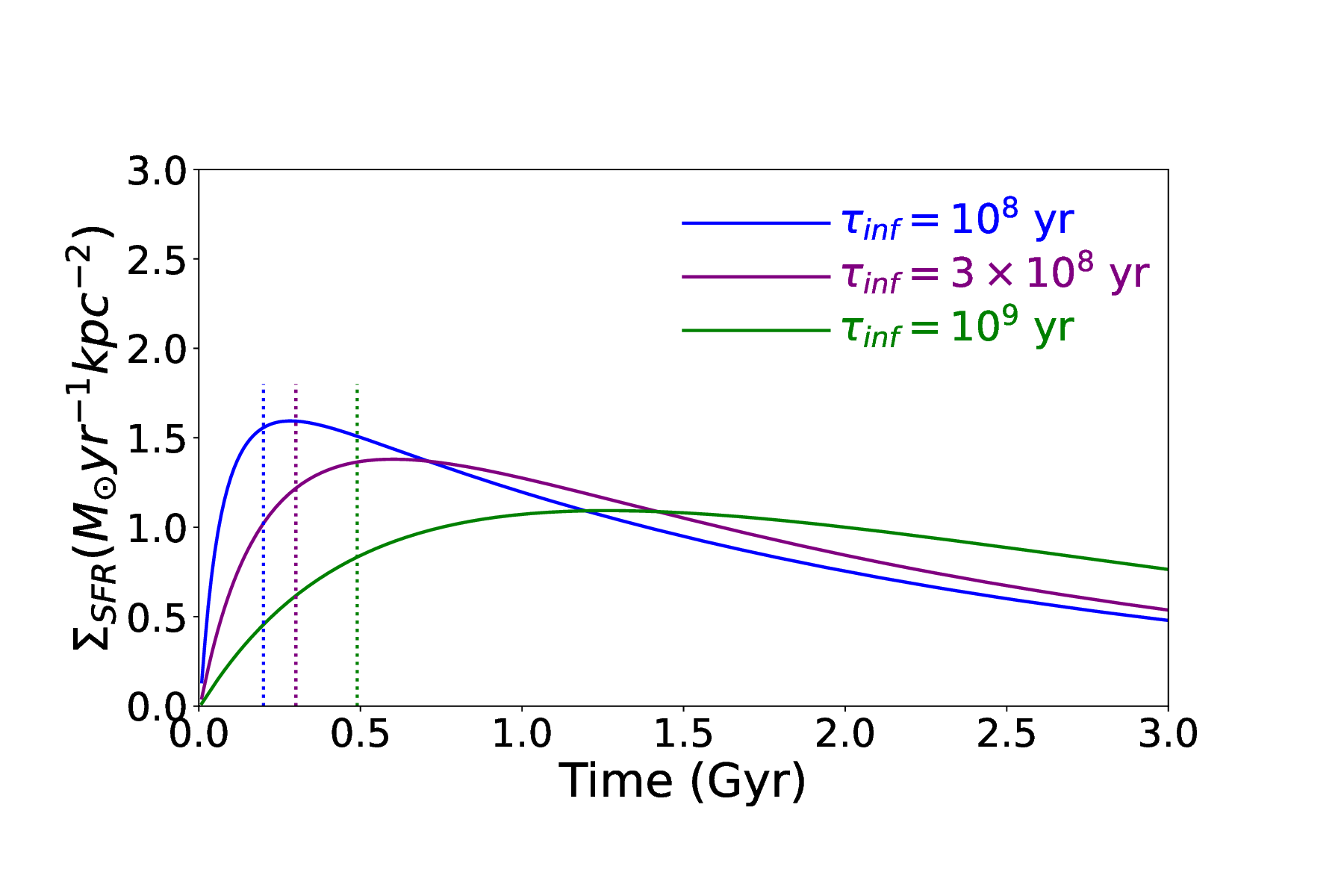,width=8.5cm} 
\caption{
$\Sigma_{\rm SFR}$ as a function of time ($T$) in the same models as adopted
in Fig. C1.
The vertical dotted line for each model  indicates $T$  when the total mass of new
stars becomes $3 \times 10^7 {\rm M}_{\odot}$.
These results can be used in discussing the observed 
relation between ages of GCs and $F_{\rm 2P}$ (ME20).
For example, if GCs can be formed only after $T=0.2$ Gyr in dwarfs
with $\tau_{\rm inf} \approx 10^8$ yr, 
th GCs with older ages are likely to have higher $F_{\rm P}$.
}
\label{Figure. 29}
\end{figure}

\section{A sweet spot for chemical enrichment by massive  AGB stars}

In the SCI scenario, only massive AGB stars ($4 \le m_{\rm agb}/M_{\odot} \le 10$) must 
chemically pollute pristine gas to form MPs that exhibit correlations and anticorrelations (e.g., O-Na) between various elements. This implies that GCs with MPs cannot form during the 
later phases of galaxy formation, when low-mass AGB stars—which outnumber high-mass ones—begin to pollute GMCs. Therefore, the present study predicts a "sweet spot" where only massive AGB stars pollute GMCs 
within their host dwarf galaxies. Using the one-zone chemical evolution models adopted by BT12, we investigate the total mass of stars ($M_{\rm TO}$) that have just left the main sequence at a given time step ($T$) as a function of the turn-off mass ($m_{\rm TO}$) in our galaxy models.

Fig. D1 shows $M_{\rm TO}$ as a function of $m_{\rm TO}$ in models with $C_{\rm SF}=0.1$
for IMF slopes of $\alpha=2.35$ (Salpeter) and $1.5$ (top-heavy) at 
$T=100$, $300$, and $800$ Myr. Here, $M_{\rm TO}$ is normalized to the total mass of infalling gas (which is set to unity) in each model, and $M_{\rm TO}$ is calculated every Myr. Obviously, $m_{\rm TO}$ 
decreases as $T$ increases, meaning that stars with lower masses can chemically enrich GMCs during later phases of galaxy evolution.

Although GMCs are polluted exclusively by massive AGB stars ($m_{\rm agb} \ge 4.5 {\rm M}_{\odot}$) 
at $T=100$ Myr, they can be polluted by all AGB stars with $m_{\rm agb} \ge 3 {\rm M}_{\odot}$ at $T=300$ Myr. 
At $T=800$ Myr, the dominant stellar population polluting GMCs consists of $m_{\rm agb} \approx 2 M_{\odot}$ AGB stars. This implies that 2P stars formed in these GMCs would have enhanced C+N+O abundances, which is a feature not observed in most 2P stars. These results suggest that GCs with MPs can form well before $T=800$ Myr in 
steadily star-forming galaxies. Since AGB stars with 
low $m_{\rm agb}$ begin to pollute GMCs after $T=800$ Myr, 
it is  unlikely that GCs with MPs can form $\approx 800$ Myr after the onset of star formation in 
steadily star-forming galaxies.

These results are independent of $\alpha$ across all six models. However, models with a top-heavy IMF yield a larger $M_{\rm TO}$ for massive AGB stars, implying that GMCs can be more heavily polluted by massive AGB stars, ultimately leading to a larger $F_{\rm 2P}$. If a secondary starburst occurs a few to several Gyr after the onset of star formation in a galaxy, a sharp peak in $M_{\rm TO}$ could emerge around $m_{\rm TO} \approx 5 M_{\odot}$ tens of Myr after the starburst. Depending on the strength of this starburst, massive SCs formed during the event could potentially host MPs.

\begin{figure}
\psfig{file=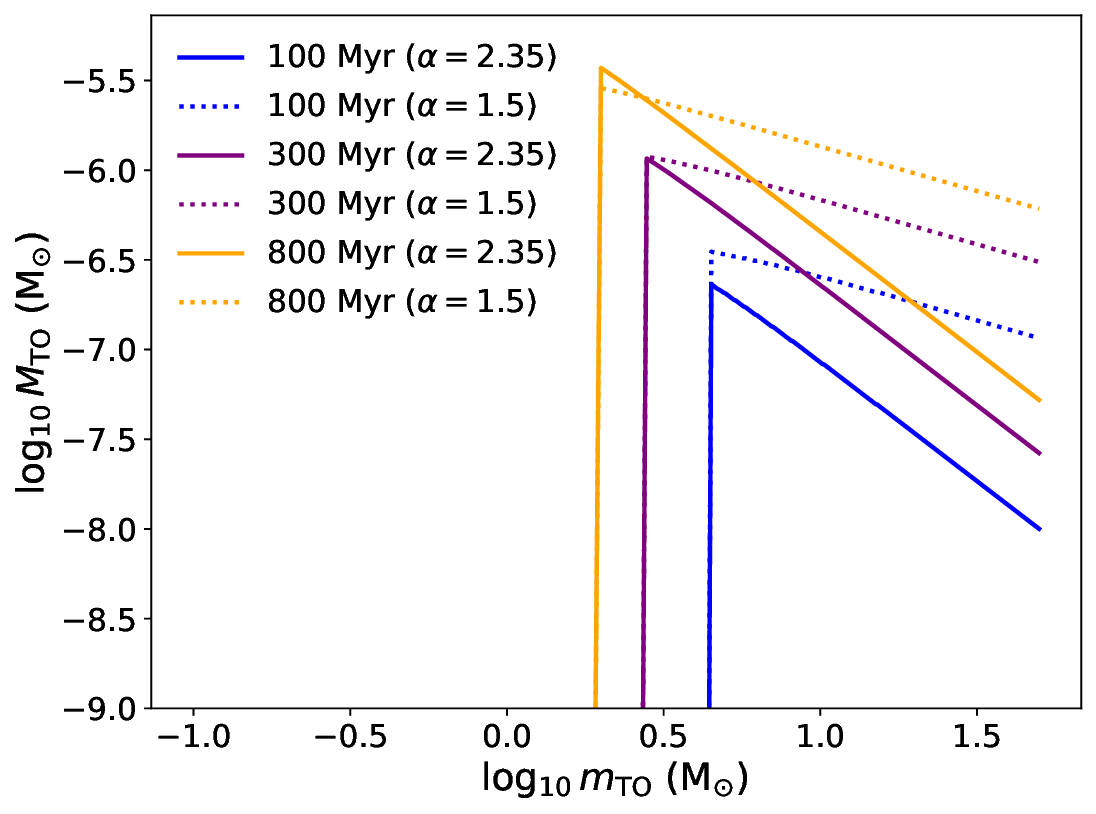,width=8.5cm} 
\caption{
$M_{\rm TO}$ as a function of the turn-off mass ($m_{\rm TO}$) at $T=100$ 
Myr (blue), 300 Myr (purple), and 800 Myr (orange) for the six models with $\alpha=2.35$ (solid lines) and 
1.5 (dotted lines). 
The total mass of stars that have just left the main sequence ($M_{\rm TO}$) at a given time step
is calculated every Myr in these models.
}
\label{Figure. 30}
\end{figure}

\end{document}